\documentclass{aa}  
\usepackage{graphicx}
\usepackage{txfonts}
\begin{document}
   \title{The Nature of Mid-Infrared Excesses From Hot Dust Around
   Sun-like Stars}

   \author{R. Smith
          \inst{1,2}
          \and
          M. C. Wyatt \inst{1} 
          \and 
	  W.R.F. Dent \inst{3}  }

   \offprints{R. Smith}

   \institute{Institute of Astronomy, University of Cambridge,
	     Madingley Road, Cambridge CB3 0HA, UK \\
             \email{rsed@ast.cam.ac.uk}\\
        \and
              Institute for Astronomy, Royal Observatory Edinburgh,
              Blackford Hill, Edinburgh EH9 3HJ, UK\\
        \and
             U.K. Astronomy Technology Centre, Royal Observatory Edinburgh,
             Blackford Hill, Edinburgh EH9 3HJ, UK}

   \date{Accepted : 25th April 2008 }

 
  \abstract
   {}
   {Studies of the debris disk phenomenon have shown that most systems
  are analogous to the Edgeworth-Kuiper Belt (EKB). However 
  a rare subset of sun-like stars
  possess dust which lies, in contrast, in the terrestrial planet
  region.  In this study we aim to determine how many sources with
  apparent mid-infrared excess are truly hosts of warm dust, and
  investigate where the dust in these systems must lie. }
   {We observed using ground-based mid-infrared imaging with TIMMI2,
  VISIR and MICHELLE a sample of FGK main
  sequence stars previously reported to have hot dust.  A new
  modelling approach was developed to determine the constraints that
  can be set on the radial extent of excess emission in such
  observations by demonstrating how the detectability of a disk of a
  given flux as a fraction of the total flux from the system
  ($F_{\rm{disk}}/F_{\rm{total}}$) depends primarily on the ratio of disk
  radius to PSF width and on the uncertainty on that PSF width.}
   {We confirm the presence of warm dust around three of the
  candidates; $\eta$ Corvi, HD145263 and HD202406. For $\eta$ Corvi
  modelling constrains the dust to lie in regions smaller than
  $\sim$3.5 AU. The modelling constrains the dust to regions
  smaller than 80-100AU for HD145263 and HD202406, with SED fitting
  suggesting the dust lies at a few tens of AU. Of two alternative
  models for the $\eta$ Corvi excess emission, we find that a model
  with one hot dust component at less than 0\farcs164 ($<$3 AU)
  (combined with the known submm dust population at $\sim$ 150 AU)
  fits all the data better at the 2.6 $\sigma$ level than an
  alternative model with two populations of dust emitting in the
  mid-infrared: hot dust at less than 0\farcs19 ($<$ 3.5 AU)
  and a mid-temperature component at $\sim$ 0\farcs66 (12 AU). We
  identify several systems which have a companion (HD65277 and
  HD79873) or background object (HD53246, HD123356 and HD128400)
  responsible for their mid-infrared excess, and for three other
  systems we were able to rule out a point-like mid-infrared source
  near the star at the level of excess observed in lower resolution
  observations (HD12039, HD69830 and HD191089).}
   {Hot dust sources are either young and possibly primordial or
     transitional in their
  emission, or have relatively small radius steady-state planetesimal
  belts, or they are old and luminous with transient emission. High
  resolution imaging can be used to constrain the location of the disk
  and help to discriminate between different models of disk
  emission. For some small disks, interferometry is needed to resolve
  the disk location. }

   \titlerunning{The Nature of Mid-Infrared Excesses Around
   Sun-like Stars}
   \authorrunning{R. Smith et al.}
   \maketitle

%

\section{Introduction}

Analysis of the IRAS database over the last 20 years has shown that
there are over 300 main sequence stars that have dust disks around
them.  This material is thought to be the debris left over at the end
of the planet formation process (e.g. Mannings and Barlow 1998). 
The spectral energy
distribution (SED) of this excess in the best studied cases 
(e.g., Vega, $\beta$ Pictoris, Fomalhaut, $\epsilon$ Eridani)
peaks longward of 60$\mu$m implying that this dust is cool ($<$80K), and so
resides in Edgeworth-Kuiper belt (EKB)-like regions in the systems. 
The EKB-like location and analogy is confirmed in the few cases where
these disks have been resolved (e.g., Holland et al. 1998, Greaves
et al. 2005, see also scattered-light imaging, e.g. Kalas et
  al. 2007), since the dust is shown to lie $>$40AU from the stars, and
its short lifetime means that it must be continually replenished by
the collisional destruction of km-sized planetesimals (Wyatt \& Dent
2002). The inner 40AU radius hole is thus thought to arise from
clearing by an unseen planetary system, the existence of which is
supported by the presence of clumps and asymmetries seen in the
structure of the dust rings (e.g., Wyatt et al. 1999; Wyatt 2003). Of
the four archetypal objects, only
$\beta$ Pictoris also has (a relatively small amount of) resolved dust in this
inner region (Lagage and Pantin 1994, Telesco et al. 2005), 
thought to be there because this is
a young (12Myr, Zuckerman et al. 2001) transitional system in which
these regions have yet to be fully cleared by planet formation
processes. However Absil et al. (2006) have recently presented
interferometric data showing Vega (thought to be around 380-500Myr old,
Peterson et al. 2006) is likely to possess extended dust
emission within 8AU, and Di Folco et al. (2007) have also recently
presented evidence for hot dust around the 10Gyr old $\tau$ Ceti.

Zuckerman (2001) noted that out of a large sample of main
sequence stars that exhibit excess mid-infrared emission (Mannings and
Barlow 1998) half of these showed an excess at 25 $\mu$m only.  While
this seems to suggest a large fraction of systems host warm dust at a
few AU, in fact the vast majority of such warm dust systems are around
the more luminous A and B stars (see e.g. Rieke et al. (2005),
Beichman et al. (2006)), and so despite the fact 
that the dust is warm, it usually still resides at 10s of AU in
regions analogous to the EKB.  In contrast hot dust seems to be rare
around the less luminous F, G and K-type stars.
Four surveys have searched for hot dust around such stars by
examining the emission at 25$\mu$m above photospheric levels as measured
by IRAS (Gaidos 1999), ISO (Laureijs et al. 2002) and Spitzer (Hines
et al. 2006; Bryden et al, 2006).  All these surveys found evidence
for hot dust with fractional luminosity $f = L_{IR}/L_\ast > 10^{-4}$
in only $2 \pm 2$\% of stars (for comparison the luminosity of the
zodiacal cloud is $10^{-8}-10^{-7} L_\odot$, Dermott et al. 2002). 
These systems
could represent a departure from the canonical picture that extrasolar
debris systems are analogous to our own Kuiper belt, since the
temperature of the dust implies that most lies in the region
2-20AU. Thus this dust is predicted to lie at distances from their
stars that would be between our asteroid and Kuiper belts, and so in
the region where we expect gas giant planets to form - just where we
expect no dust. These disks pose several fundamental questions about
the outcome of planet formation in these systems. Are these the Kuiper
belts of systems in which planet formation failed beyond
$\sim$ 10AU (e.g., due to a stellar flyby, Larwood and Kalas 2001, or the
rapid dispersal of the protoplanetary gas disk, Hollenbach et al. 2000)?
Or are we witnessing the collisional destruction of massive asteroid
belts or the sublimation of comets in the middle of fully formed
planetary systems? Or dust from a more distant belt trapped in
resonance with a giant planet (Moran et al. 2004)? Or perhaps these are
systems in a transitional (mid-planet formation) stage (Kenyon \&
Bromley 2002)?

To begin to tackle these issues, we need to know the true dust
distribution in these systems. This can be determined from SED fitting
to multi-wavelength infrared photometry, and from constraints provided
by resolved imaging.  There are several
uncertainties regarding these putative disks, in addition to the
temperature of the dust emission.  Most importantly, the excesses
taken from the IRAS database cannot be used at face value.  It was
noted by Song et al. (2002), who searched the IRAS database for excess
emission towards M-type stars, that when searching a large number of
stars for excesses close to the detection threshold, a number of false
positives must be expected due to noise.  Also, there have been a few
instances in which the IRAS excess has been shown to be attributed to
background objects that fall within the relatively large IRAS beams
($>$30''). Such objects range from highly reddened carbon stars or
Class II YSO's \cite{lisse}, to distant galaxies
\cite{sheret}. Another possible source of mid-infrared excess
emission is reflection nebulosity (Kalas et al. 2002). 
Indeed it is now routine for papers discussing the excess sources
found by IRAS to address the possibility that some of these are bogus
debris disks (Moor et al. 2006, Rhee et al. 2007). Thus
it is imperative that we determine if the excesses are real and
centred on the stars.

This paper is structured as follows: In \S \ref{s:sample} the sample
selection is described. In \S \ref{s:obs} the various observational
and analysis techniques employed for each instrument are outlined, and
in section \S \ref{s:ext} a new method of placing quantifiable
extension limits on unresolved disk images is described.   
The results, analysis and discussion of individual sources 
are presented in \S \ref{s:res}, and the implications of these results
discussed in \S \ref{s:disc}.  Conclusions are in \S \ref{s:conc}.

%

\section{The Sample}
\label{s:sample}

The sample consists of F, G and K stars with IRAS published detections
of excess emission at 12 and/or 25 $\mu$m. \footnote{The sample stars
  are listed in the Debris Disk Database at
http://www.roe.ac.uk/ukatc/research/topics/dust.}
A first-cut was applied to the list of all published detections 
consisting of the following analysis to determine if the
excess identified by IRAS was likely to be real.

For all stars in the sample  J, H and K fluxes are obtained from 2MASS
and V and B magnitudes from Tycho2.  The Michigan Spectral Catalogues
or SIMBAD were used to 
determine the stellar spectral type.  This was used to model the
photospheric emission based on a Kurucz model atmosphere appropriate
to the spectral type and scaled to the K band flux. This allowed
determination of the photospheric contribution to the emission
assuming there is no excess emission at K.

The IRAS fluxes were taken from the Faint Source Catalogue,
and the Point Source Catalogue when FSC fluxes were not available
(i.e. for sources in the Galactic plane). This information was then
compared with fluxes extracted using SCANPI (the Scan Processing and
Integration tool) \footnote{http://scanpi.ipac.caltech.edu:9000/}
which results in much reduced errors.  This tool scans the raw IRAS
data and averages individual scans to determine the point source flux
and error of the object in question (as determined by coordinates) in
each of the IRAS bands (12, 25 60 and 100 $\mu$m). 
The fluxes using different extraction methods could thus be analysed
to give an independent determination of the significance of any excess
measured to see if e.g., problems with background subtraction were
affecting the results.  Colour-correction was applied to the fluxes at
the levels described in the IRAS Explanatory Supplement \footnote{The
  IRAS Explanatory Supplement is available at
  http://irsa.ipac.caltech.edu/IRASdocs/exp.sup/}. Specifically
colour-correction applied to the 12, 25, 60 and 100 $\mu$m fluxes was
1.43, 1.40, 1.32 and 1.09 respectively.  For stars with effective
temperatures greater than 7000K (as determined by Kurucz profile
fitting), colour corrections of 1.45 and 1.41 at 12 and 25 $\mu$m
respectively were applied. The colour-correction was applied only to
the stellar component of emission, through multiplication of the
expected stellar flux by the colour-correction factor before
subtraction to determine the excess emission. No further
colour-correction was applied to the excess emission.  The proximity
of the IRAS sources to the stars was also checked given the quoted
uncertainty error ellipse, since some surveys allowed excess sources
to be up to 60 arcsec offset and have since been shown to not be related
(Sylvester and Mannings 2000).

\begin{table*}
\caption{The Sample}
\label{sample}
\centering                          
\begin{tabular}{|c|c|c|l|l|l|} \hline Star name & Stellar type & Age &
Distance & \multicolumn{2}{|c|}{IRAS fluxes (mJy)$^a$} \\ 
HD &  & Gyr & pc & 12$\mu$m & 25$\mu$m  \\ \hline
10800 & G1/2V & 7.6$^b$ & 27.1 & 479 + 15 (20) & 113 + 82 (18)  \\
12039 & G3/5V & $0.03^c$ & 42.4 & \multicolumn{2}{|c|}{Not in IRAS
  database$^c$} \\
53246$^d$ & G6V & O(0.1)$^e$ & 36.5 & 82 + 293 (30) & 19 + 143 (26)  \\
65277$^d$ & K4V & 4.2$^f$ & 17.5 & 184 - 46 (27) & 43 + 83 (29)  \\
69830 & K0V & 2$^g$ & 12.6 & 603 + 77 (26) & 142 + 171 (33)  \\
79873$^d$ & F3V & 1.5$^b$ & 68.9 & 157 - 21 (25) & 37 + 95 (38)  \\
$\eta$ Corvi$^h$ & F2V & 1.3$^b$ & 18.2 & 1212 + 412 (42) & 283
+ 420 (50)  \\
123356$^d$ & G1V & O(0.1)$^e$ & 20.9 & 14 + 1270 (53) & 3 + 615 (56)  \\
128400 & G5V & 0.3$^i$ & 20.4 & 260 + 178 (24) & 61 + 64 (23)  \\
145263 & F0V & 0.009$^j$ & 116.3 & 19 + 422 (50) &
4 + 583 (35)  \\
191089 & F5V & 0.1$^k$ & 53.5 & 101 - 34 (29) & 24 + 287 (55)  \\
202406 & F2IV/V & 0.002 & 429.2 & 53 + 233 (33) & 13 + 272 (48)  \\ \hline
\end{tabular}

Notes: $^a$=Fluxes are shown as star +
 excess (error), for HD65277 the 12 $\mu$m IRAS photometry suggests a
 lower flux than is expected from the photosphere, and so the excess
 is shown as negative;  $^b$=Age taken from Geneva-Copenhagen Survey; $^c$=
 Identified as having excess by Hines et al. 2006; 
$^d$= Binary object, see individual object descriptions, section \S
 \ref{s:res}; 
$^e$=Age estimated by placing on colour-magnitude diagram following
 the work of Song et al. 2000;
$^f$=Age taken from Valenti \& Fischer (2005);
$^g$=Beichman et al. (2006); 
$^h$=HD 109085 also has excess at 60 and 100 $\mu$m;
$^i$= Age from Gaidos (1999);
$^j$=Honda et al.(2004); $^j$=Age from Zuckerman \& Song (2004). 
\end{table*}

The final sample consisted of 11 stars of spectral types F, G and K  
and these are listed in Table \ref{sample}. HD12039, not included in
the IRAS catalogues, was identified as a warm dust host by Hines et
al. (2006), and included in the later stages of this study.
  
%

\section{Observations and Data Reduction}
\label{s:obs}

\subsection{Observations}

The observations were performed using a combination of: the Thermal
Infrared MultiMode Instrument TIMMI2 on the ESO 3.6m telescope at La
Silla; VISIR, the VLT Spectrometer and imager for the mid-infrared
on the ESO VLT; and MICHELLE on Gemini North.

All of the observations employed a chop throw of 10\arcsec in the
North-South direction (except for the MICHELLE observations for which
the chop is 15\arcsec, and the chop throw was at 30$^\circ$). 
Telescope and sky emissions were removed by an additional nod throw of
the same size, taken in the perpendicular direction for TIMMI2 and
VISIR, and in the parallel direction to the chop for the MICHELLE
observations. 

For the observations performed in perpendicular mode, this means that
a straight co-addition of the data results in an image with two
positive and two negative images of the source.  The parallel chop-nod
technique results in one central positive image and one negative image
at half the level of the central image on either side in the throw
direction.  A residual dc (dark-current) offset was removed by
subtracting the median 
in each column of the array and then in each row (the areas around the
source images are masked off when determining these medians).  The
resulting images showed statistical uncertainty varying by just a few
percent across the central 20 square arcsecond region around the
images for all instruments.  Bad pixels were determined by
looking at the variations in individual chop frames, first creating `empty'
images in which only the half of a frame not containing the source
would be used, together with the opposite half of the frame from the
following nod position (which would also be empty). Pixels with a
variance across the frames of 10 times more than the average were
labelled `bad' and 
masked off. Regions towards the edge of the array were
found to be particularly prone to such variations, and were masked
more frequently. Typically this stage would remove a few percent of
pixels ($\sim 1000$, array 320x240 or 256x256).  This was also used to
determine the variation of the sky during the observation, and in turn
to determine the responsivity of individual pixels, so creating a gain
map (in a perfect detector gain for all pixels would be 1). Note that
in determining the gain map the regions on which the source emission
fell on the detector would be masked off, as due to the chop and nod
pattern the pixels would be unevenly illuminated in different nod
frames and this would lead to inaccuracies in determining the gain
map. Any pixels showing a particularly high or low gain ($<2/3$ or
$>3/2$) were masked off.  This would on average remove a few tens of
pixels in addition to the previous masking.  In total an average of
around 7\% of pixels were removed in the TIMMI2 observations, and
around 4\% of pixels in the MICHELLE and VISIR observations. This
level was much reduced within the on-source apertures used to
$\lesssim 1$\%,  as most of the problem pixels were confined to the
edges of the arrays, or to other regions which were avoided when
deciding where to have the objects' images on the array.  

In order to
minimise the effects of changing conditions and airmass, calibration
observations were taken of standard stars within a few degrees of
the science object, immediately before and after each science
observation whenever time constraints permitted.  The standards were
chosen from the list of K and M giants identified by Cohen et al. (1999).  
In addition to photometric calibration, these standards were used to
characterise the PSF and used for comparison with the science sources
to detect any extension (see section 3.2). 

\subsubsection{TIMMI2}

The observations on TIMMI2 were taken over three runs on 11-12
September 2003, 19-21 November 2003 and 24-26 January 2005 (proposals
71.C-0312, 72.C-0041 and 74.C-0700).  The conditions on these nights
were very different.  In particular observations performed in January
demonstrated poor photometric accuracy. For the nights in which
accurate photometry was not possible, it was still possible to place
constraints on possible companion/background sources and extension
with the data.

A wide range of the instruments filters were used to study this sample
(M, N1, N2, 9.8, 11.9, 12.9).  The pixel scale was 0\farcs3 for the M
band and 0\farcs2 for the longer wavelengths, giving fields of view of
96\arcsec \, x 64\arcsec \, and 64\arcsec \, x 48\arcsec \,
respectively. The FWHM was 0\farcs80 $\pm$ 0\farcs12 in the N
band. 

Absolute pointing of the telescope is accurate to 5-10\arcsec
\,. However, pointing accuracy of 1\arcsec could be achieved by performing
acquisition at M (which almost always detects the stars) and
accounting for offsets between the filters by observations of the
standards. 

\subsubsection{VISIR}

The VISIR observations were carried out over three nights in December
2005 (proposal 076.C-0305).  The conditions were good over all three
nights, and allowed good photometric accuracy. The seeing was somewhat
variable, with FWHM for standards in N band of 0\farcs465 $\pm$
0\farcs161, and in the Q band 0\farcs597 $\pm$ 0\farcs166 over all
observations. The PSF showed typical ellipticities of 0.18 and 0.1 in
N and Q respectively. The same ellipticity was seen at the same
position angle (regardless of on-sky chop angle) in the science and
standard images and this instrumental artifact was well accounted for
using the standard star images as model PSFs (see section \S \ref{s:ext}). 

Two filters were used for the observations; the N band filter SiC with
central wavelength 11.85 $\mu$m (bandwidth 2.34 $\mu$m) and Q band
filter Q2 with central wavelength 18.72 $\mu$m (bandwidth 0.88
$\mu$m). The pixel scale used was 0\farcs075, giving a
19\farcs2x19\farcs2 field of view. Observations of standards were
performed before and after each observation, and standard
observations were used throughout the night to determine an airmass
correction. Calibration accuracy was 4\% and 8\% in N and Q
respectively. Acquisition was performed in the N band for all stars. 
Chopping and nodding were performed in perpendicular mode as
described above.  The detector array for the instrument had several
regions of very poor gain that were masked out by both the pipeline
and our own reduction procedures, which required careful positioning
of the stellar image on the array, particularly when also trying to
image companion objects. 

\subsubsection{MICHELLE}

MICHELLE observations of $\eta$ Corvi were performed in
service mode and 
taken on December 31st 2005 under proposal GN-2005B-Q15 with filter
Si-5 ($11.6\mu$m, bandwidth $1.1\mu$m).  The detector array is 320x240
pixels, with pixel scale 0\farcs099 (resulting field of view is
31\farcs68x23\farcs76).  The FWHM of the standards was 0\farcs35$\pm$
0\farcs02.  

An average of the two observations of the standard was used for
calibration, with an uncertainty of $\pm 5.5 \% $
in calibration factor found between them. No airmass correction
was necessary as the objects were observed at very similar airmasses
(1.3-1.25). As guiding is only possible in one of the chopped
positions with MICHELLE, one of the chop beams was always much 
less resolved than the other, giving an image of roughly twice the
Full Width at Half Maximum (FWHM) found for the guided beam.  
Only the guided beams were included in our analysis. 

\subsection{Photometry and Background/Companion Objects}

The result of the data reduction was an image for each observation
consisting of four images of the target star (two positive, two negative)
if observed in perpendicular mode, or three images of the target (one
positive, two negative at half the level of detection) if observed in
parallel mode.  The multiple images were
co-added to get a final image by first determining the centroid of each
of the individual images. Photometry was then performed using a
1\arcsec \, radius aperture for the TIMMI2 images and a 0\farcs5
radius aperture for the VISIR and MICHELLE images.  These sizes were
chosen to just exceed the full-width at half-maximum (FWHM) found for
each instrument (as described in section 3.1). This minimises noise
inclusion whilst including all the flux from an unextended source. 
Note that the filters used in these observations were narrow band and so
no colour-correction was applied.  Residual
statistical image noise was calculated using an annulus centred on the
star with inner radius the same as that used for the photometry, and
outer radius of twice the inner radius (so 2\arcsec \, for TIMMI2 and
1\arcsec \, for VISIR and MICHELLE).  Typical levels for statistical
noise at the 1 $\sigma$ level in a half hour observation were
44mJy total in the 1\farcs0  radius aperture of TIMMI2, 4 mJy and
12mJy for the 0\farcs5 aperture of VISIR in N and Q respectively, and
6mJy in the 0\farcs5 aperture of MICHELLE.

Smaller apertures were used to search for background sources and to
place limits on detected sources.  The aperture sizes were chosen to
maximise the signal to noise of a point source in the aperture as
determined by testing the standard star observations. The sizes of
aperture used were 0\farcs8 in radius for the TIMMI2 observations, 
0\farcs4 for
MICHELLE, and  0\farcs32 and 0\farcs35 for the N and Q filters for VISIR.
Apertures systematically centred on each pixel of the array in turn
were searched for significant signal at the 3 $\sigma$ level or
above (based on the statistical noise). Where none were found, the
limits placed on the background object were based on the 3 $\sigma$
uncertainty in the aperture plus calibration uncertainty.  For the
non-photometric nights, limits were based on calibration to the IRAS
flux of the object.  The upper limits to background sources are listed
in Table \ref{observations}.

%

\section{Extension testing}
\label{s:ext}

An important part of this study was to look for evidence of extension
in the observation images, or use the lack of extension to place limits on
possible disk structure around the stars. For all observations we
fitted a two-dimensional Gaussian to detected sources and compared the
science image fit to the found for the standard stars. In addition for
all observations the sources surface brightness profile was determined
by calculating the average surface
brightness in a series of annuli centered on the source of 2 pixel
thickness by increasing inner radius from 0 to 3\arcsec \,. The
resulting sizes and profiles for all science observations were
compared with those of the standards observed immediately before and
after the science observations to search for any discrepancies in
width.

To assess whether there is any evidence of extension in the science
image the images of the point-like standard stars scaled to the peak
of the science observation were used to model what an unextended
source would be expected to look like. A straight-forward subtraction
of the model from the science image was then performed and the
resulting image subjected to a test to check for
consistency with noise levels as measured on the pre-subtraction stellar
image. Tests optimised for varying disk geometries were applied,
choosing those that would give the highest signal-to-noise detection
should such disks exist, as outlined in the following section. Note
that since the PSF is scaled to the peak, then if the disk
contributes to the peak some of the disk flux has been
removed. Essentially we are testing the null hypothesis that the
source is unextended. 

\subsection{A new method of determining extension limits}

\begin{figure*}
\begin{minipage}{2.5cm}
PSF 
\end{minipage}
\begin{minipage}{2.5cm}
Model 
\end{minipage}
\begin{minipage}{2.5cm}
Convolved image 
\end{minipage}
\begin{minipage}{2.5cm}
PSF subtraction 
\end{minipage}
\begin{minipage}{2.5cm}
Optimal region 
\end{minipage}\\
\vspace{0.2cm}
\begin{minipage}{2.5cm}
\includegraphics[width=2.3cm]{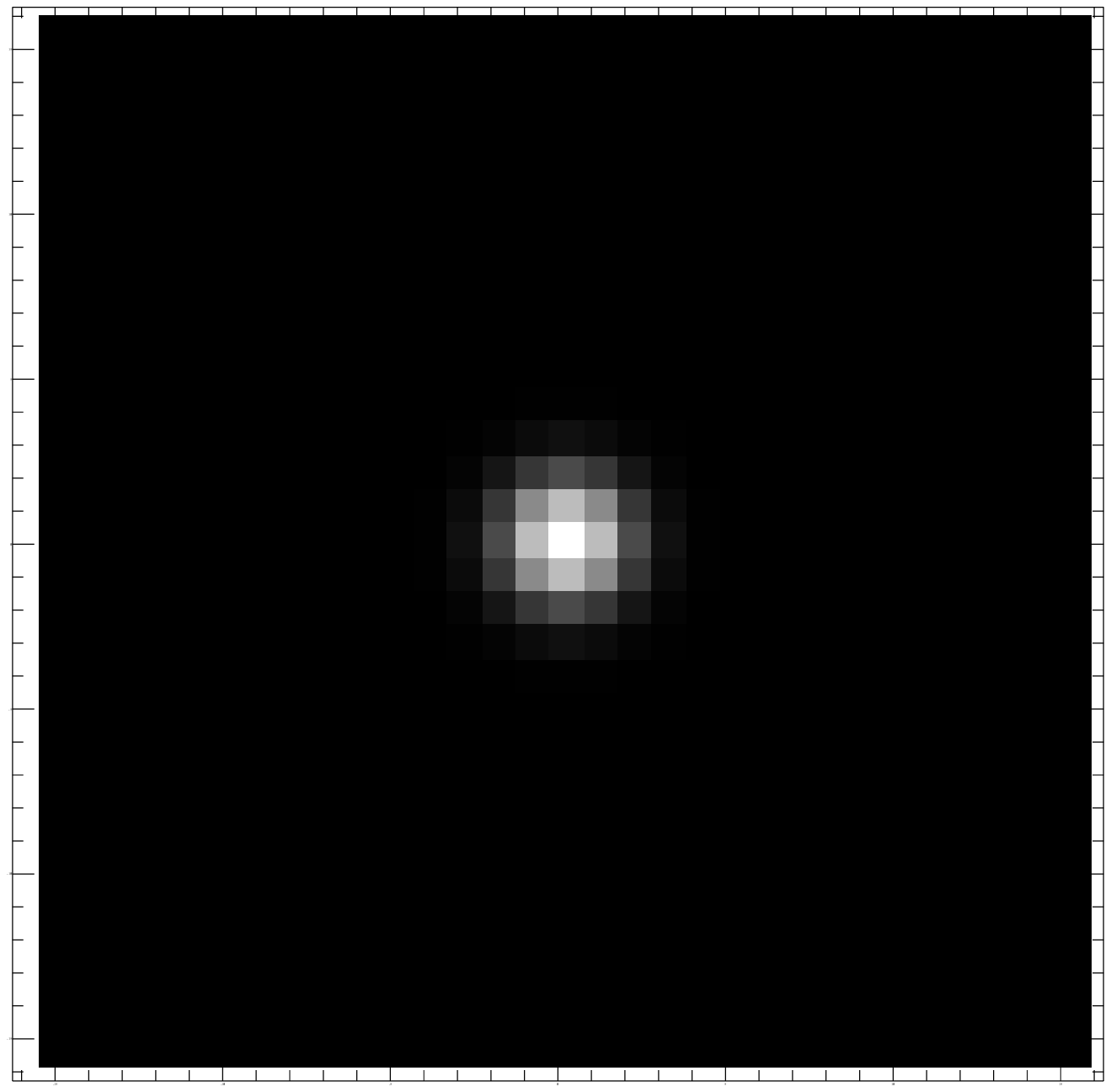}
\end{minipage}
\begin{minipage}{2.5cm}
\includegraphics[width=2.3cm]{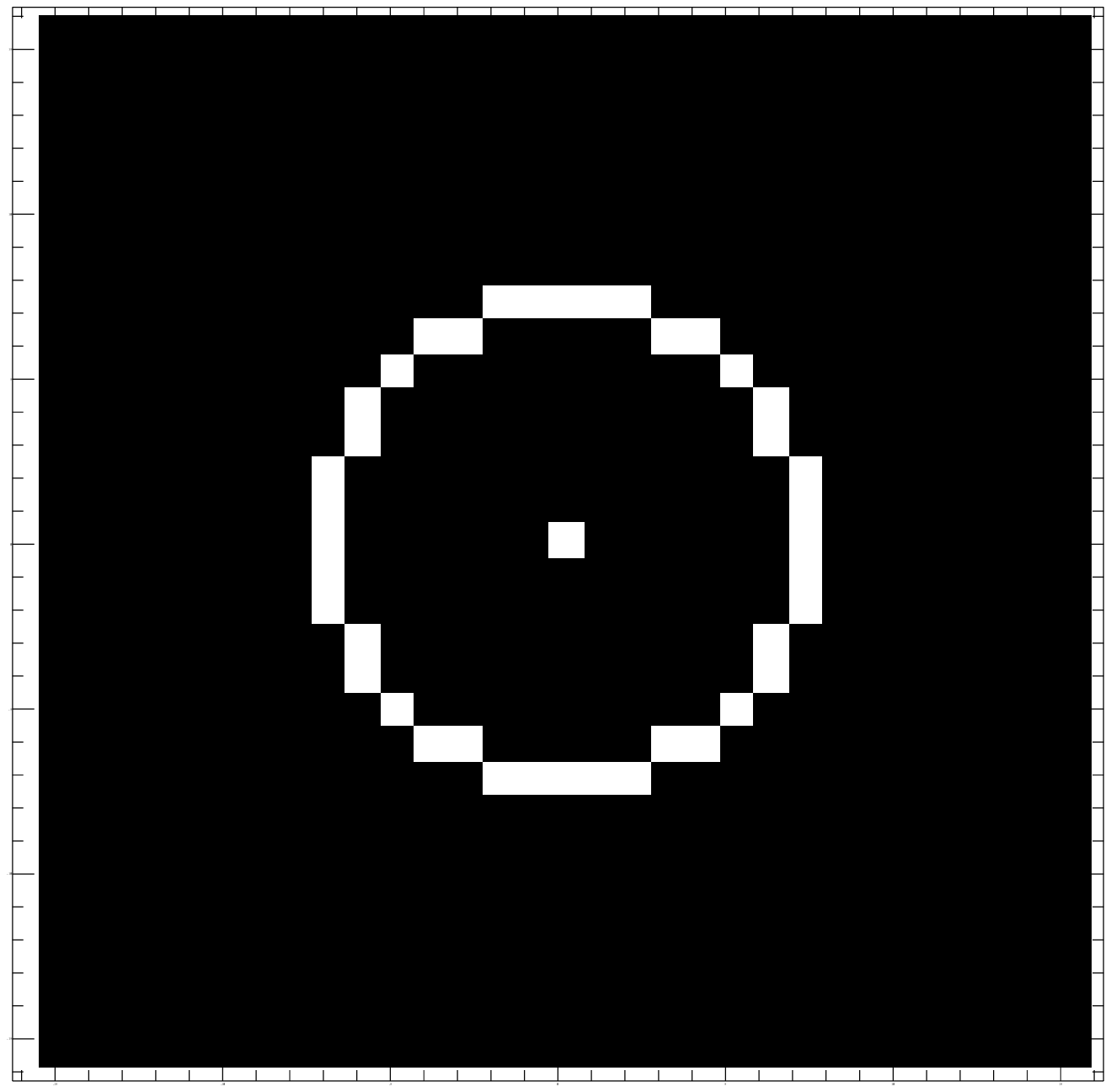}
\end{minipage}
\begin{minipage}{2.5cm}
\includegraphics[width=2.3cm]{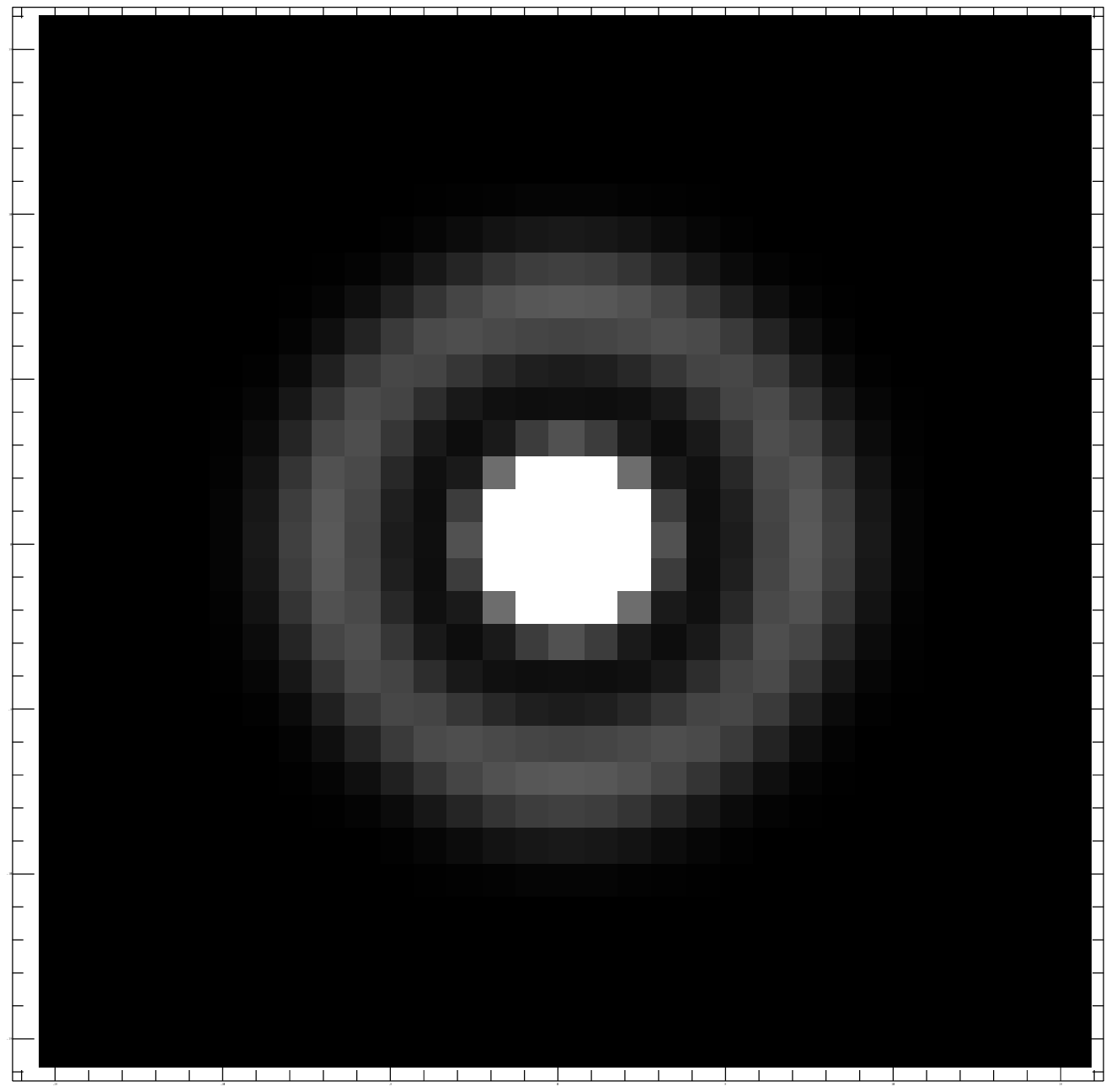}
\end{minipage}
\begin{minipage}{2.5cm}
\includegraphics[width=2.3cm]{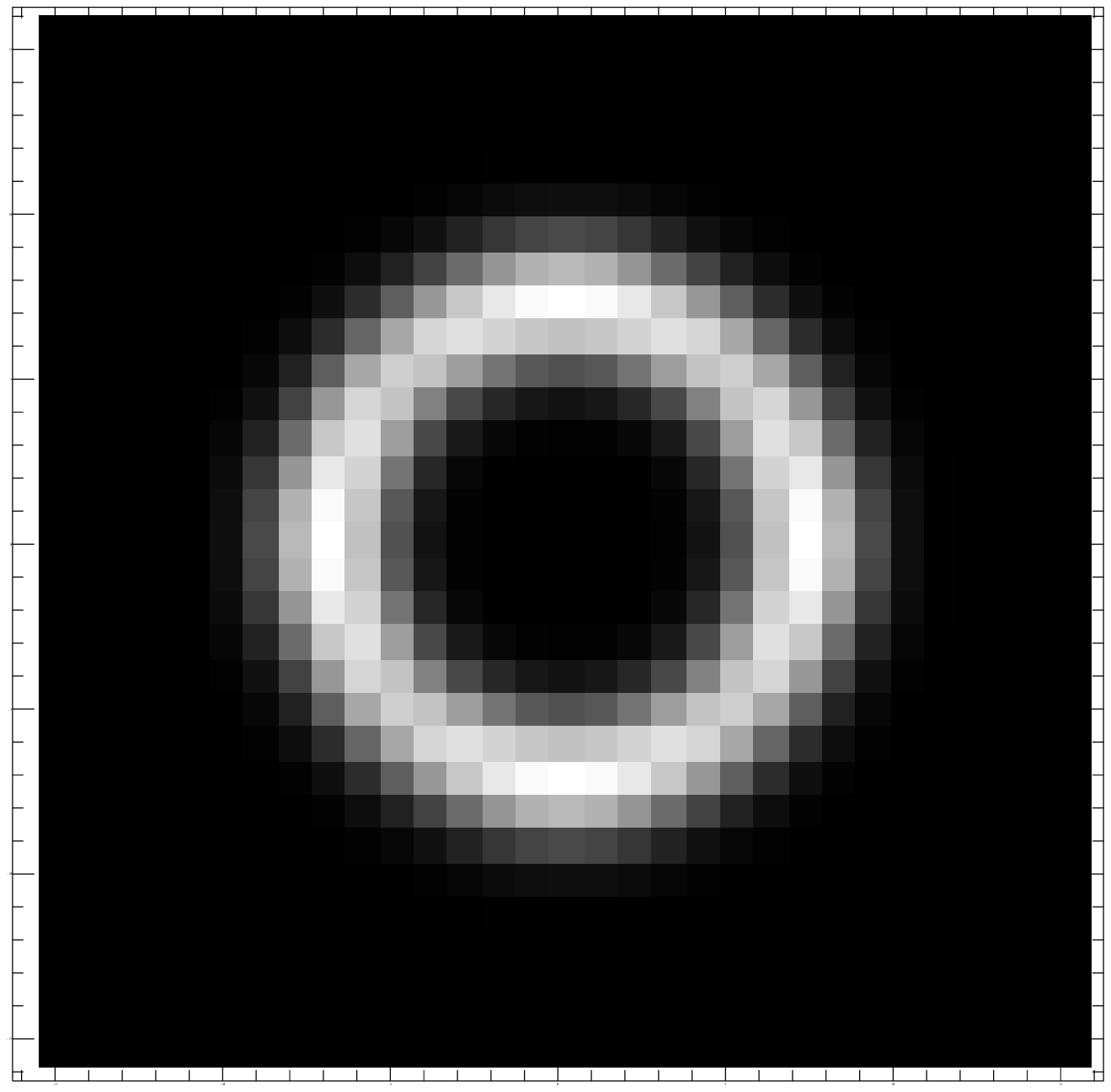}
\end{minipage}
\begin{minipage}{2.5cm}
\includegraphics[width=2.3cm]{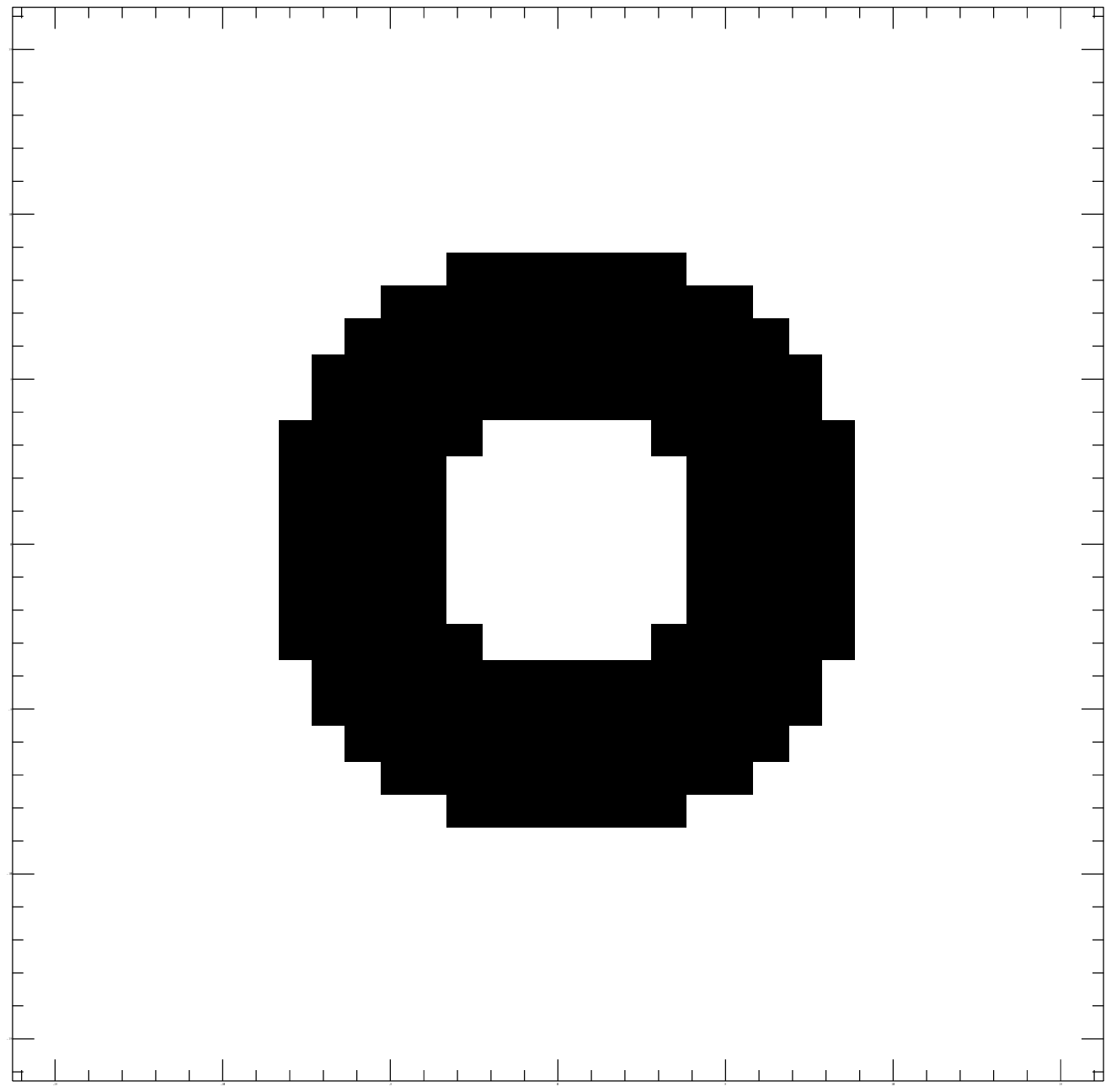}
\end{minipage}\\
\vspace{0.2cm}
\begin{minipage}{2.5cm}
\includegraphics[width=2.3cm]{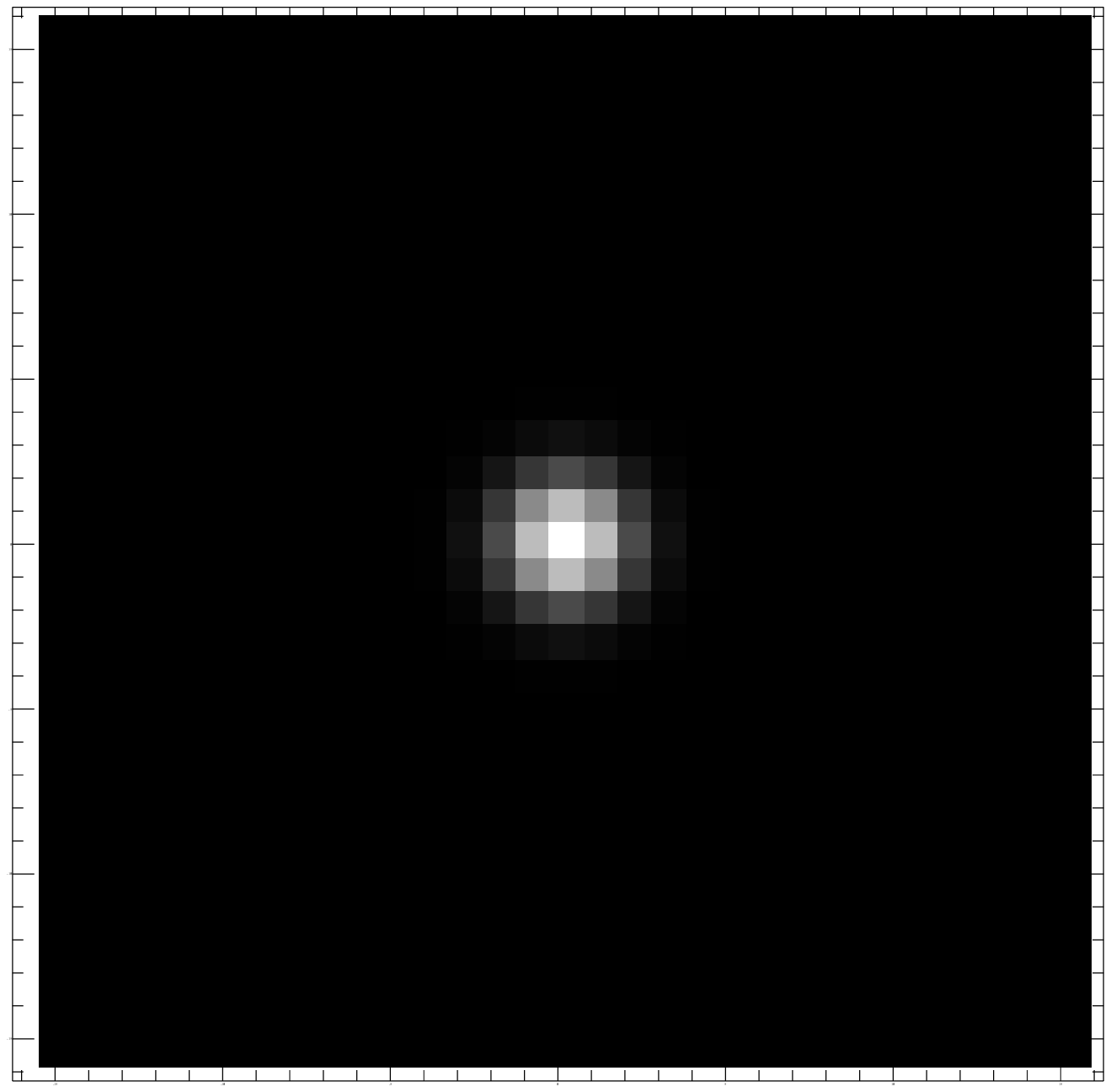}
\end{minipage}
\begin{minipage}{2.5cm}
\includegraphics[width=2.3cm]{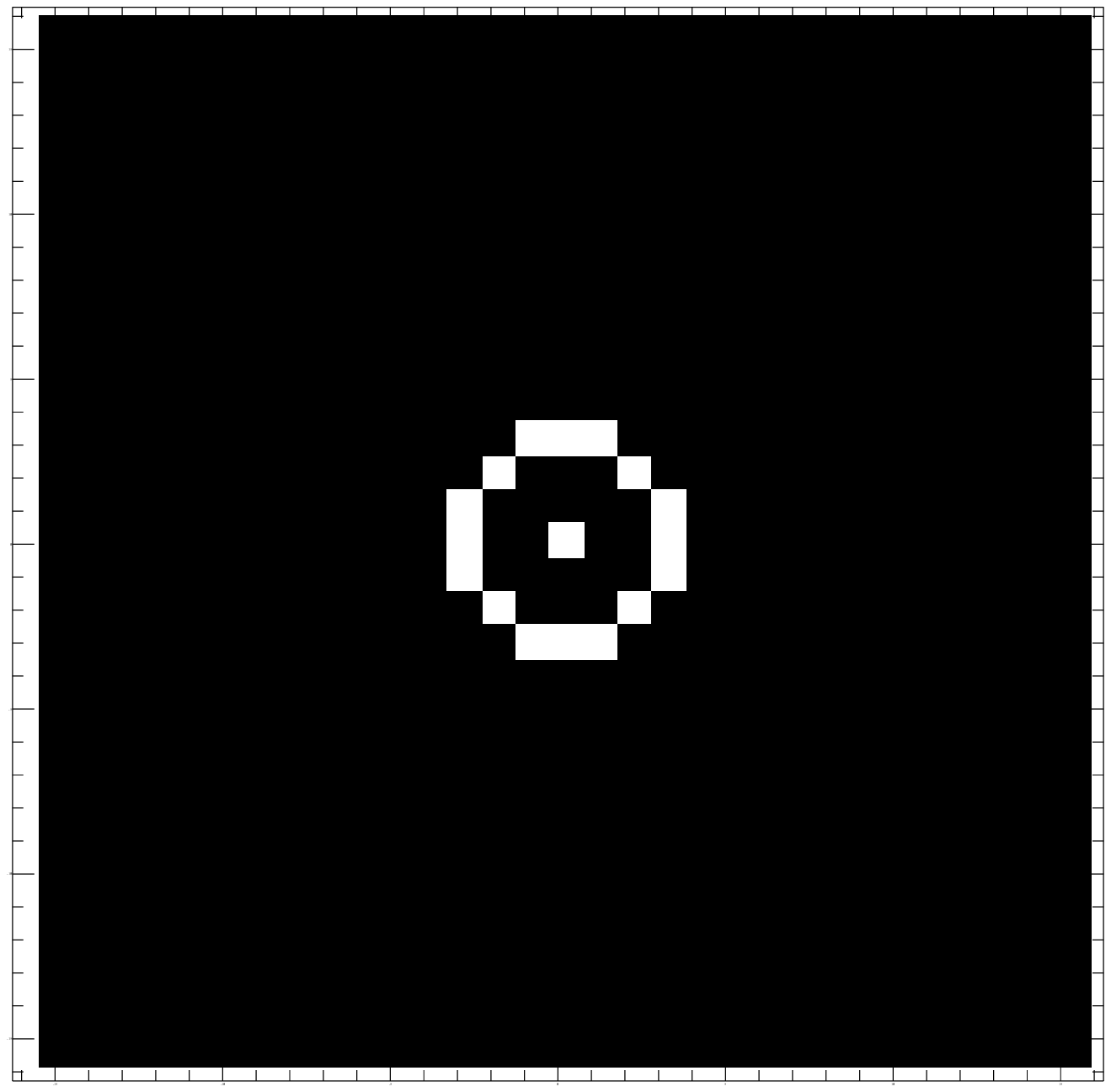}
\end{minipage}
\begin{minipage}{2.5cm}
\includegraphics[width=2.3cm]{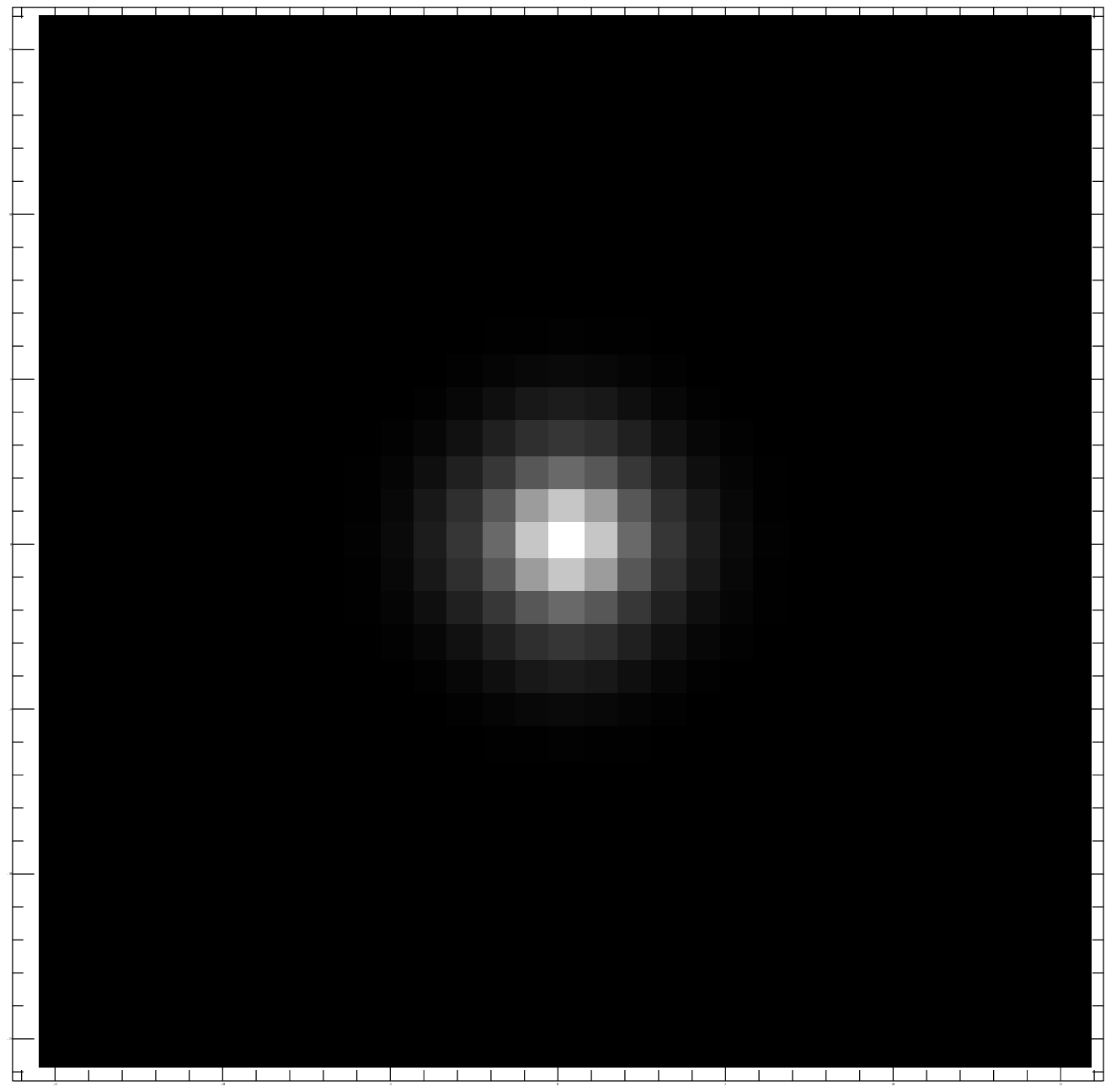}
\end{minipage}
\begin{minipage}{2.5cm}
\includegraphics[width=2.3cm]{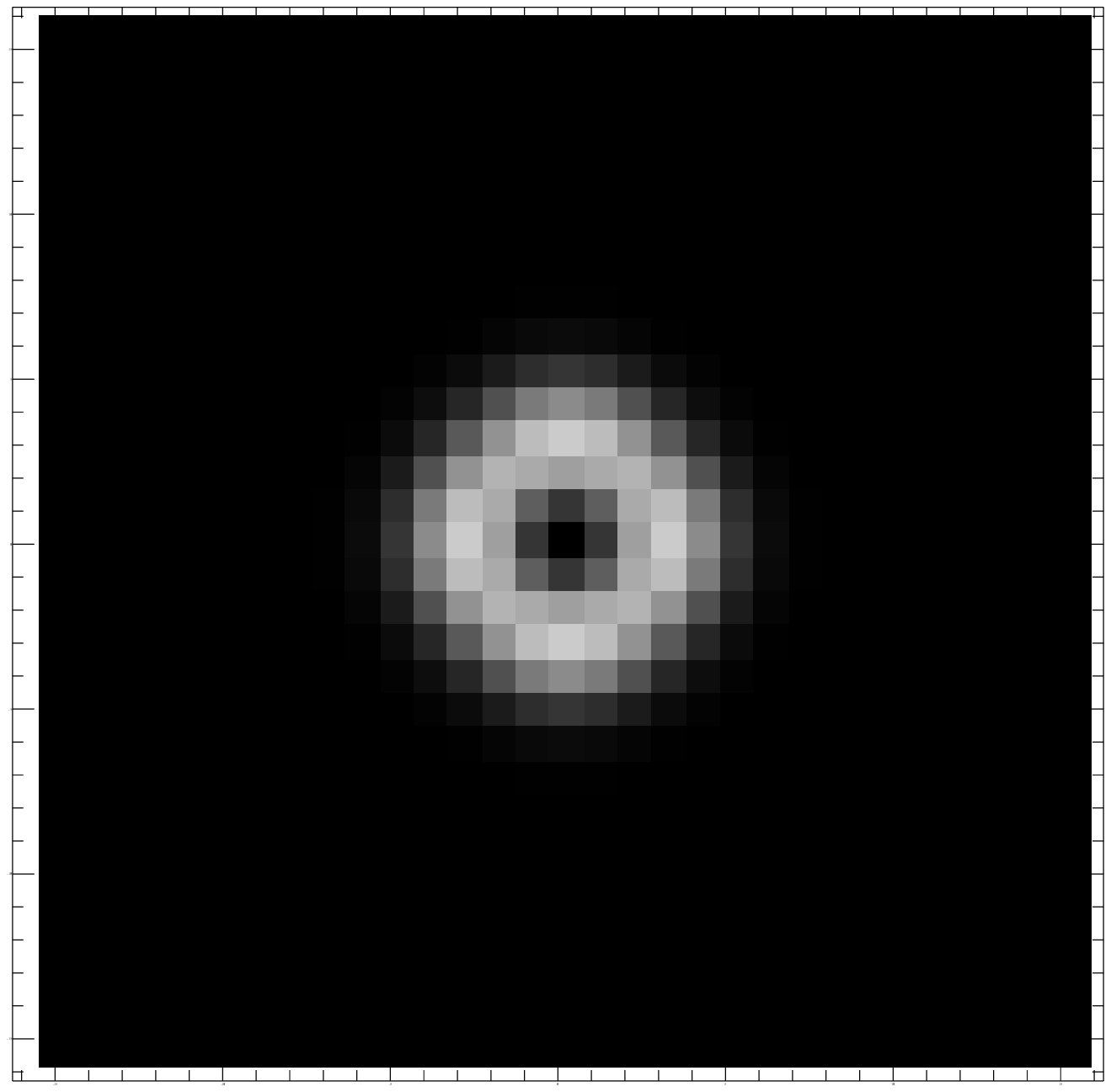}
\end{minipage}
\begin{minipage}{2.5cm}
\includegraphics[width=2.3cm]{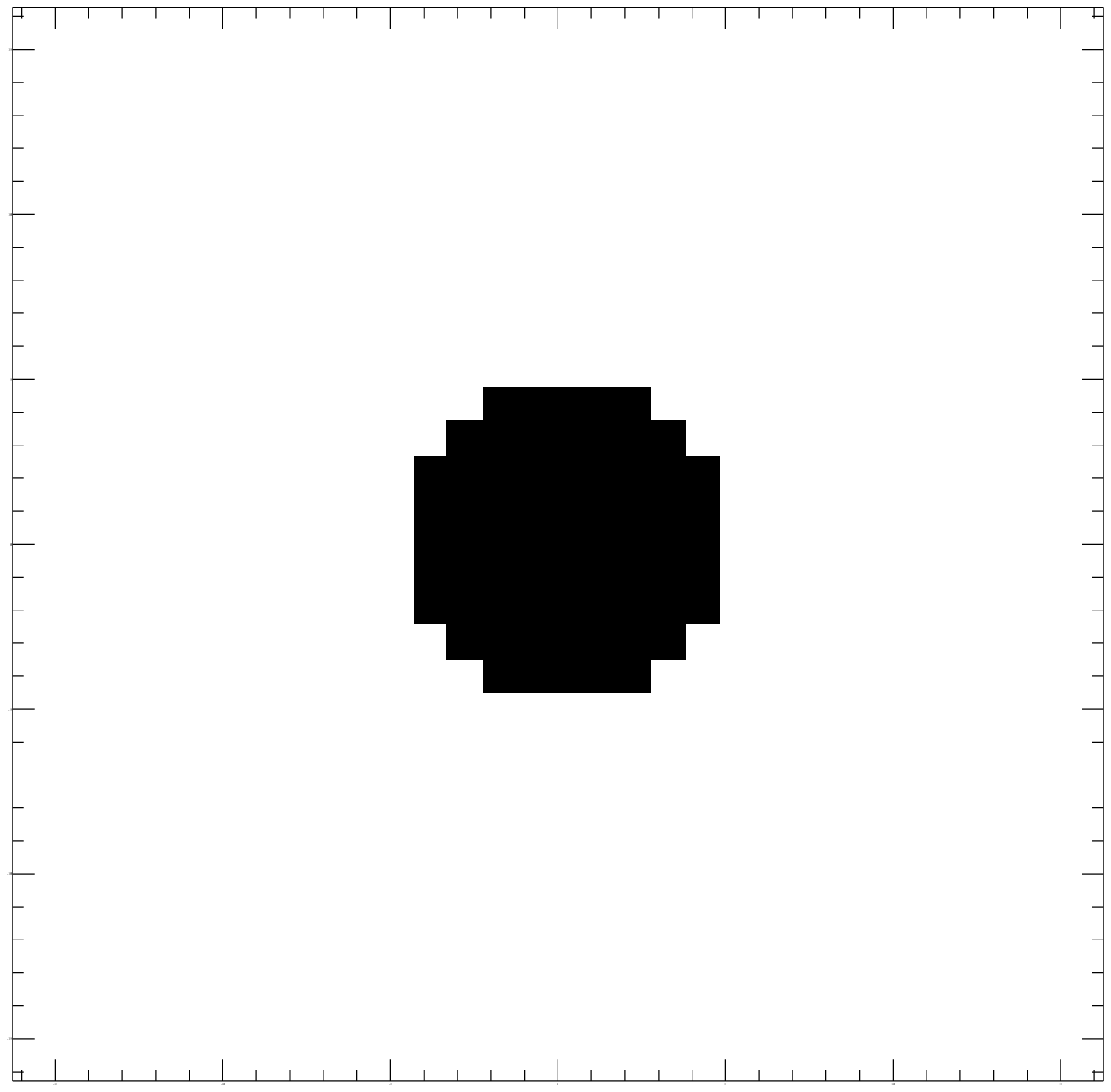}
\end{minipage}\\
\vspace{0.2cm}
\begin{minipage}{2.5cm}
\includegraphics[width=2.3cm]{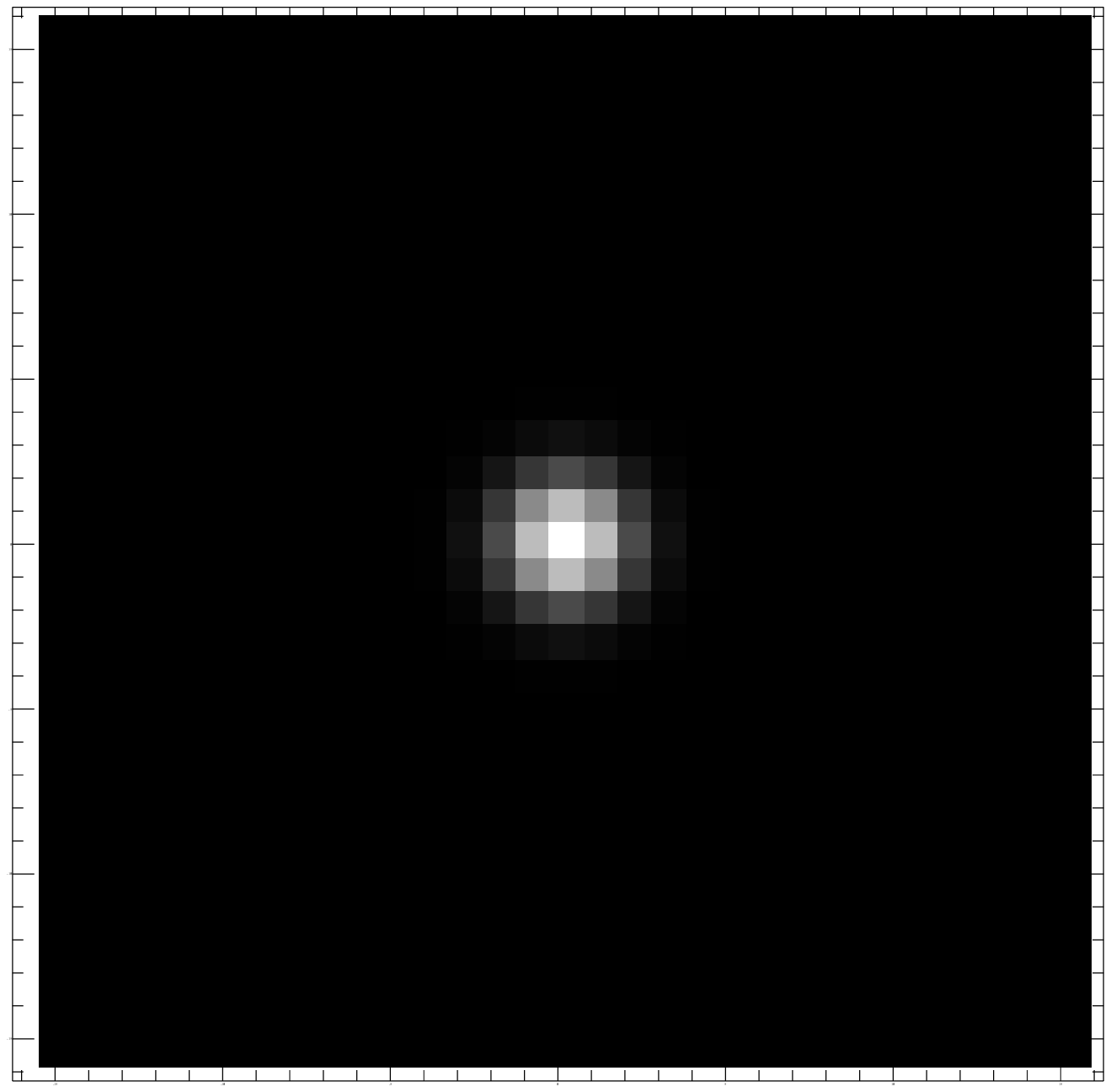}
\end{minipage}
\begin{minipage}{2.5cm}
\includegraphics[width=2.3cm]{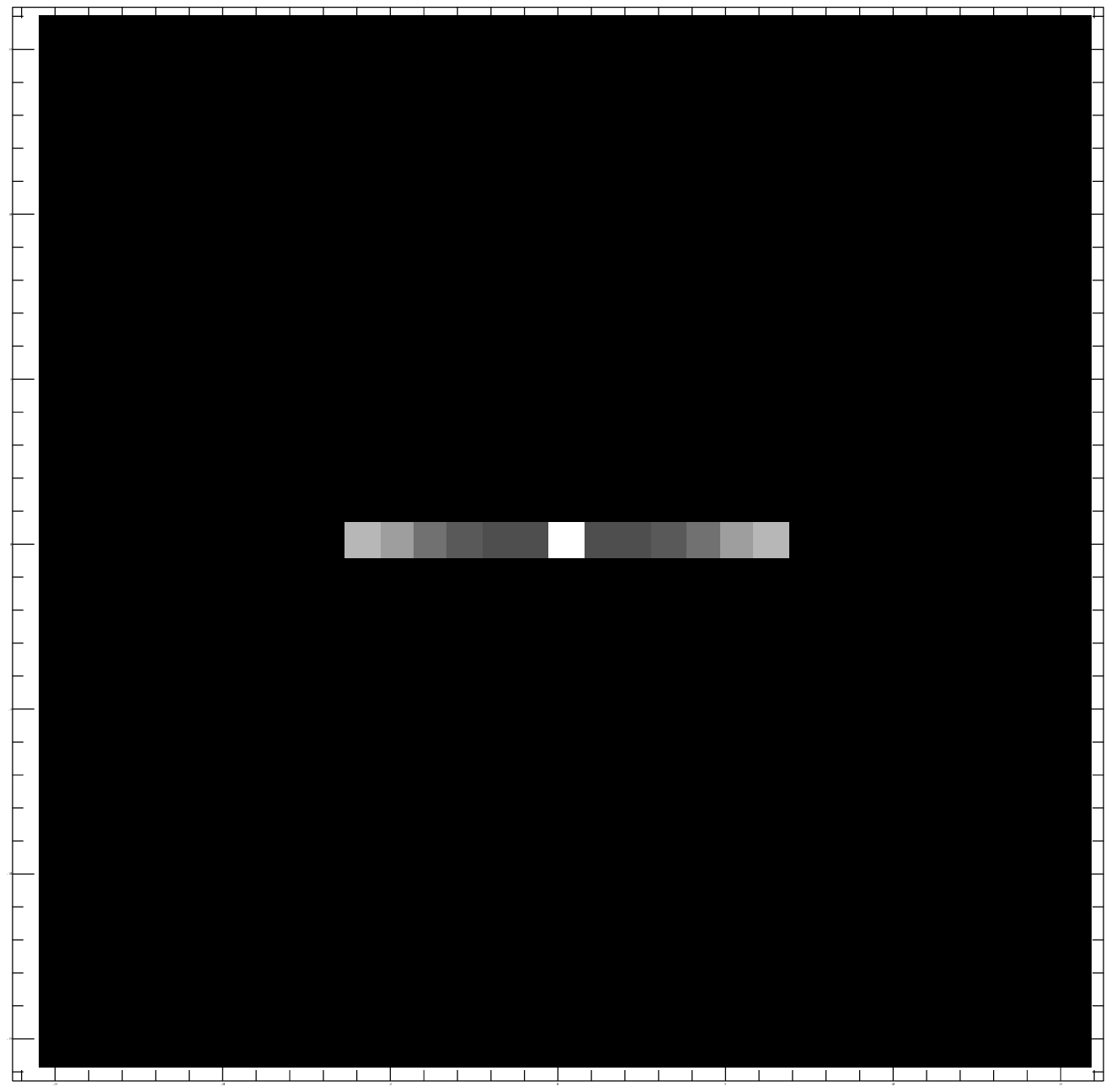}
\end{minipage}
\begin{minipage}{2.5cm}
\includegraphics[width=2.3cm]{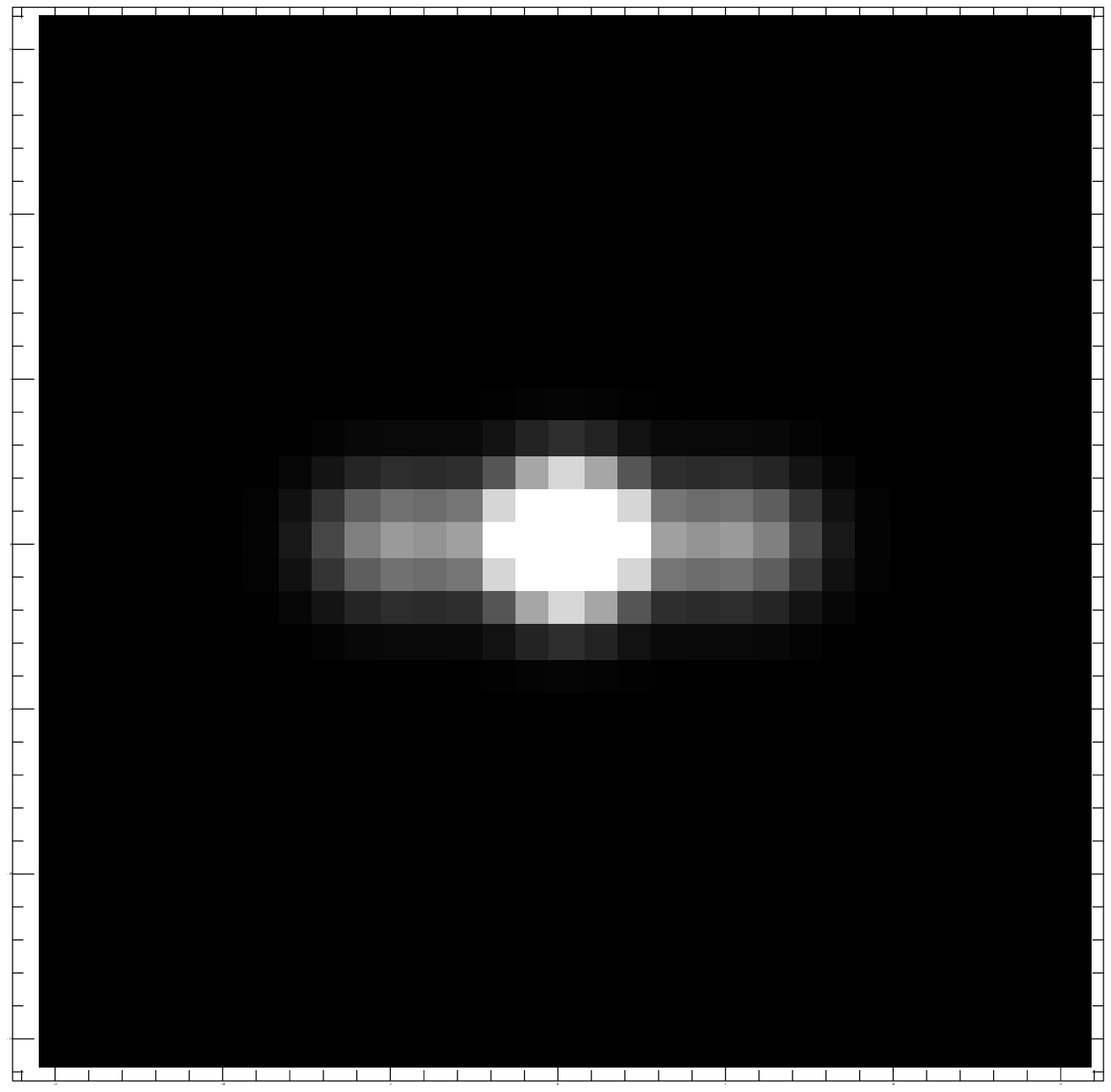}
\end{minipage}
\begin{minipage}{2.5cm}
\includegraphics[width=2.3cm]{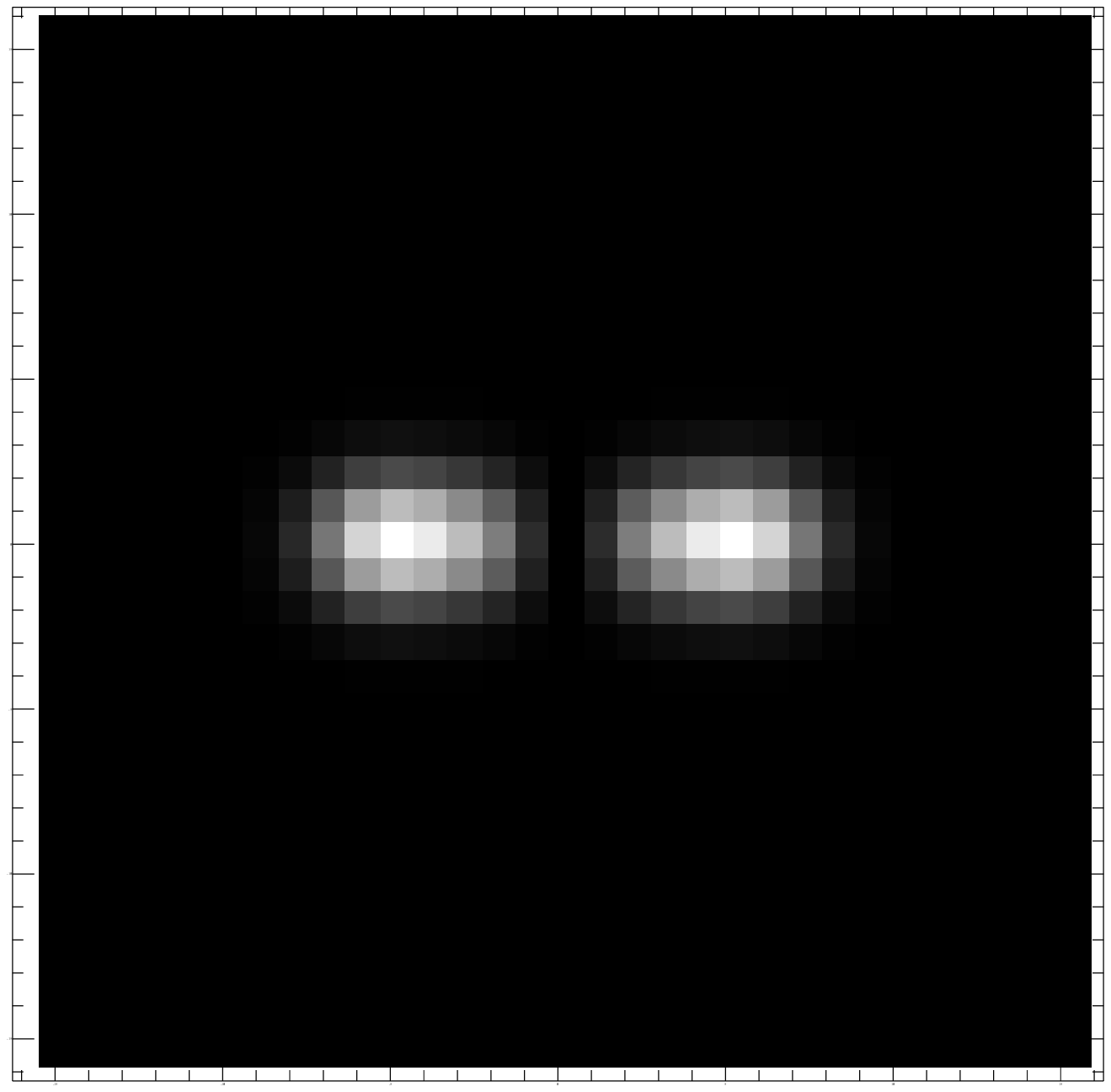}
\end{minipage}
\begin{minipage}{2.5cm}
\includegraphics[width=2.3cm]{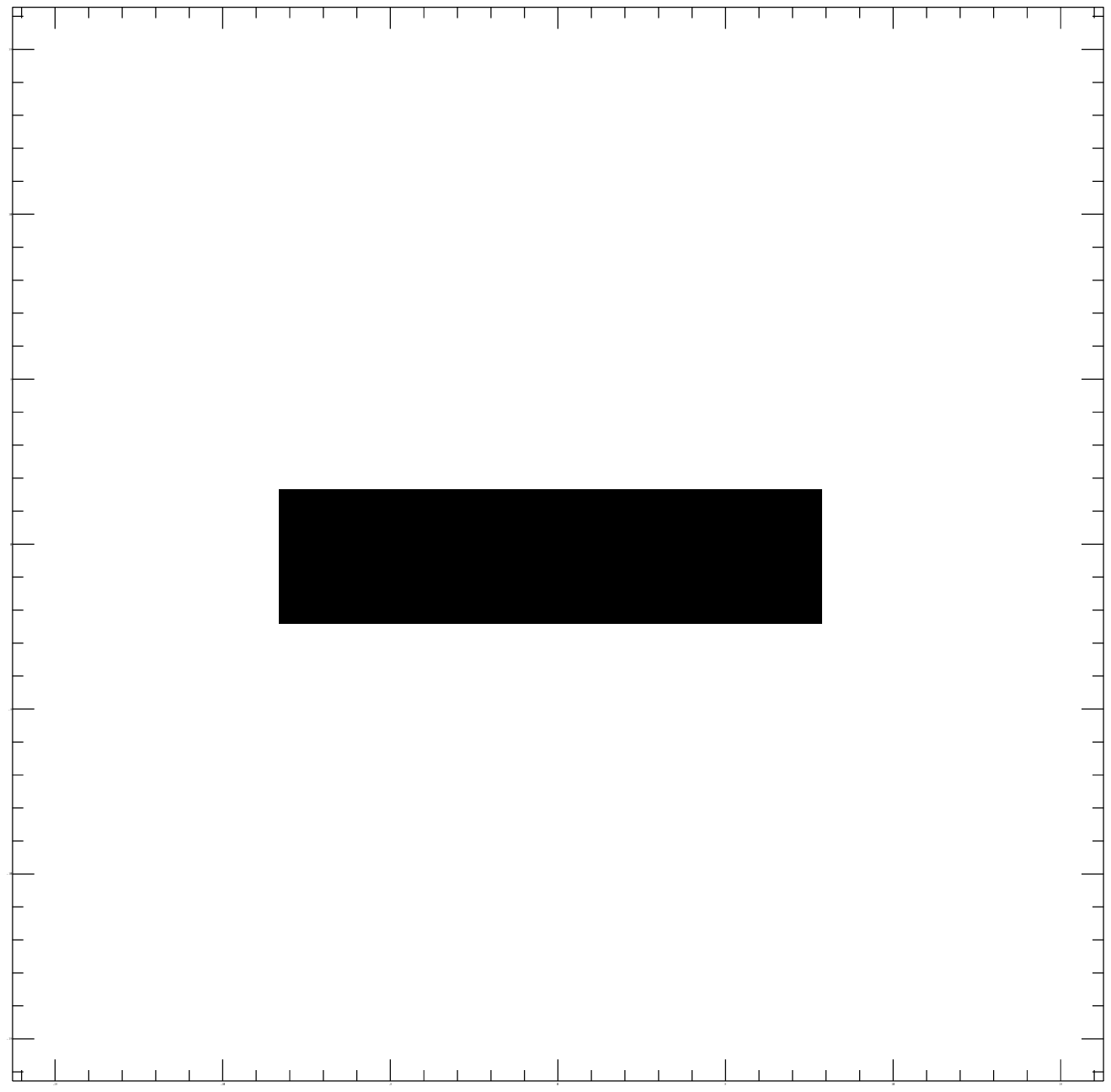}
\end{minipage}
\caption{\label{fig:ext_models} Examples of the models tested and the
  various stages used to determine the optimal testing regions for
  detecting extended emission. The rows show three different
  models:\emph{Top:} Face-on, large radius $r$; \emph{middle:} Face-on
small radius $r$; \emph{bottom:} Edge-on large radius $r$.  The model
images (second column) are convolved with a PSF (first column), here
approximated by a Gaussian to give the convolved images (third
column).  The point-like component of the final image is then removed
by subtracting the PSF scaled to the peak of the convolved image
(forth column).  Finally a range of possible regions to test for
residual emission are determined by finding the shape and size of a
region that maximises the S/N on any residual emission on the array
(black region, fifth column).}
\end{figure*}

Here we consider what levels of disk flux could be detected in an
observation, given its geometry. To do so we made model images of an
unresolved star, at a level $F_\star$, and a disk at a level
$F_{\rm{disk}}$, which we characterised by the parameter
$R_{\rm{\lambda}} = F_{\rm{disk}}/(F_\star + F_{\rm{disk}}) =
  F_{\rm{disk}}/F_{\rm{tot}} $ (see Figure \ref{fig:ext_models} second
  column).   
The disk was assumed to be an annulus of radius $r$ and width $dr$ (so
with inner radius $r-dr/2$, outer radius $r+dr/2$), with
uniform surface brightness, at an inclination to our line of sight of $I$.
These images were convolved with model PSFs (Figure
\ref{fig:ext_models} first and third columns). 
In this section we approximate the PSF by a Gaussian of FWHM $\theta$, but
in later sections we use the true observed PSFs.
Models with $dr/r \in [0.2,2.0]$, $r/\theta \in [0.083,6.67]$, $I \in
[0, 90]$, and $R_\lambda \in [0.001,0.99]$ were tested. 
A best estimate of the unresolved contribution to the image was removed
by subtracting a PSF scaled to the peak surface brightness (centered
on the star, Figure \ref{fig:ext_models} forth column). The optimum
aperture that would be able to detect the 
residual disk emission given the uncertainties inherent in the
observing process is then determined (Figure \ref{fig:ext_models}
fifth column). This optimal region has area
$A_{\rm{op}}$. 

We considered two sources of noise that hinder a detection.
The first is the background noise on the array, which we assumed is
Gaussianly distributed and which increases $\propto t^{0.5}$ for
longer integrations.  This leads to an increase in the S/N on the
source $\propto t^{0.5}$ (signal increasing $\propto t$). 
This was characterised by $S_\star$, the signal to noise achieved on
a flux $F_\star$ within an aperture of radius $\theta$, and area
$A_\theta = \pi \theta^2$, where the noise per pixel is assumed to be
the background noise that is found across the array.  Note that here
we have ignored the photon noise contribution to the statistical noise
term. This is because when searching for residual emission after the point
source subtraction the flux is likely to be faint and thus background
limited. Any noise caused by incorrect subtraction of the point-source
is included in the second component of the noise described below.
The definition of $R_\lambda$ thus implies
that the signal to noise on the disk flux in the same aperture is
$S_\star(R_\lambda^{-1}-1)^{-1}$ if $F_{\rm{disk}}$ lies entirely
within the aperture. Note that $S_\star$ does not necessarily equate
exactly with quoted instrumental sensitivities for which the region
used for optimum detection must be considered. 
The second is the uncertainty in the PSFs due, e.g., to changes in the
atmosphere which we characterise by the uncertainty in the FWHM $d\theta$
leading to uncertainties in the flux in an optimal region of size
$A_{\rm{op}}$ of $N_{d\theta}$. These uncertainties were quantified as
the difference in the flux in that optimal region when the PSF was changed
from $\theta$ to $\theta + d\theta$. We tested $d\theta/\theta \in
[0., 0.1]$.
These noise sources were added in quadrature so that the final signal
to noise in a region of area $A_{\rm{op}}$ is
\begin{equation}\label{eq:sop}
S_{\rm{op}} = F_{\rm{op}}/N_{\rm{op}} = F_{\rm{op}}/\sqrt{
  (A_{\rm{op}}/A_\theta)N_\star^2 + N_{d\theta}^2}
\end{equation}
where $N_\star = F_\star/S_\star$ is the background statistical noise in the
aperture ($A_\theta$) used on the point source. Here $F_{\rm{op}}$ is
the flux in the optimal region, which assuming accurate subtraction of
the stellar component in the PSF subtraction should be some fraction
of $F_{\rm{disk}}$, and $N_{\rm{op}}$ is the noise in this same
optimal region. 

For any given geometry, a broad range of aperture parameters was
considered and the one that gave the highest signal-to-noise detection as
defined in equation \ref{eq:sop} was chosen.  We consider a detection
to be where $S_{\rm{op}} > 3. $

\subsubsection{Face-on ring}

\begin{figure*}
\begin{minipage}{8cm}
\includegraphics[width=8cm]{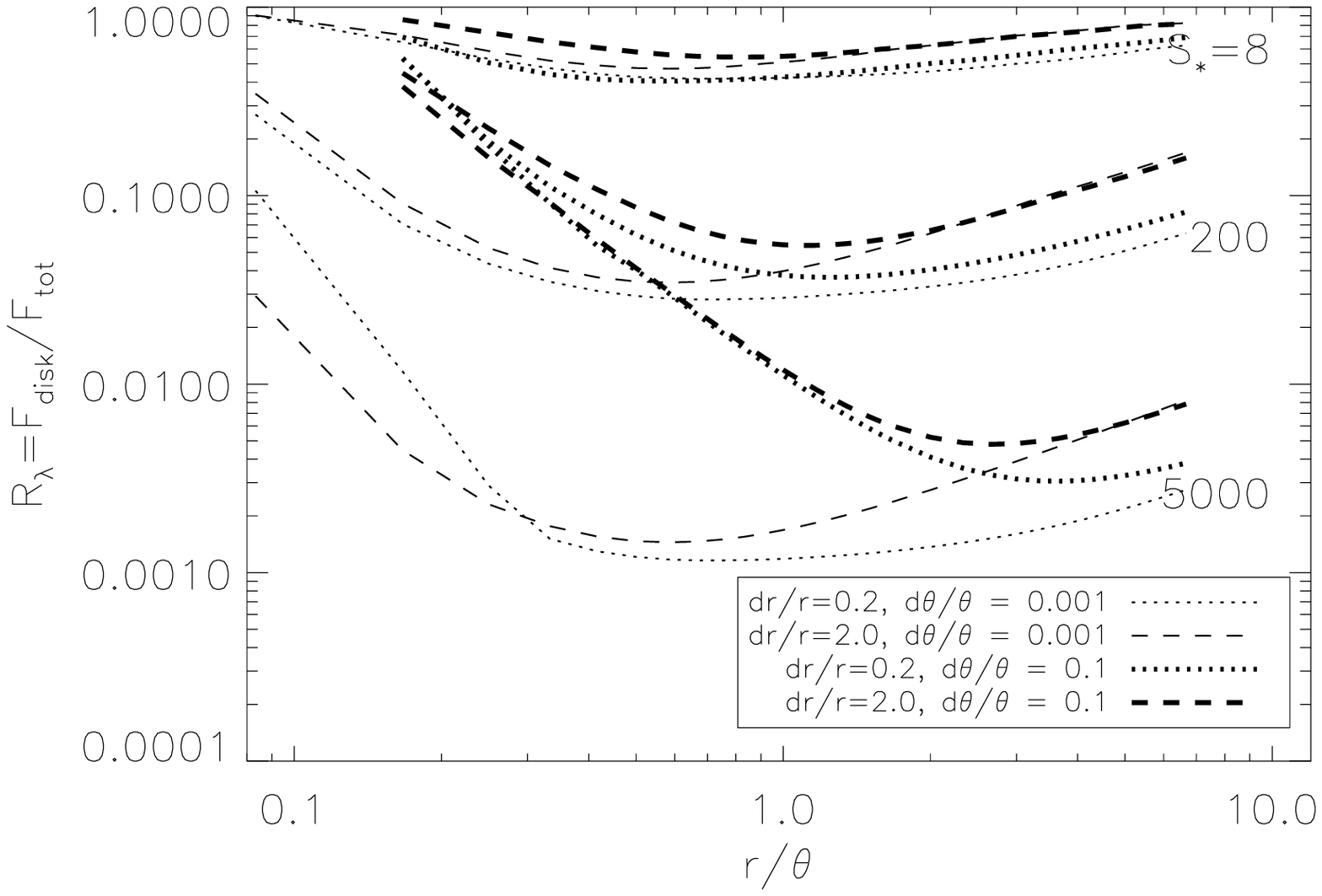}
\end{minipage}
\hspace{0.5cm}
\begin{minipage}{8cm}
\includegraphics[width=8cm]{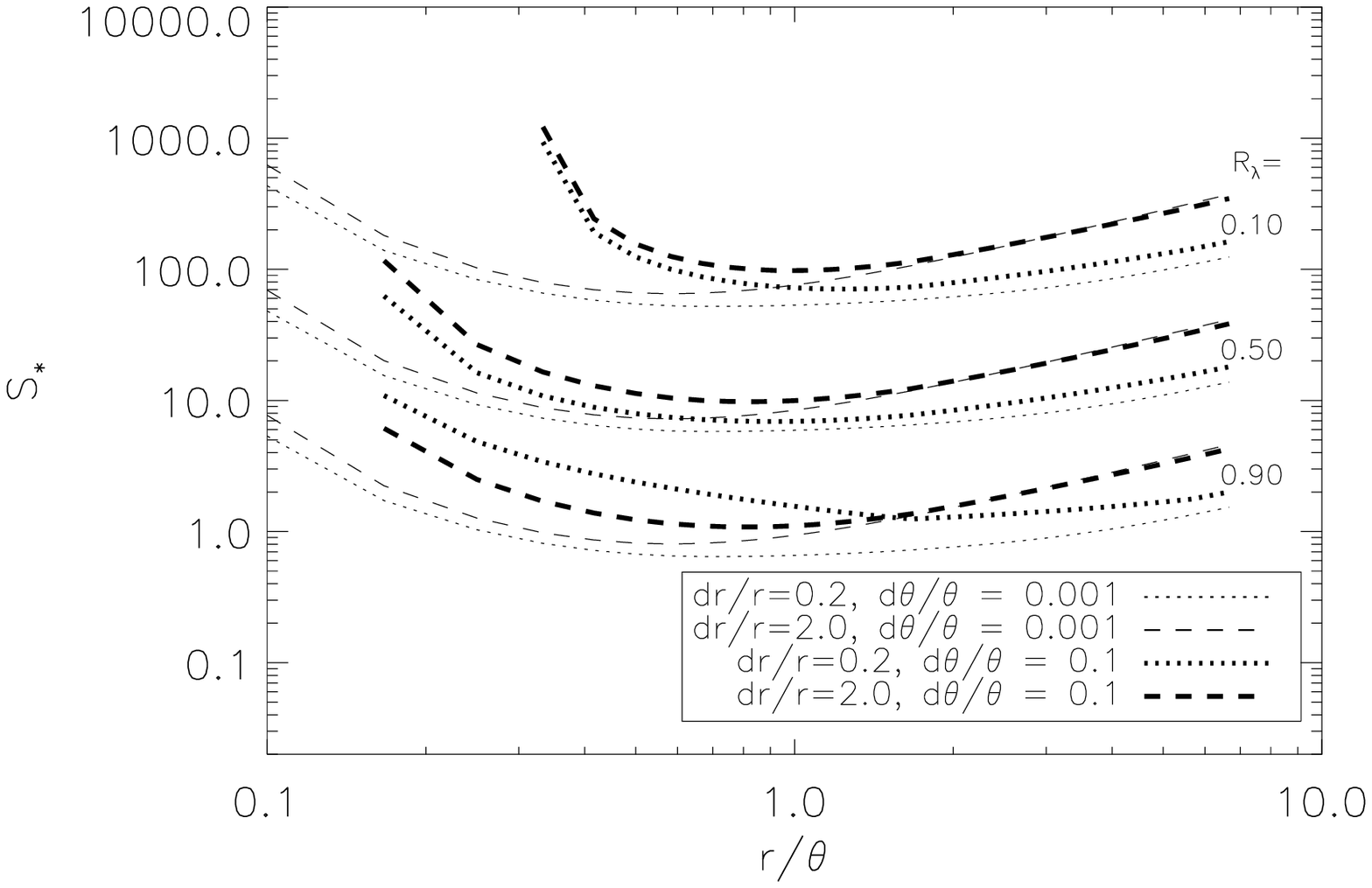}
\end{minipage}
\vspace{-0.2cm}
\caption{\label{limface} Limits on detectable face-on disks for varying disk
  parameters.  The region above the lines represents the region of
  detectability. Left: The disk flux required to get a 3 sigma
  detection of extension for disks of varying geometry in a face-on
  orientation (disk flux given in terms of $R_\lambda$, $F_\star$ = 10
  in these plots)
  Right: The signal-to-noise required for a significant
  detection for varying $R_\lambda$.}
\end{figure*}

Here we consider the results of the modelling when applied to face-on
rings. For large disks the symmetrical nature of a face-on ring means
that the optimum region will be a ring of radius $R$, and width
$\Delta$, so that $A_{\rm{op}} \approx 2\pi R \Delta$. However, for
disks close to or smaller than the size of the PSF 
($r/\theta \ll 1$), we find that PSF accuracy is often the limiting
factor.  The optimal region for detecting residual extended
emission would tend to a circular aperture. Using the $R, \Delta$
notation we note that when $ R - \Delta / 2 <0 $ the inner radius of
the annulus becomes zero and the optimal region becomes a circular
aperture of radius $R+\Delta/2$.  For the face-on disk case we find 
\begin{eqnarray}\label{faceon}
R/r & = & 1+0.5(r/\theta)^{-1}(d\theta/\theta)(dr/r) \nonumber \\ 
\Delta/r & = &  
  \sqrt{(dr/r)^{2-5(d\theta/\theta)}+(1.47-(d\theta/\theta))(r/\theta)^{-2.3
  + 10 (d\theta/\theta)}}  \nonumber \\ 
F_{\rm{op}} & = & (1 - (d\theta/\theta)^{2} (r/\theta)^{-2}
R_\lambda^{-1}) 10^{-0.1(r/\theta)^{-1}} F_{\rm{disk}} \nonumber \\
N_{\rm{op}} & = & \sqrt{(A_{\rm{op}}/A_\theta)N_\star^2 + N_{d\theta}^2} \\
N_{d\theta} & = & (d\theta/\theta)
(dr/r)^{-1}10^{-10(r/\theta)}R_\lambda^{-0.5} F_{\rm{tot}}\nonumber 
\end{eqnarray}
with $S_{\rm{op}}$ determined from equation \ref{eq:sop}.  With these
equations we can fit the numerical results for $S_{\rm{op}}$ to better
than $\pm$ 50\% for 85\% of the disks models tested. 

Notice that if the disk is large ($r/\theta \gg 1$) or the PSF
perfectly known ($d\theta/\theta = 0$) then $N_{d\theta} = 0$ and
$N_{\rm{op}} = \sqrt{(A_{\rm{op}}/A_\theta)N_\star^2}$.  Also when
$r/\theta \gg 1$, $F_{\rm{op}} \simeq F_{\rm{disk}}$ and 
\begin{equation}\label{sop_face}
S_{\rm{op}}
\simeq S_{\star}(\theta/r)(2\Delta/r)^{-0.5}(R_\lambda^{-1} -
1)^{-1}.
\end{equation}

The required levels of $R_\lambda$ as a function of $r/\theta$
and disk geometry to get a significant detection ($S_{\rm{op}} > 3$), 
as well as the signal required for a detection for a given $R_\lambda$
are shown in Figure \ref{limface}. These plots show the fitted
functions given in equations \ref{faceon}.  As mentioned above these
functions fit the numerical results to better than $\pm$ 50\% for 85\%
of the disk models tested. The main features of the plots can
be understood as follows: As can be seen in the equation for
$F_{\rm{op}}$ (equations 
\ref{faceon}) the signal falls to zero when $r/\theta <
(d\theta/\theta)R_\lambda^{-0.5}$. Thus even when the disk emission
completely dominates the signal ($R_\lambda \simeq 1$) we cannot detect
an extended disk to a smaller size than the uncertainties on the PSF. 
The optimal size of a disk in terms of ease
of detectability (minimal required $R_\lambda$ and $S_\star$) is
$r/\theta \simeq 1$. This is easily understood from an intuitive point
of view, as larger disks $r/\theta>1 $ have their flux dispersed over
a wider area and so have reduced surface brightness making them harder
to detect ($S_{\rm{op}} \propto (r/\theta)^{-1}$, equation
\ref{sop_face}), and smaller disks are more adversely affected by 
errors in PSF subtraction ($N_{d\theta} \propto 10^{-10(r/\theta)}$,
equations \ref{faceon}), as well as by losing a large percentage
of the disk flux in the peak-scaled point source subtraction
($F_{\rm{op}}/F_{\rm{disk}} \propto 10^{-0.1(r/\theta)^{-1}}$, 
equations \ref{faceon}).  Similarly in the large disk case wider
disks are more difficult to detect as they have a lower surface
brightness (the statistical noise over the optimal region will be
higher as $(\Delta/r)^2 \propto (dr/r)^{2}$).  
The sharp fall-off of $N_{d\theta}$
with $r/\theta$ also explains why this error term can be neglected in
the case of large face-on disks, and why for $r/\theta \gg 1$ the
required $R_\lambda$ (or $S_\star$) for detecting extension with large
or small $d\theta/\theta$ tend to the same limits.  The dependence of
$N_{d\theta} \propto d\theta/\theta$ (equations \ref{faceon})
means that for smaller disks a higher uncertainty in the PSF has a
strong effect in reducing the detectability of a disk (disks of a
given geometry require much higher $R_\lambda$ or alternatively higher
$S_\star$ to be detected). Notice also that as $N_{d\theta}
\propto (dr/r)^{-1}$, when PSF error dominates over statistical noise
wider disks are easier to detect as less of the disk flux is lost in
PSF subtraction and more disk flux may fall outside the region of PSF
uncertainty.

In the small disks limit there are two contributions to the noise term
$N_{\rm{op}}$, $N_{d\theta}$ from the PSF uncertainty and
$\sqrt{A_{\rm{op}}/A_\theta}N_\star$ from the statistical noise in the
optimal region. A high signal to noise
will mean that $N_{\rm{tot}} \simeq N_{d\theta}$ for small disks, as
can be seen by the convergence of the disk detectability limits with 
$S_\star = 200$ and 5000 when $r/\theta$ is small. Conversely when
$S_\star$ is low the statistical errors can dominate even in the small
disk limit and there is little difference in the detectable disk
limits for small or large $d\theta/\theta$, as can be seen in the
limits for $S_\star = 8$.  The dominance of $N_{d\theta}$ for small
disks and large $S_\star$ means that for small disks there is a limit
at which detectability cannot be improved by increased observation
time (increased $S_\star$).  We can identify this point by considering
when $N_{d\theta} > \sqrt{A_{\rm{op}}/A_\theta} N_\star$, i.e. when
$N_{\rm{tot}}$ is dominated by PSF errors.  Using $N_{d\theta}$ as
given in equations \ref{faceon} we can see that $N_{d\theta}$
dominates when 
\begin{eqnarray}\label{ndthdom_face}
S_\star & > & \frac{\sqrt{A_{\rm{op}}/A_\theta}(dr/r)
  10^{10(r/\theta)}}{d\theta/\theta} R_\lambda^{0.5}(1-R_\lambda). 
\end{eqnarray}

\subsubsection{Edge-on ring}

\begin{figure*}
\begin{minipage}{8cm}
\includegraphics[width=8cm]{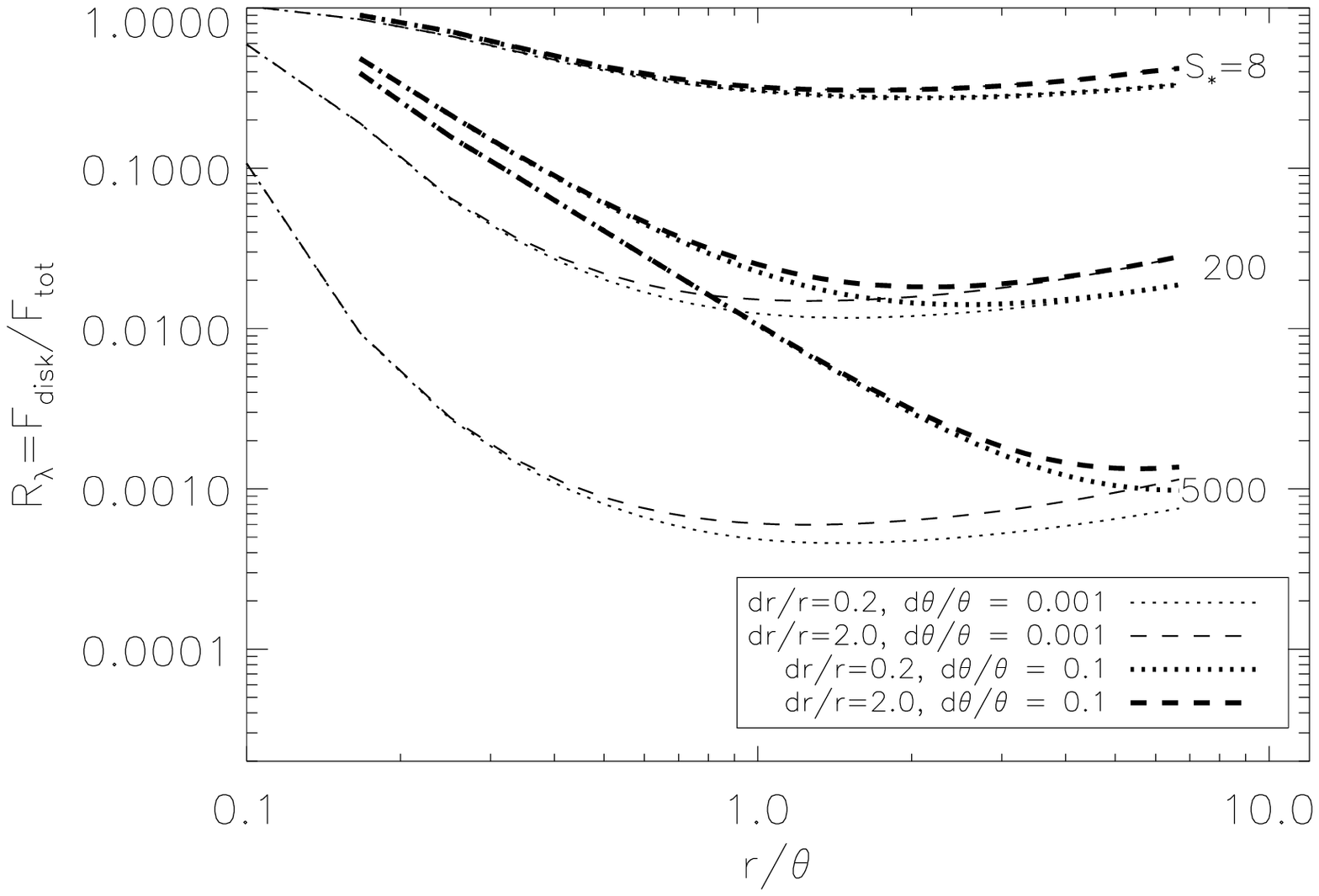}
\end{minipage}
\hspace{0.5cm}
\begin{minipage}{8cm}
\includegraphics[width=8cm]{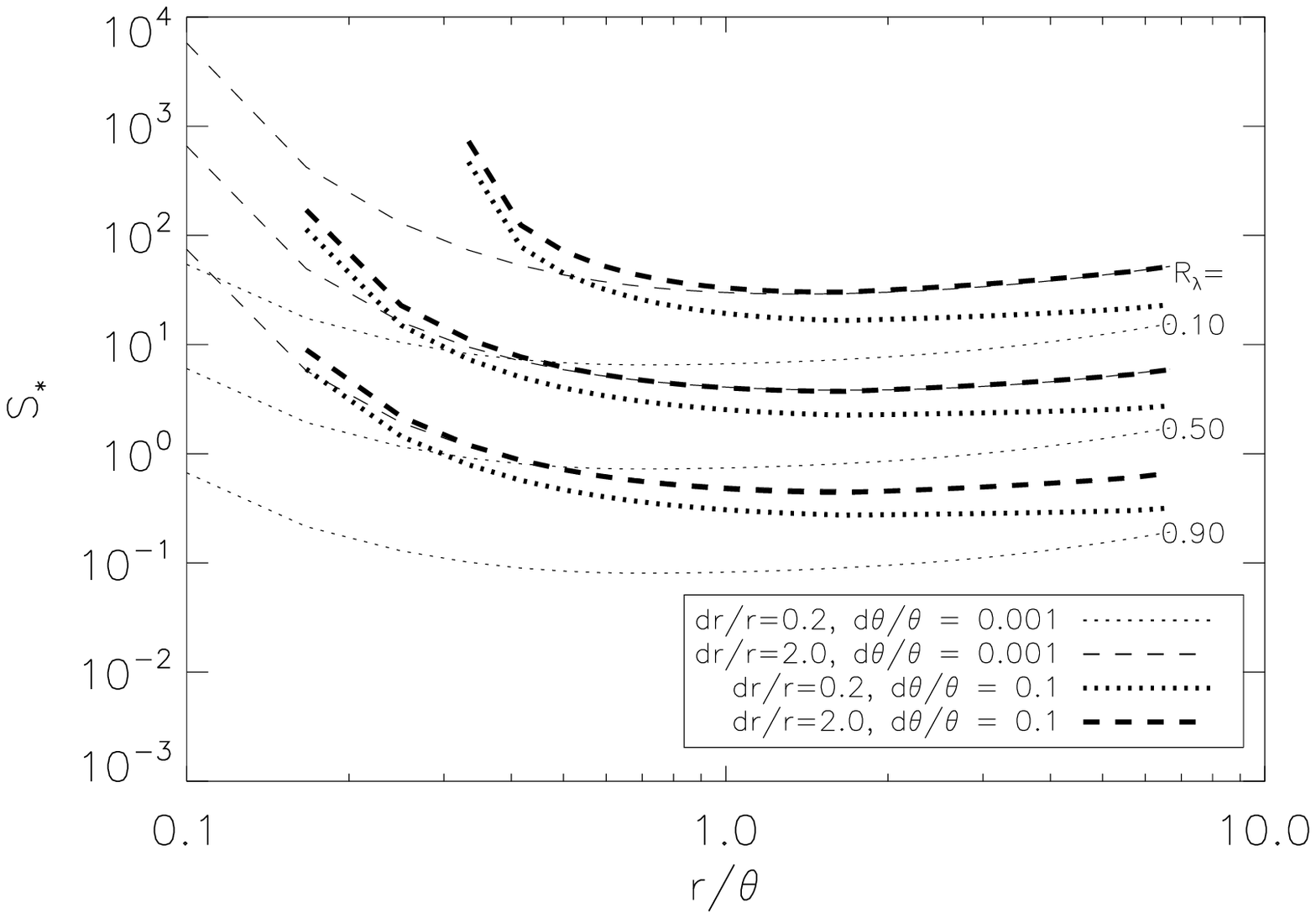}
\end{minipage}
\vspace{-0.2cm}
\caption{\label{limedge} Limits on detectable edge-on disks for various
disk parameters. The region above the lines represents the region of
detectability.   Left: The disk flux required to get a 3 sigma
  detection of extension for disks of varying geometry in an edge-on
  orientation (disk flux given in terms of $R_\lambda$). Any disk
  above the line of detection would be detected at a significant
  level. Right: The signal-to-noise required for a significant
  detection for varying $R_\lambda$.}
\end{figure*}

For an edge-on ring the optimum region can be modelled by a 
rectangular box with side lengths in the major and minor directions of
$L_{\rm{maj}}$ and $L_{\rm{min}}$ respectively.  The orientation of the
major axis is that of the edge-on disk, which in testing the model
limits is known as we know the input model. In the testing of actual
source images for a disk, all orientations of major axis should be
tested.  We find 
\begin{eqnarray}\label{edgeon}
L_{\rm{maj}}/r & = & 2
\sqrt{2(dr/r)^{0.5}+0.5(r/\theta)^{-2.7} + 10R_\lambda(d\theta/\theta)
  (1+r/\theta)^{-2} } \nonumber \\
L_{\rm{min}}/r & = & 2\sqrt{0.007(dr/r)^{0.5} + (0.3-0.8d\theta/\theta
  R_\lambda)(r/\theta)^{-2}} \nonumber \\
F_{\rm{op}} & = & (1-(d\theta/\theta)^{2}(r/\theta)^{-2}R_\lambda^{-1})
10^{-0.25(r/\theta)^{-1}} F_{\rm{disk}}  \\
N_{d\theta} & = & 0.1(d\theta/\theta)10^{-10(r/\theta)}
R_\lambda^{-0.5}F_{\rm{tot}} \nonumber
\end{eqnarray}
with $N_{\rm{op}}$ and $S_{\rm{op}}$ determined from equation
\ref{eq:sop}.  With these equations we can fit the numerical results
for $S_{\rm{op}}$ to better than $\pm$ 50\% for 80\% of the disk
models tested.  Notice that as with the face-on disks when $r/\theta$
is large $F_{\rm{op}} \simeq F_{\rm{disk}}$ and we have 
\begin{equation}\label{sop_edge}
S_{\rm{op}} \simeq 
S_\star (\theta/r)(R_\lambda^{-1}-1)^{-1} / \sqrt{L_{\rm{maj}}
  L_{\rm{min}} / \pi r^2}. 
\end{equation}

The required levels of $R_\lambda$ as a function of $r/\theta$
and disk geometry to get a significant detection, as well as the
signal required for a detection for a given $R_\lambda$ are shown in
Figure \ref{limedge}. In general the detectability limits for an
edge-on disk follow a similar pattern to the limits for a face-on
disk, as can be seen in the 
similarity between figures \ref{limface} and \ref{limedge}.
The differences can be understood as follows: 
The increased $R_\lambda$ or $S_\star$ required for a significant
detection is less steep in $r/\theta$ than for the face-on case
because in the edge-on case the loss in surface
brightness with increasing disk radius is slower than for a face-on
disk. Thus for a fixed $\theta$, the signal to 
noise will be generally higher in the edge-on geometry than for a
face-on disk. Also in the edge-on case there is no dependence
of $N_{d\theta}$ on $dr/r$, and so for small disks there is little
difference between the detectability of wide and narrow disks. Errors
will be dominated by PSF uncertainty (through $N_{d\theta}$) in the
small disk case provided  
\begin{equation}\label{eq:sastedge}
S_\star > 
\frac{\sqrt{A_{\rm{op}}/A_\theta}10^{10(r/\theta)}}{10(d\theta/\theta)}
R_\lambda^{0.5}(1-R_\lambda). 
\end{equation}

\subsubsection{Inclined Ring}
The case of an inclined disk, not edge on, falls between these two
extrema, and the optimal region can be determined by interpolation
between the two models dependent on the sine of the disk inclination,
$\sin(I)$.  The signal to noise for an inclined disk, and thus the
disk flux required for a detection for a given observation, also
follows a smooth transition between the two extremes.

\subsubsection{Summary}

The equations and figures in this section can be used as a guide to
what disks may 
be detectable as extended sources in single dish imaging. The plots of
$R_\lambda$ vs $r/\theta$ for different sensitivity of observation
(characterised by $S_\star$) can be used to provide guidelines as to
how bright a disk must be compared to the star to be detected for
different geometries. Any disks lying below the lines shown cannot be
detected as extended sources, thus if an observation shows no evidence
of extension, the area below the lines of detectability give the
region of the parameter space in which the disk can lie. The plots of
$S_\star$ vs $r/\theta$ can be used to determine the required
observational time to resolve a disk in terms of the signal to noise
required on the point-like star (combined with knowledge of the
instrumental sensitivity and an approximation of the PSF) if the disk
parameters are known or can be approximated (for example from SED
fitting).  Predictions based on these models for the
  resolvability limits acheivable with 8m telecsopes and comparison
  with already resovled disks will be included
  in a forthcoming paper (Smith \& Wyatt in prep.). 

The limits that can be placed on the extension of a disk for a
given observation are dependent upon having a measure of $R_\lambda$.
Often the disk flux is poorly constrained by the photometry, and so
this limits the accuracy to which the possible extent of the disk can
be constrained.  If the disk flux is well known, then 
there are essentially two regimes when determining the
detectability of disk extension.  When $r/\theta > 1$, variations in
the PSF have little effect on the optimal region and the signal to
noise therein, and extension detection is limited purely by the
background statistical noise on the array.  When $r/\theta < 1$, the
variation in 
the PSF dominates the noise through the $N_{\rm{d\theta}}$ term, and
thus disk detections are limited by the degree of certainty
to which the PSF can be characterised.  A disk cannot be detected to a
smaller size than the absolute errors in the PSF, or obviously to
smaller than the pixel scale of the images, regardless of the signal
strength of the observation. We acknowledge that we are in affect
talking about super-resolution of the disks, as in our models we can
detect extension just larger than a single pixel scale if the PSF is
perfectly known.  In reality however, variation in the PSF both in
terms of absolute width and variation in shape will severely restrict
the possibility of resolving disks of this size. 
Figures \ref{limface}
and \ref{limedge} also show that the optimal disk size for detectability
changes from $\sim r/\theta$ when $ d\theta/\theta$ is very small to
larger radii with larger and more realistic values of $d\theta/\theta$.
It is worth
reiterating that the value of $\theta$ does not encompass all of the
information about the PSF, in particular any asymmetries or
ellipticity $\neq 0$ can affect extension limits, therefore when
determining the limits placed on the observed sources in this paper,
we used the PSF determined for each source.
For disks smaller than the limits to which we may reasonably expect a
stable and unvarying PSF, single aperture imaging will be unable to
resolve the disk and interferometric observations will be needed. 

\begin{table*}
\caption{The Observations}
\label{observations}
\centering                          
\begin{tabular}{*{9}{|c}|} \hline Star name &
  \multicolumn{3}{|c|}{Observation} & Exp. Photospheric & 
  \multicolumn{4}{|c|}{Results$^{a}$} \\ HD &
  $\lambda$, $\mu$m & Int. time, s & Instrument &  Flux, mJy & 
 Flux, mJy & Tot. Error, mJy & Stats. Error & Background limit,
  mJy$^b$ \\ \hline
10800 & 11.59 & 1800 & TIMMI2 & 513 & 477 & 54 & 15 & $\le$ 39 \\ 
      & 18.72 & 3762 & VISIR & 200 & 186 & 29 & 6 & $\le$ 14 \\ \hline
12039 & 11.85 & 3588 & VISIR & 72 & 77 & 3 & 1 & $\le$ 2 \\ \hline
53246 & 11.59 & 1800 & TIMMI2 & 87 & 111 & 30 & 25 & $\le$ 62 \\ \hline
65277 & 11.59 & 2400 & TIMMI2 & 197 & 197 & 38 & 11 & $\le$ 31 \\
      & 11.85 & 1794 & VISIR & 188 & 182 & 4 & 2 & $\le$ 5 \\
      & 18.72 & 3762 & VISIR & 77 & 78 & 14 & 4 & $\le$ 10 \\
Binary  & 11.59 & 2400 & TIMMI2 & 55 & 33 & 17 & 11 & /  \\ 
      & 11.85 & 1794 & VISIR$^d$ & 53 & 32 & 5 & 2 & /  \\ 
      & 18.72 & 3762 & VISIR$^d$ & 21 & 14 & 6 & 4 & /   \\ \hline 
69830 & 9.56 & 1980 & TIMMI2 & 941 & 1255 & 135 & 32 & $\le$ 84 \\ \hline
79873 & 11.59 & 1800 & TIMMI2 & 167 & 160 & 18 & 11 & $\le$ 28 \\ 
      & 18.72 & 1881 & VISIR & 65 & 39 & 9 & 5 & $\le$ 12 \\ 
Binary  & 11.59 & 1800 & TIMMI2 & 14 & 0 & 11 & 11 & /  \\
      & 18.72 & 1881 & VISIR & 6 & 0 & 5 & 5 & /  \\ \hline
$\eta$ Corvi & 9.56 & 1620 & TIMMI2 & 1896 & 2883 & 240 & 63 & $\le$ 162 \\
(109085) & 10.54 & 3600 & TIMMI2 & 1565 & 2451 & 373 & 48 & $\le$ 84 \\
       & 11.59 & 840 & TIMMI2 & 1298 & 2151 & 127 & 40 & $\le$ 76 \\
       & 11.6 & 1244 & MICHELLE & 1296 & 1626 & 184 & 5 & $\le$ 33 \\ 
       & 11.85 & 1076 & VISIR$^\ast$ & 1243 & 1951 & 216 & 19 & $\le$ 28 \\ 
       & 18.72 & 1881 & VISIR & 505 & 814 & 76 & 10 & $\le$ 23 \\ \hline
123356 & 10.54 & 660 & TIMMI2 & [18]$^c$ 681 & 207 & 78 & 64 & $\le$
  164 \\ \hline
128400 & 8.60 & 600 & TIMMI2 & 498 & 469 & 92 & 41 & $\le$ 109 \\
       & 9.56 & 661 & TIMMI2 & 406 & 507 & 118 & 61 & $\le$ 162 \\ \hline
145263 & 8.60 & 1380 & TIMMI2 & 37 & 426 & 57 & 25 & $\le$ 64 \\ \hline
191089 & 12.21 & 1440 & TIMMI2 & 98 & 92 & 27 & 16 & $\le$ 43 \\ \hline 
202406 & 9.56 & 1800 & TIMMI2 & 83 & 270 & 43 & 12 & $\le$ 30 \\
       & 11.59 & 1560 & TIMMI2 & 57 & 278 & 54 & 16 & $\le$ 43 \\ \hline
\end{tabular}

The expected photospheric emission is determined by a Kurucz model
profile appropriate to the spectral type of the star and scaled to the
K band 2MASS magnitude as outlined in section 2 unless otherwise
stated in the individual source description. Errors are 1$\sigma$.  M
band TIMMI2 observations were largely non-photometric and primarily
used to improve pointing accuracy and thus are not listed in this
table.  
Notes: $^a$ Errors are total errors (inclusive of calibration
uncertainty and image noise). $^b$ Limits are 3$\sigma$ upper limit to
undetected object including calibration errors, or scaled to IRAS
fluxes when conditions were non-photospheric. These limits are valid
to within 28\arcsec of the detected source for TIMMI2 observations,
12\farcs6 for MICHELLE observations and 11\farcs4 of the source for
VISIR observations.  $^c$ Here the companion
object is brighter than the primary; we show the primary flux in
brackets; $^\ast$ This observation was
affected by rising cirrus, and so levels of noise on the image are
much higher than other observations taken with this filter. 
\end{table*}
\begin{table*}
\caption{The fits for stars with confirmed excesses}
\label{results}
\centering                          
\begin{tabular}{*{7}{|c}|} \hline Star name &
 \multicolumn{3}{|c|}{Fit as dust disk} & Limit on extension & 
$f_{\rm{IR}} = L_{\rm{dust}}/L{\ast}$ & $f_{\rm{max}}^{\,a}$
\\ HD & Temp, K &
 Radius, AU & Radius, \arcsec  &  Radius & $\times 10^{-5}$ & $\times
 10^{-5}$ \\ \hline 
$\eta$ Corvi & 320 & 1.7 & 0.09 & $<$0\farcs164 ($^{+0.014}_{-0.009}$) & 26 &
0.042   \\  
(109085) & 360 + 120$^b$ & 1.3 + 12 & 0.07 + 0.66 & - & 22 + 6 & 0.022
+ 4.01  \\ \hline 
145263$^c$ & 290 & 1.8 & 0.015 & $<$0\farcs69 $^{+0.31}_{-0.21}$ & 2033 &
7.0  \\  
202406$^d$ & 290 & 7.4 & 0.025 & $<$0\farcs33$^{+0.21}_{-0.13}$  & 371
& 22.9 \\ \hline
12039 & 120$^e$ & 5.05 & 0.12 & - & 8.9 & 23.3 \\ 
69830 & 390$^f$ & 0.33 & 0.026 & - 
& 25.4 & 0.0006  \\ 
191089 & 110 & 11.5 & 0.21 & - & 233 & 47.4  \\ \hline
\end{tabular}

Note that the objects with no extension limits have too low a
fractional excess for the extension to have been detected in the
images regardless of size. Estimates of radius are based on blackbody
fits and could be up to three times larger than suggested (Schneider
et al. 2006).  Limits shown here are for a narrow face-on
disk. Errors arise from 3 sigma photometric errors - see section
4.1.4.  Horizontal lines indicate division into photometrical
confirmed debris disks, suspected pre-main sequence stars, and sources
for which our results provide constraints on the disks (sections 5.1,
5.2 and 5.3 respectively). \\
Notes: $^a$ see section 6.2 for details of this limit; $^b$ Fit
suggested by Chen et al. (2006); $^c$ HAeBe Star ; $^d$ Possible HAeBE
star, see section 5.2 ; $^e$ Fit from Hines et al. (2006); $^f$
Beichman et al. (2006) suggest Hale-Bopp type cometary material. 
\end{table*}

%

\section{Results and Analysis}
\label{s:res}

The observed sample can be divided into several sub-groups: main
sequence stars with confirmed hot dust; hot dust hosts that have been
incorrectly identified as main sequence objects; and those with no
excess or whose infrared excesses are actually due to
background/companion objects or statistical anomaly.  
Table \ref{observations} gives a brief description of the results, and
Table \ref{results} gives the best fits to the objects for which the
excesses are confirmed.   Sources are discussed individually below.

\begin{figure*}
\begin{minipage}{8cm}
\includegraphics[width=8cm]{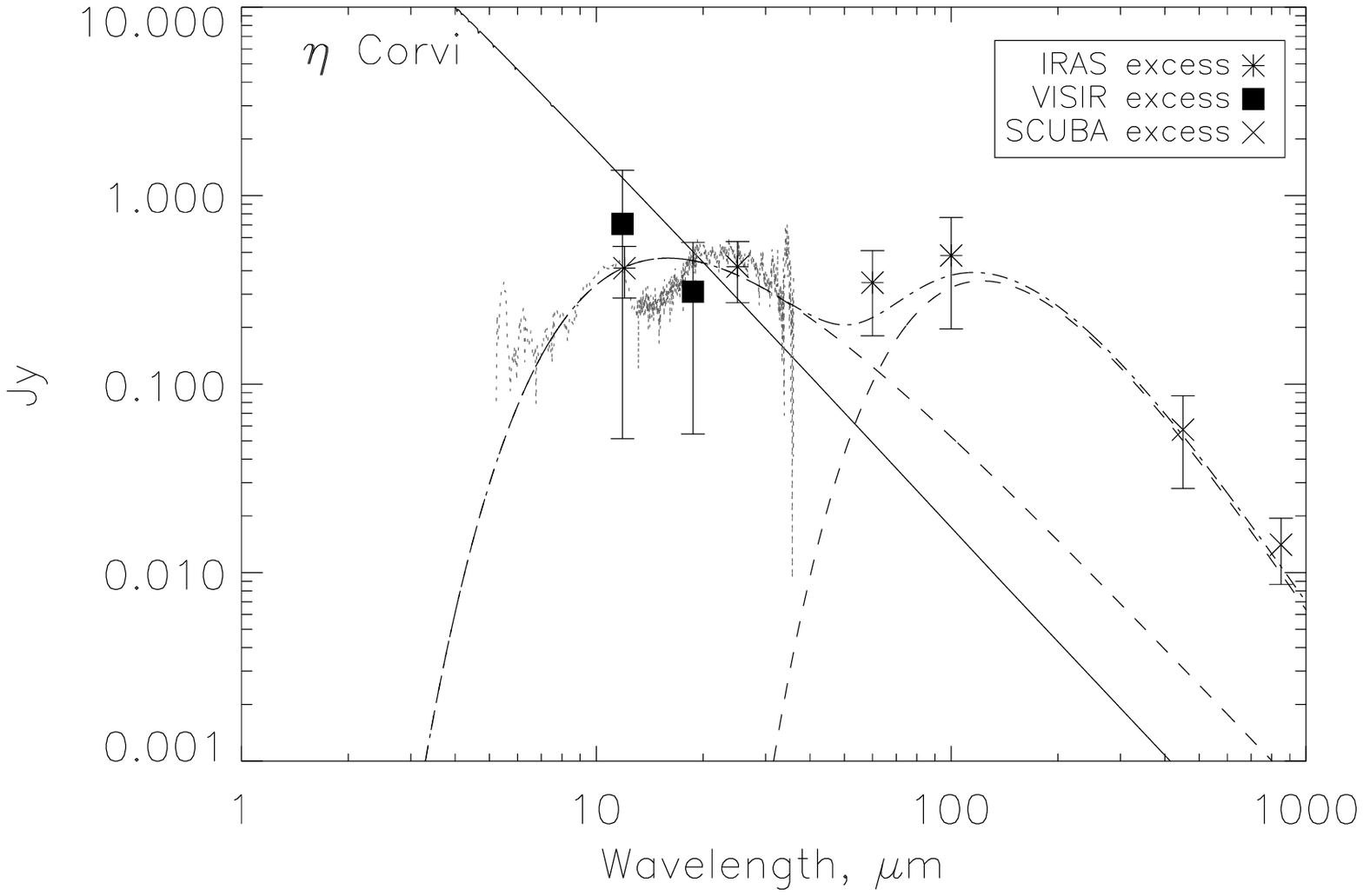}
\end{minipage}
\hspace{0.2cm}
\begin{minipage}{8cm}
\includegraphics[width=8cm]{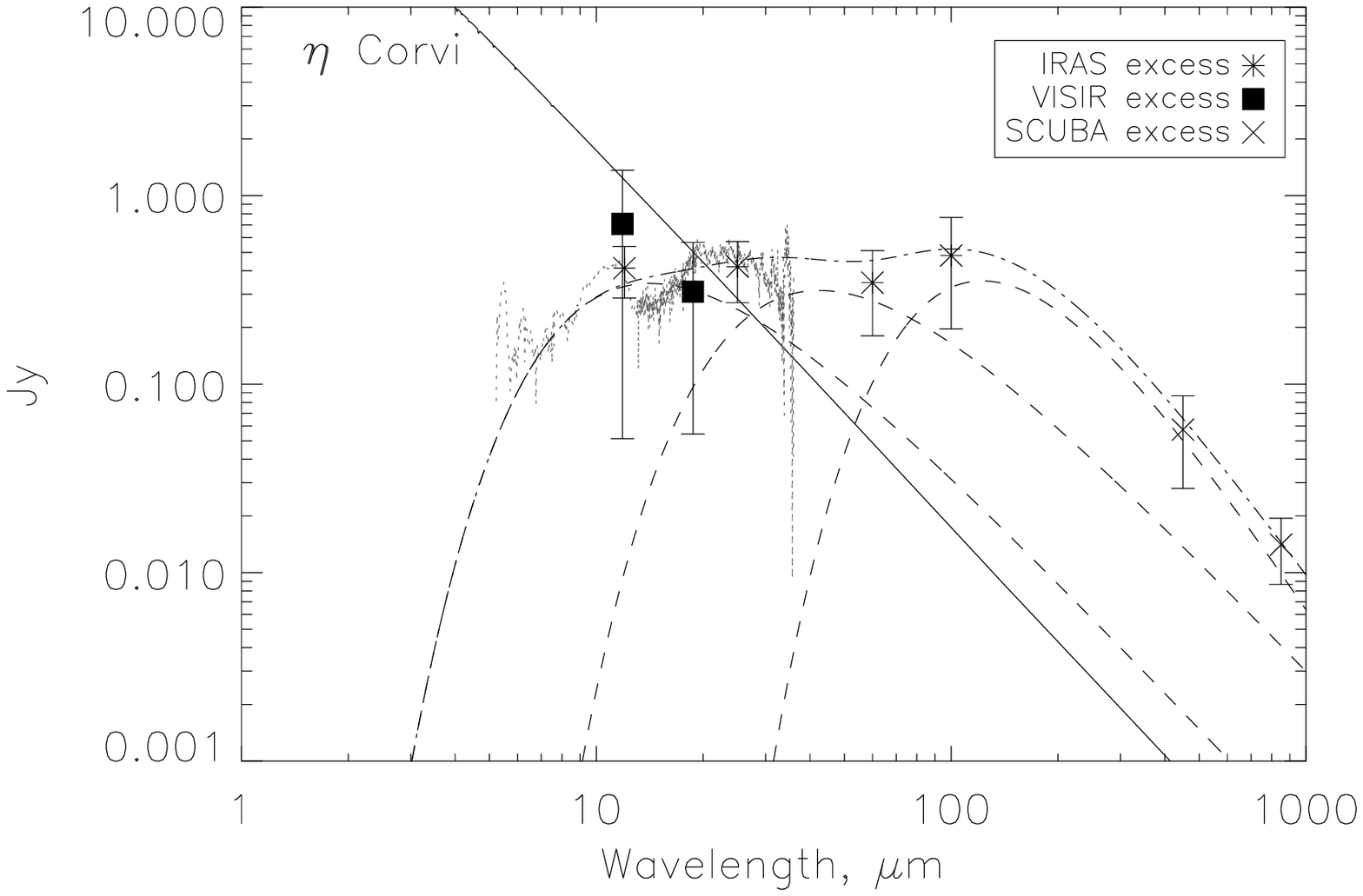}
\end{minipage}
\caption{\label{etaSED} The two alternative fits to the excess
  emission of $\eta$ Corvi.  The symbols $> 10 \mu$m represent
  calibrated flux  after subtraction of photospheric emission. Error bars are 3
  $\sigma$.  The grey dotted line represents the IRS spectra of Chen et al.
  (2006) after subtraction of the photosphere. The dashed lines
  indicate blackbody emission modelling of the disk flux, left: model
  A; and right: model B;  and the dot-dashed lines the total emission from the
  multi-temperature disk. }
\end{figure*}

\begin{figure*}
\begin{minipage}{1cm}
N 
\end{minipage}
\begin{minipage}{4cm}
\center{$\eta$ Corvi}
\includegraphics[width=4cm]{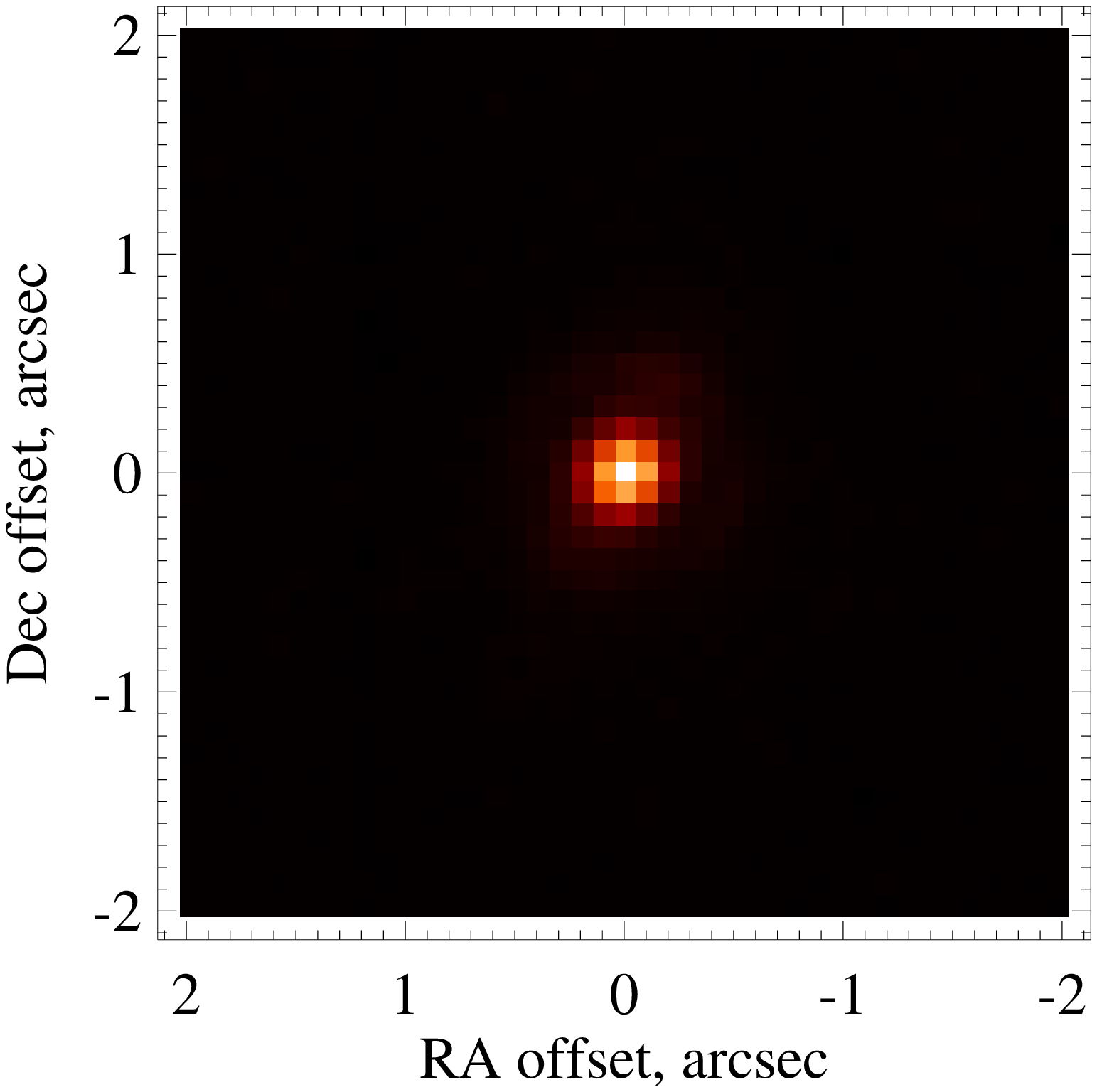}
\end{minipage}
\begin{minipage}{4cm}
\center{Average PSF}
\includegraphics[width=4cm]{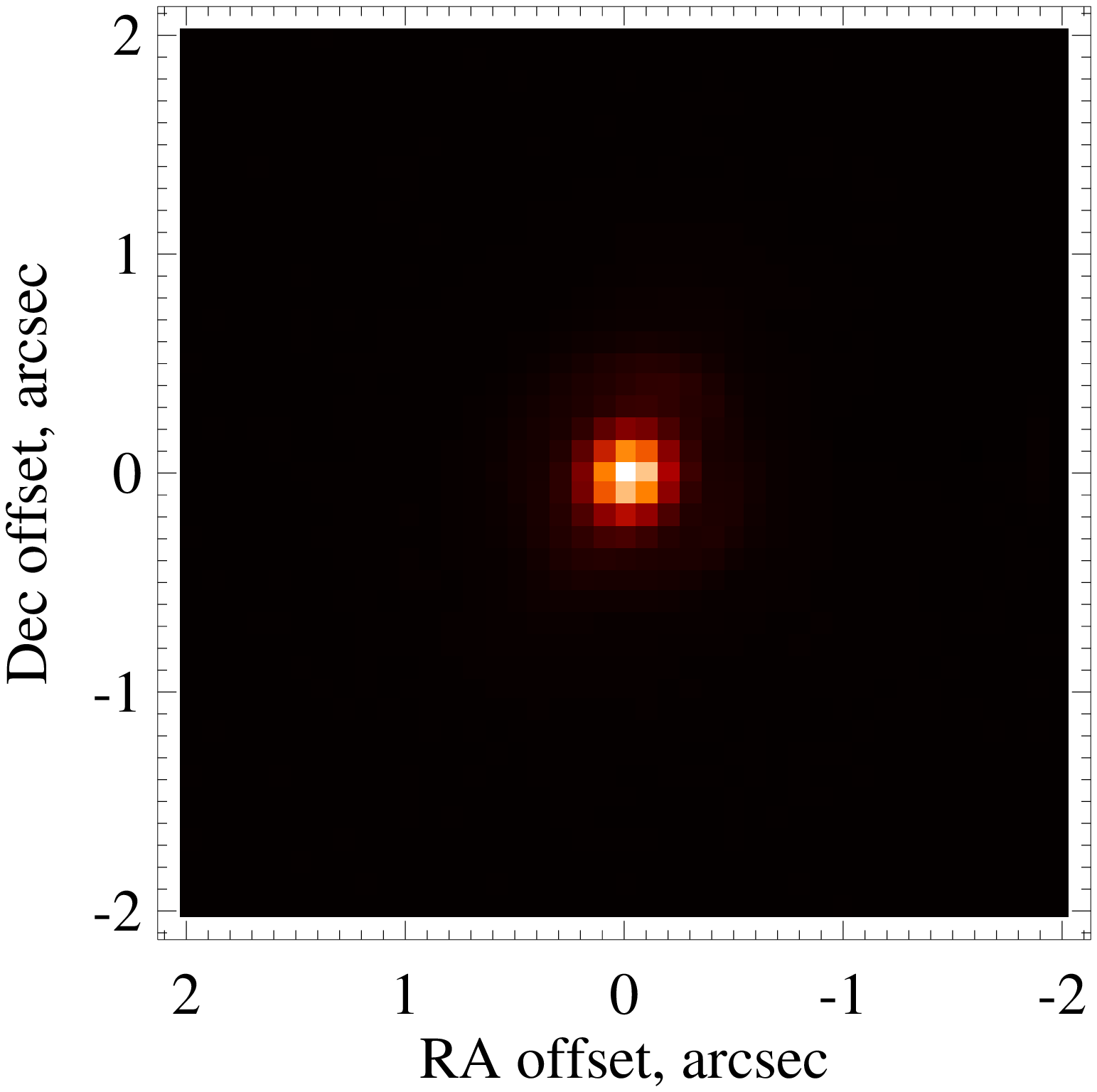}
\end{minipage}
\begin{minipage}{4cm}
\center{Residuals}
\includegraphics[width=4cm]{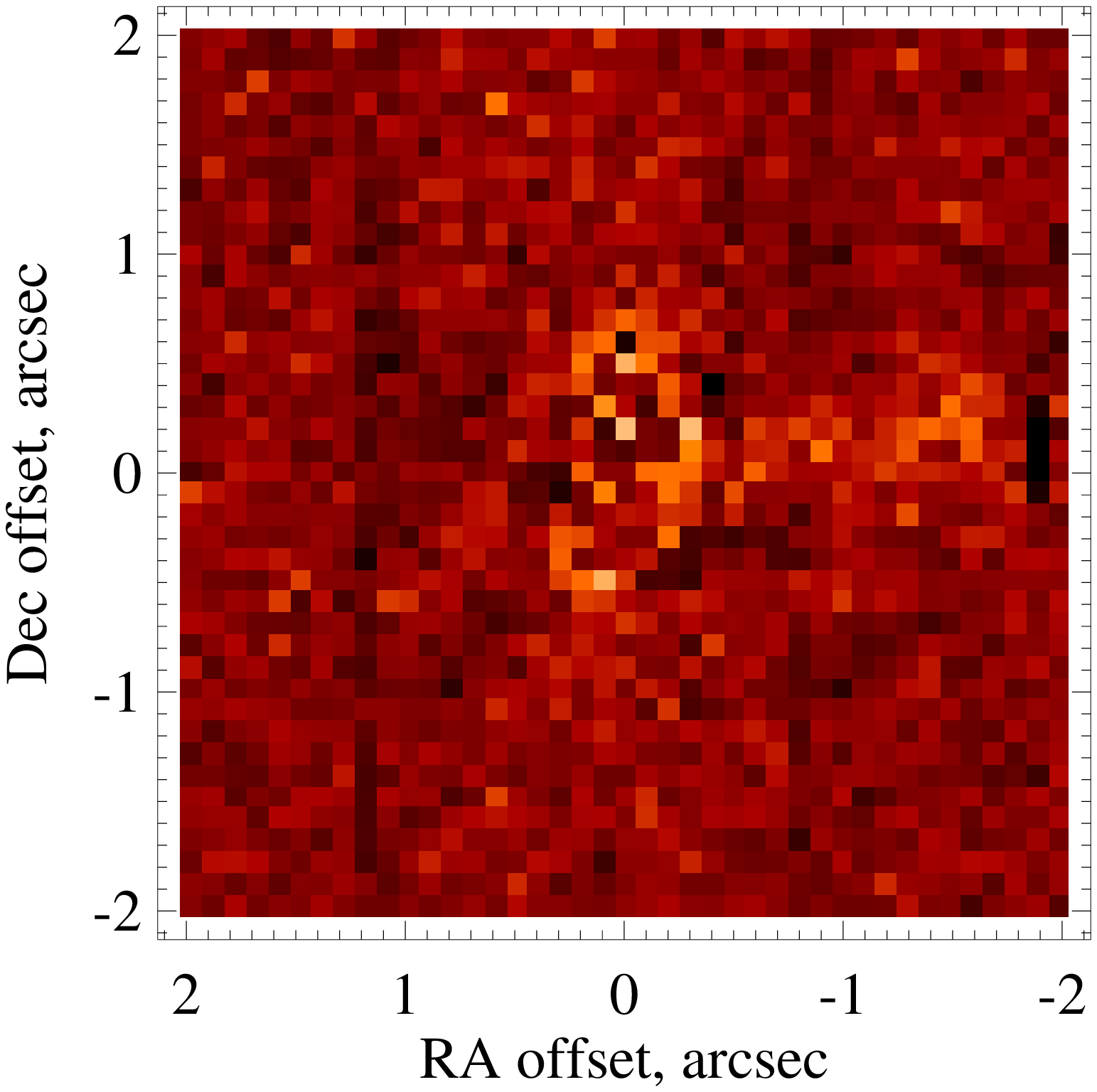}
\end{minipage} \\
\begin{minipage}{1cm}
Q
\end{minipage}
\begin{minipage}{4cm}
\includegraphics[width=4cm]{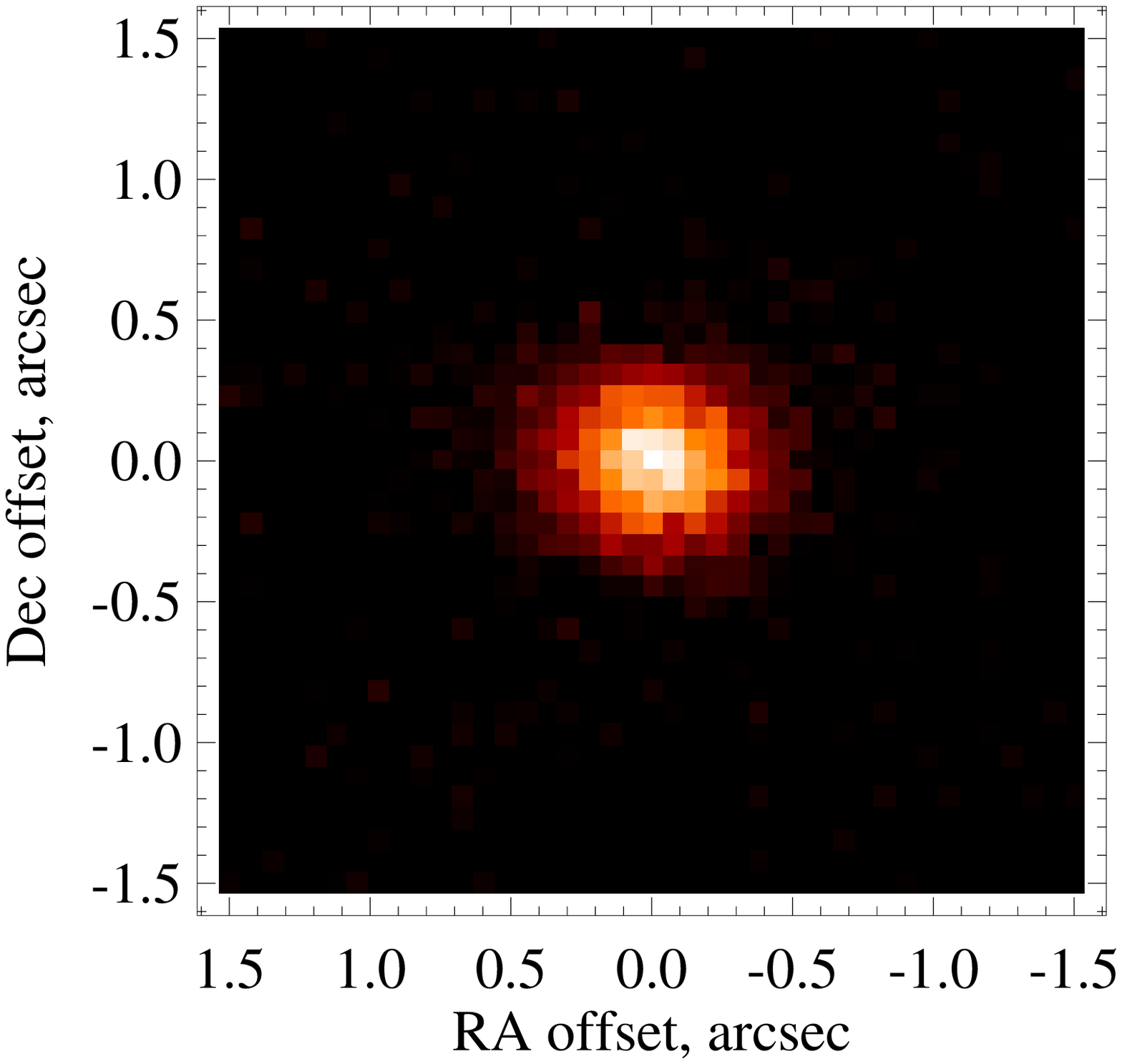}
\end{minipage}
\begin{minipage}{4cm}
\includegraphics[width=4cm]{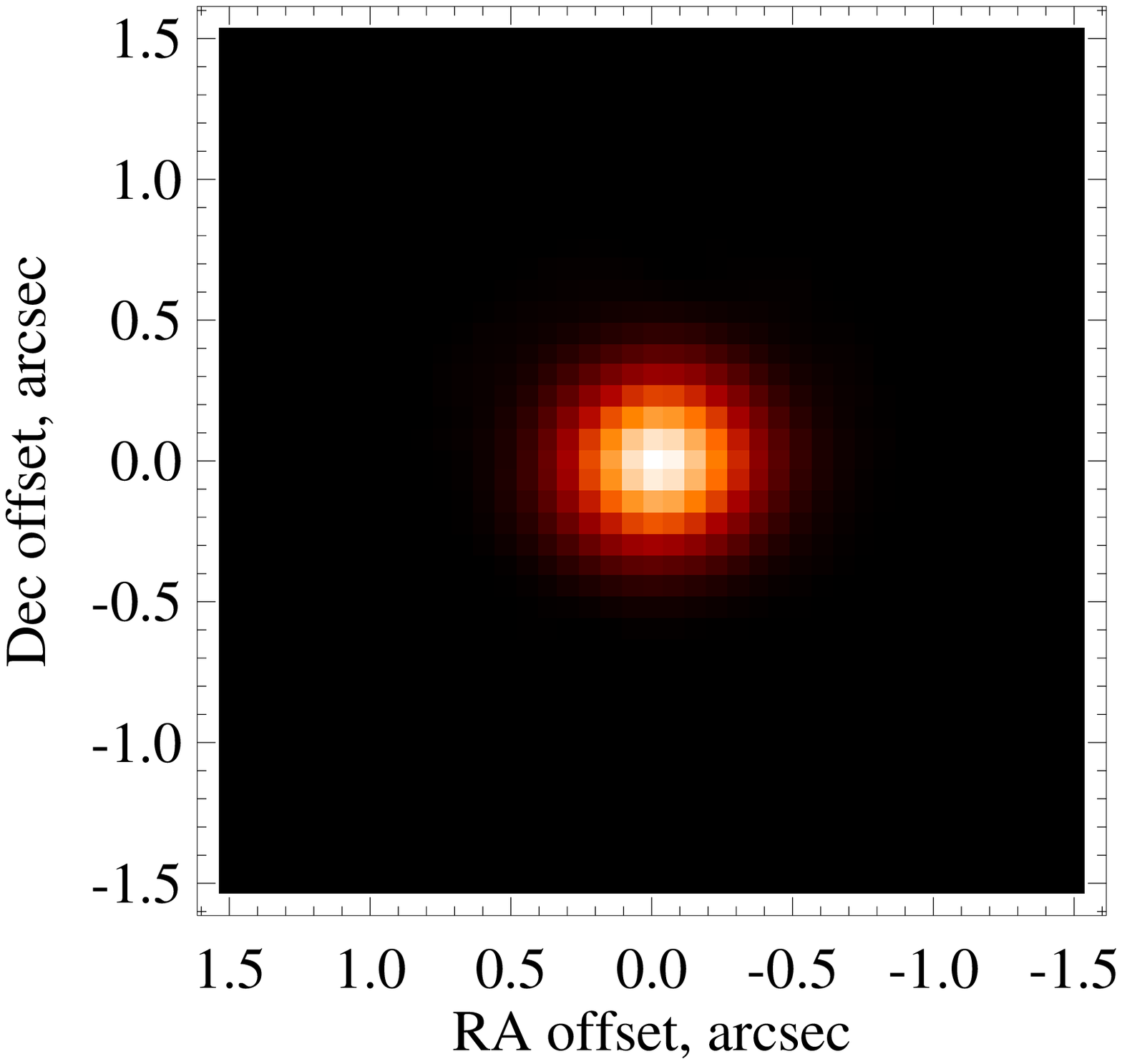}
\end{minipage}
\begin{minipage}{4cm}
\includegraphics[width=4cm]{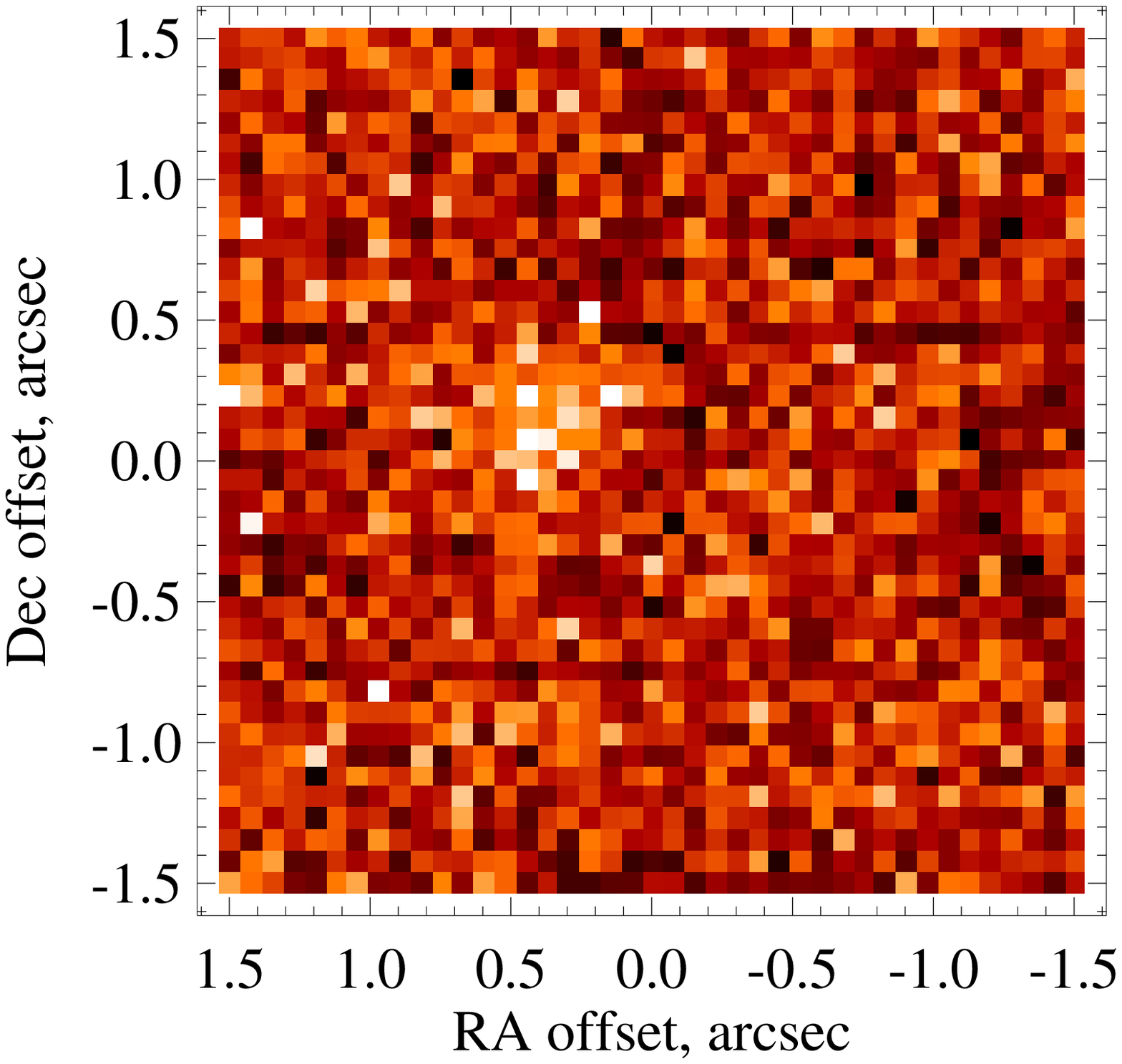}
\end{minipage}
\caption{\label{etaimages} The final images of $\eta$ Corvi and 
the standard star associated with it, and the $\eta$ Corvi
image after subtraction of the scaled standard star image which is
examined for residuals indicative of extended disk
emission. \emph{Top:} The MICHELLE N band images; \emph{bottom:} the
VISIR Q band images. All scales are linear.  The images of the
residual emission are shown with minimums (black) of -3 $\sigma$ and
maximum (white) of +3 $\sigma$ (where $\sigma$ is the background noise
level per pixel). While the Q band residuals appear to show a
12$\sigma$ peak to the East (determined in a 0\farcs35 radius
aperture, see section 3.2), this emission does not appear after
subtraction of the first standard star image only, which is broader
than the standard star observation taken after observing $\eta$
Corvi (signal in same aperture is 0.7$\sigma$). }
\end{figure*}

\begin{figure*}
\begin{minipage}{7cm}
\includegraphics[width=7cm]{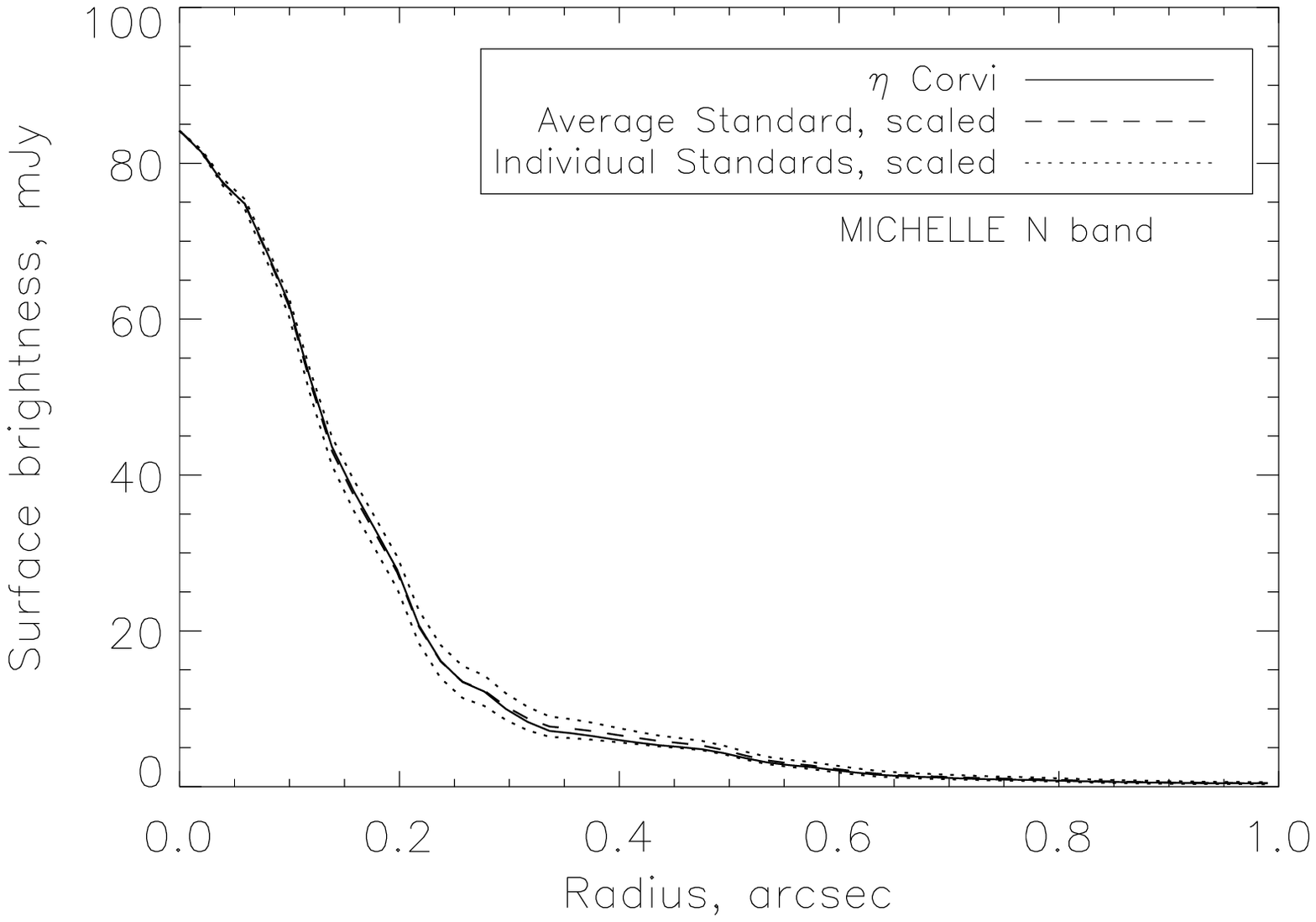}
\end{minipage}
\begin{minipage}{7cm}
\includegraphics[width=7cm]{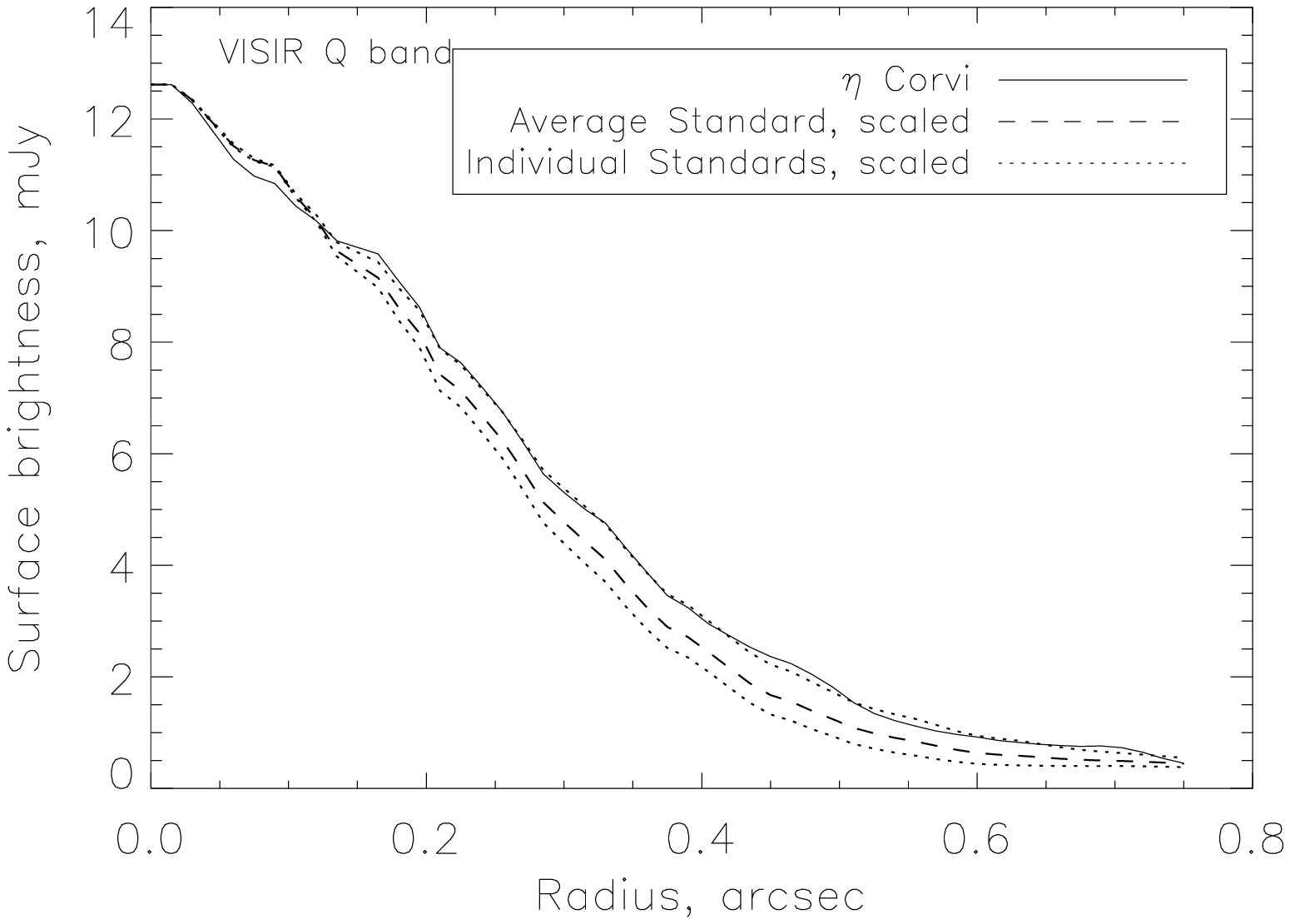}
\end{minipage}
\caption{\label{etasufb}The surface brightness profiles of $\eta$
  Corvi and the standard star images associated with the
  observations. Standard star profiles are scaled to the $\eta$ Corvi
  profile. 
 The profiles of $\eta$ Corvi are consistent with those of the
 point-like standard stars. }
\end{figure*}

\begin{figure*}
\begin{minipage}{8cm}
\includegraphics[width=8cm]{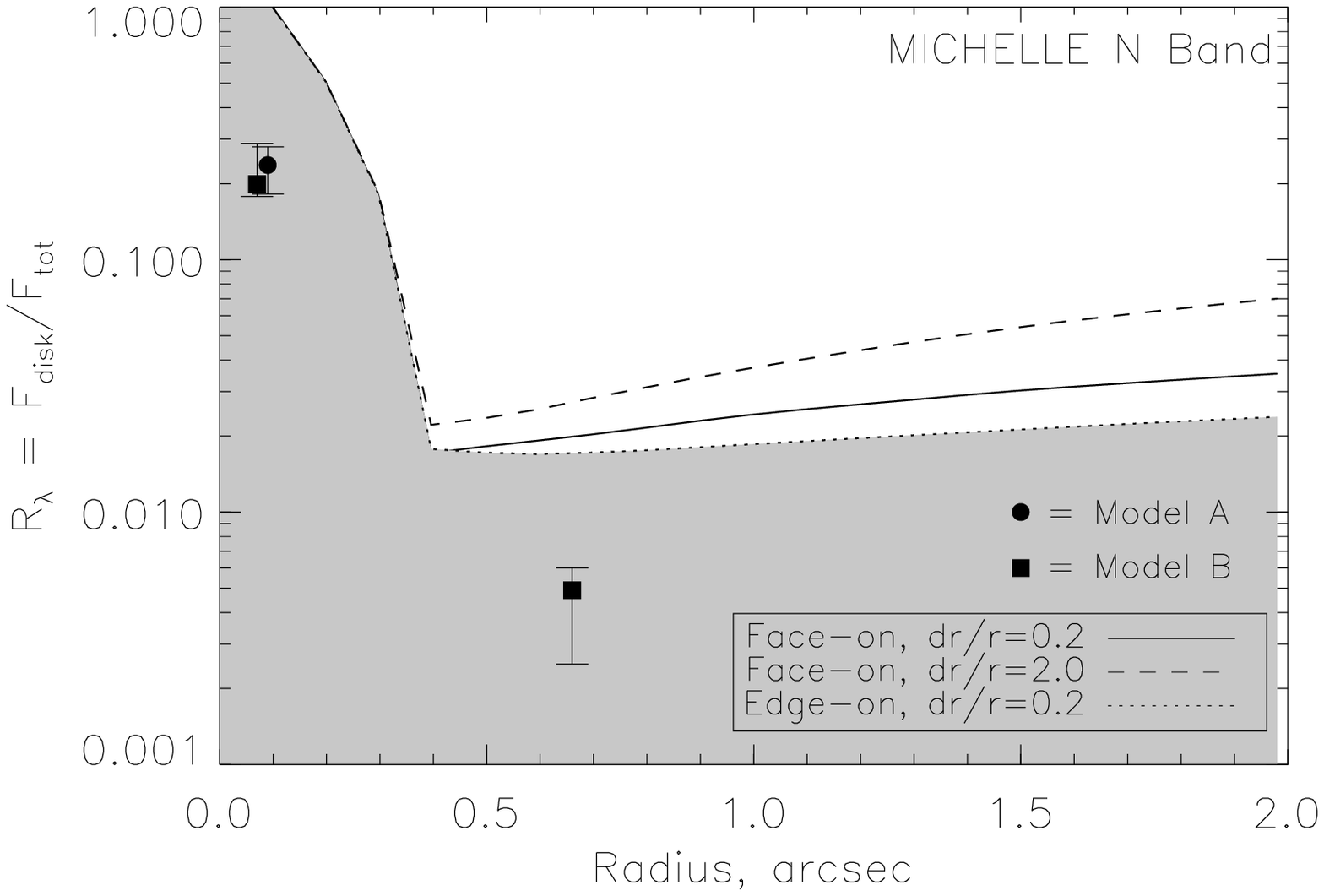}
\end{minipage}
\hspace{0.2cm}
\begin{minipage}{8cm}
\includegraphics[width=8cm]{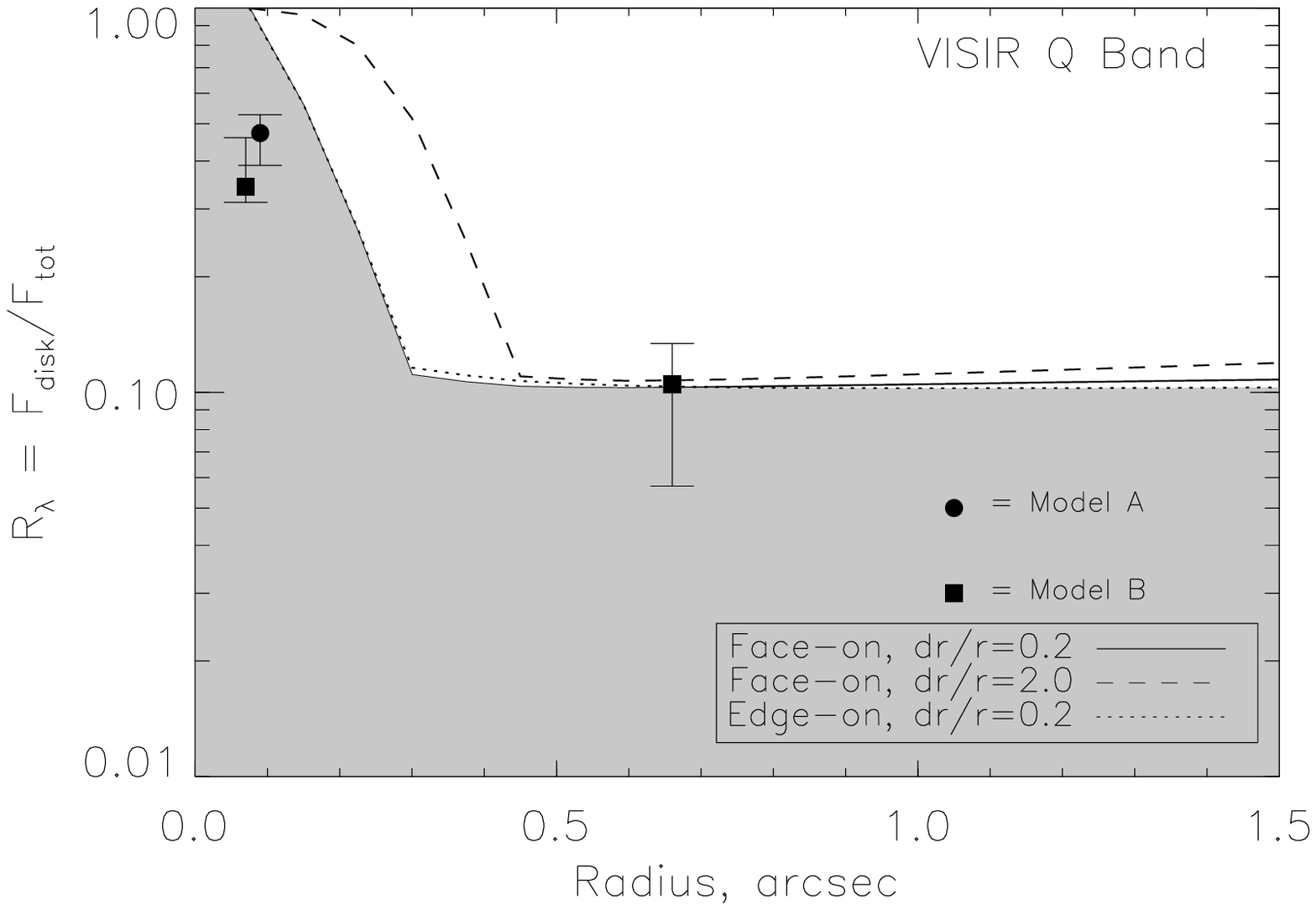}
\end{minipage}
\vspace{-0.2cm}
\caption{\label{exteta} The 3 $\sigma$ limits placed on the disk models by
  non-detection of extension in the images.  \emph{Left:}
  The most stringent limits placed on possible disk radius at N are
  achieved with MICHELLE due to the poor seeing of the VISIR
  observation. The parameters used in determining these limits are
  $S_\star$ = 325, $\theta$ = 0\farcs36, $d\theta/\theta$ = 0.1.
  \emph{Right:} The extension limits given by the VISIR Q
  band imaging. The parameters used in determining these limits are
  $S_\star = 81$, $\theta$ = 0\farcs58 and $d\theta/\theta = 0.08$. 
  In both plots error bars represent the 3 $\sigma$
  errors on $R_{\lambda}$ due to photometric uncertainty.  The shaded
  area is the area in which disk populations could lie given the
  non-detection of extension. Note that the cold dust at 40K imaged in
  the sub-milimetre lies at approximately 45$^\circ$ to the line of
  sight, so between the edge-on and face-on limits presented here. 
  See text for model details and the implications of these limits. }
\end{figure*}

\subsection{Confirmed hot dust around $\eta$ Corvi}

The results confirm the presence of excess emission centred on the star
toward $\eta$ Corvi which was originally shown to have an infrared
excess by Stencel \& Backman (1991) based on the large infrared flux
in the IRAS catalogue. The excess is 412 $\pm$ 42 mJy at 12
$\mu$m and 420 $\pm$ 50 mJy at 25 $\mu$m (Table \ref{sample}). 
 $\eta$ Corvi also has a sub-mm excess, at an approximate temperature
of 40K, which has been imaged by Wyatt et al. (2005) using SCUBA. 
The deconvolved size of this object is 100AU at 850
$\mu$m.  The 450 $\mu$m image can be modelled by a ring at 150AU. The
SED of this object, having a large mid-infrared excess shows evidence
for a hot component in addition to the cool 40K component. However it
is not clear if the hot component is at a single temperature of 370K,
as modelled by Wyatt et al. (2005) or at two temperatures, 360K and
120K, as suggested by Chen et al. (2006) 
 
This source was observed with TIMMI2 at 9.56 $\mu$m,
10.54 $\mu$m and 11.59 $\mu$m.  The images at 11.59 $\mu$m have the
greatest calibration accuracy and were previously reported in Wyatt
et al. (2005).  With these observations a background or companion
source within the TIMMI2 field of view can be ruled out at the level
of less than 76 mJy, indicating the excess is indeed centred on the
star.   

Further observations presented here using VISIR confirm the
presence of excess emission at N and Q centered on the star at a level
consistent with that detected by IRAS and Spitzer (Chen et al. 2006).  
The detected flux is 1951 $\pm$ 216 mJy and 814 $\pm$ 85 mJy at 11.85
and 18.72 $\mu$m respectively (photospheric emission expected to be
1243 and 505 mJy in these filters).  The N band excess emission
from the VISIR observation is higher than that of IRAS at 12 $\mu$m
and IRS, but the large calibration error means that this difference is not
significant.  The MICHELLE observation also has a high calibration
error: the detected flux is 1626 $\pm$ 184 mJy and so does not confirm
the excess at the 3 $\sigma$ level of significance (photosphere
expected to be 1298 mJy). The limit on excess is 
in line with the IRAS measurements (see Table \ref{sample}). 
These data points, together with the IRAS and SCUBA measurements of
excess and the IRS spectra presented by Chen et al. (2006) are shown
in Figure \ref{etaSED}.  The observations allow us to place limits on
possible background companions within the field of view of the
instruments to less than 28 mJy at N and less than 23 mJy at Q.

The final
images for $\eta$ Corvi from the MICHELLE and VISIR Q band imaging are
shown in Figure \ref{etaimages} together with the average PSFs
obtained from the standard star observations and the residuals after
subtracting the scaled average PSF from the science images.  The
average PSF was determined by co-addition of the individual images of
the observed standard star.  It is possible to
assess the level of PSF variability during the observations since
these bright sources can be easily characterised even in short
integrations. Therefore we can compare the FWHM measurements from
2-dimensional Gaussian fits to sub-integrations of the observations, that is
dividing the total dataset for any integration into shorter
integrations of equal length, for both the standard stars and
$\eta$ Corvi.  For the MICHELLE N band observation the standard images
have a median FWHM of 0\farcs357 and standard error 0\farcs0024 (20
sub-integrations), and $\eta$ Corvi a median FWHM of 0\farcs363 with
standard error 0\farcs003 (24 sub-integrations). Note that the  For the Q band
observation with VISIR, the standard star FWHM observations had a
median value of 0\farcs577 and standard error 0\farcs037 (4
sub-integrations) with $\eta$ Corvi having a median FWHM of 0\farcs607
and standard error of 0\farcs018 (observation divided into
  only 3 sub-integrations to have adequate S/N to determine FWHM).
Thus, based on the FWHM measurements 
the $\eta$ Corvi images are not significantly larger than the PSF
images (at either wavelength).  Futhermore, the residual images were
subjected to testing using a wide range of optimal regions as defined
in section 4.1 to search for significant residual emission indicative
of extension. No significant extension was found at either N or Q.
The residual emission in the Q band residual image which
appears to have $\sim$12sigma significance based on the ratio of signal to
pixel-to-pixel statistical noise is not interpreted as extended emission,
since such a calculation does not account for the uncertainty in the PSF.
In fact, the PSF of the standard star observed before eta Corvi looks very
similar to that of eta Corvi (see Figure \ref{etasufb}), and when using this
individual standard star observation (rather than the average) as the
model PSF, the signal in the region previously highlighted for potential
extension (centered on
$\sim$ 0\farcs47, PA 71$^\circ$), is reduced to 29 $\pm$ 37 mJy.
Thus there is no extension beyond the uncertainty in the PSF.
This illustrates the potential to mis-identify extended emission if
PSF uncertainty is not taken into account. 

The observed PSFs were then convolved with our range of disk models
described in section 4.1 and these convolved images treated in the
same way to test which set of disk parameters would have led to a
significant detection in our optimal regions.  Figure
\ref{exteta} shows the extension limits plots for the MICHELLE N band
imaging (which due to better seeing provides more stringent limits than
the VISIR N band imaging) 
and VISIR Q band imaging, which as discussed in section 4.1.4
are strongly dependent on the level of fractional excess, and thus on the
number of disk temperatures used to fit the excess emission.  We
discuss these limits in the context of two possible interpretations,
labelled model A and model B, in which the dust emitting in the
mid-infrared is at either A: one temperature, or B: two temperatures,
making the further assumption that each temperature corresponds to a
different radius in the disk.  To determine the limits the
  non-extension places on the different models the value of
  $R_{\lambda} = F_{\rm{disk}}/F_{\rm{total}}$ is crucial.  In the
  following limits discussion, the 
  value of $R_\lambda$ adopted is derived from the IRS spectrum at the
  wavelengths of the images used (11.6$\mu$m and 18.72$\mu$m for
  MICHELLE N band and VISIR Q band images respectively), as this
  spectrum provides more accurate photometry than our ground-based
  observations.  The
  blackbody fits shown on the SED plots in Figure \ref{etaSED} act as
  a guide to an approximate temperature and thus location of the dust
  populations. For model A the excess emission at both N and Q is
  assumed to come from a single component at a single location.  For
  model B the blackbody fits have
  been used to give relative contributions to the emission at
  each wavelength from the two components. In both models the cold
  disk component imaged by Wyatt et al. (2005) is fit by a 40K
  blackbody, and does not contribute to the flux in the mid-infrared. 

Model A:  The IRS photometry suggests fractional excess of
$R_\lambda = 0.24$ and 0.47 at N and Q.   The extension limits show that
assuming a face-on narrow disk geometry, a single disk component must be at
less than 0\farcs164 $\pm$ 0\farcs01 (from the tightest Q band limit,
errors from uncertainty in $R_{\lambda}$ from IRS spectra uncertainty),
which translates to a radial offset of 2.98 AU.  Assuming a
wide ring geometry the limit is 0\farcs253 (4.6 AU, see Figure
\ref{exteta}).   Using a single temperature blackbody to fit the hot
component we find that a fit of 320K is best suited to our
interpretation of the IRAS measurements, slightly lower than the 370K
found by Wyatt et al. (2005) (see Figure \ref{etaSED}, left).  The
luminosity of this F2V star as fitted by a Kurucz profile (see section
2) is 5.5 $L_\odot$, and thus assuming that the emitting grains behave
like blackbodies, dust grains emitting at 320K would be at a distance
of 1.7 AU (0\farcs09). This small radial offset is consistent with the
extension test limits.  However there can be a difference of up to a
factor of 3 between a blackbody fit and the true radial offset of a
dust population (Schneider et al. 2006), thus the limits from these
observations show that a single mid-infrared component is not likely
to be much hotter than the blackbody fit of 320K (maximum of $\sim
1.3$ or 1.6 times the blackbody temperature for narrow and wide ring
geometries respectively). 

Model B: The two components of the mid-infrared emission in this model
have $R_{11.6\mu\rm{m}}=0.20$ and $R_{18.72\mu\rm{m}} = 0.34$ for the
dust at 360K and  $R_{11.6\mu\rm{m}}=0.005$ and $R_{18.72\mu\rm{m}} =
0.105$ for the dust at 120K, based on blackbody fits (see earlier in
this subsection). The extension limits suggest an outer limit
of 0\farcs19 $\pm$ 0\farcs02 for a narrow face-on ring (Q band limit)
for the hot component assuming 3 $\sigma$ limits (see Figure
\ref{exteta}).  This is consistent with the 0\farcs07 (1.3 AU) size
predicted by a blackbody grain assumption. 
For the 120K component, the Q band limits greatly
restrict the possible location of the disk.  In fact at the predicted
0\farcs66 (12 AU) location from an assumption of blackbody dust
grains, this mid-temperature component is ruled out at the 3.5
$\sigma$ level assuming a narrow face-on ring. A narrow edge-on ring
is also ruled out at a significance of 3.4 $\sigma$, as is a wide
face-on ring at a lower significance of 2.6 $\sigma$ although a wide
edge-on ring is only ruled out at 2.3 $\sigma$. Note that from Figure
\ref{exteta} it can be seen that larger disks (within the factor of 3
expected from Schneider et al. 2006) are also ruled out at the
$\gtrsim 2 \sigma$ limit. Thus at a significance of $> 2 \sigma$ these
observations rule out this model for the mid-infrared excess emission
of $\eta$ Corvi. There remains some uncertainty in these
limits, as these limits assume $R_\lambda = 0.105$ at Q for this dust component.
Photometric errors and errors in determination of the relative
  contributions to the excess emission from the 370K and 120K
  components respectively at Q mean that this could be as high as
0.134, or as low 
as 0.057. These uncertainties include the uncertainty from the
  IRS spectra, although the dominant source of uncertainty in
  $R_{\lambda}$ in this model is the poorly constrained relative
  contributions arising from the two components emitting at Q. A
longer observation of this source at Q, with a signal 
to noise of at least double that achieved in these observations, would
either resolve this component, or allow it to be ruled out at a
more certain level of significance. 

To summarise, the observations do not allow a certain differentiation
between the two alternative models for the excess emission.  At the
2.6 $\sigma$ level (assuming a reasonably favourable disk geometry) we
rule out the middle temperature component required by model B and thus
the limits favour model A - a single hot component at 320K in addition to the
cool 40K component already known. We were also able to set constraints
on the radial extent of the model A fit and the hotter component of
model B. These limits suggest that the radial size of the disk is at
most 1.75 times that predicted from a blackbody interpretation for
model A, or 2.7 times the blackbody prediction for the hottest
component of model B.  Deeper observations at Q are required to allow
a clearer differentiation between the two models. Components at 320K
or 360K (models A and B respectively) are expected to be smaller or
comparable to the single pixel scale of VISIR and MICHELLE, and are
unlikely to be resolvable on 8m instruments. Mid-infrared
interferometry is the only tool that currently has the potential to
resolve emission on such a small spatial scale.

\subsection{Confirmed hot disks around young stars.}
\begin{figure*}
\begin{minipage}{8cm}
\includegraphics[width=8cm]{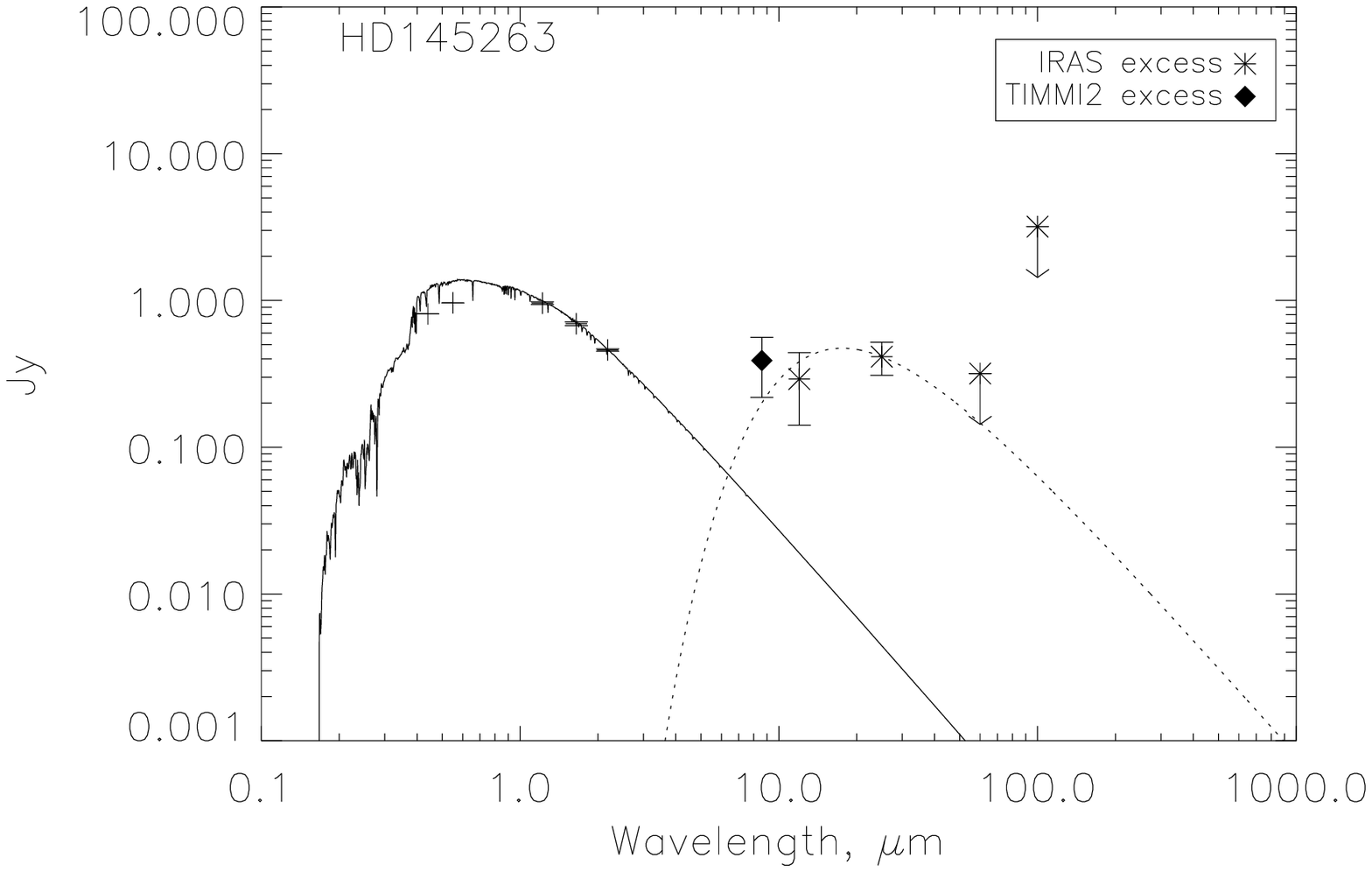}
\end{minipage}
\hspace{1cm}
\begin{minipage}{8cm}
\includegraphics[width=8cm]{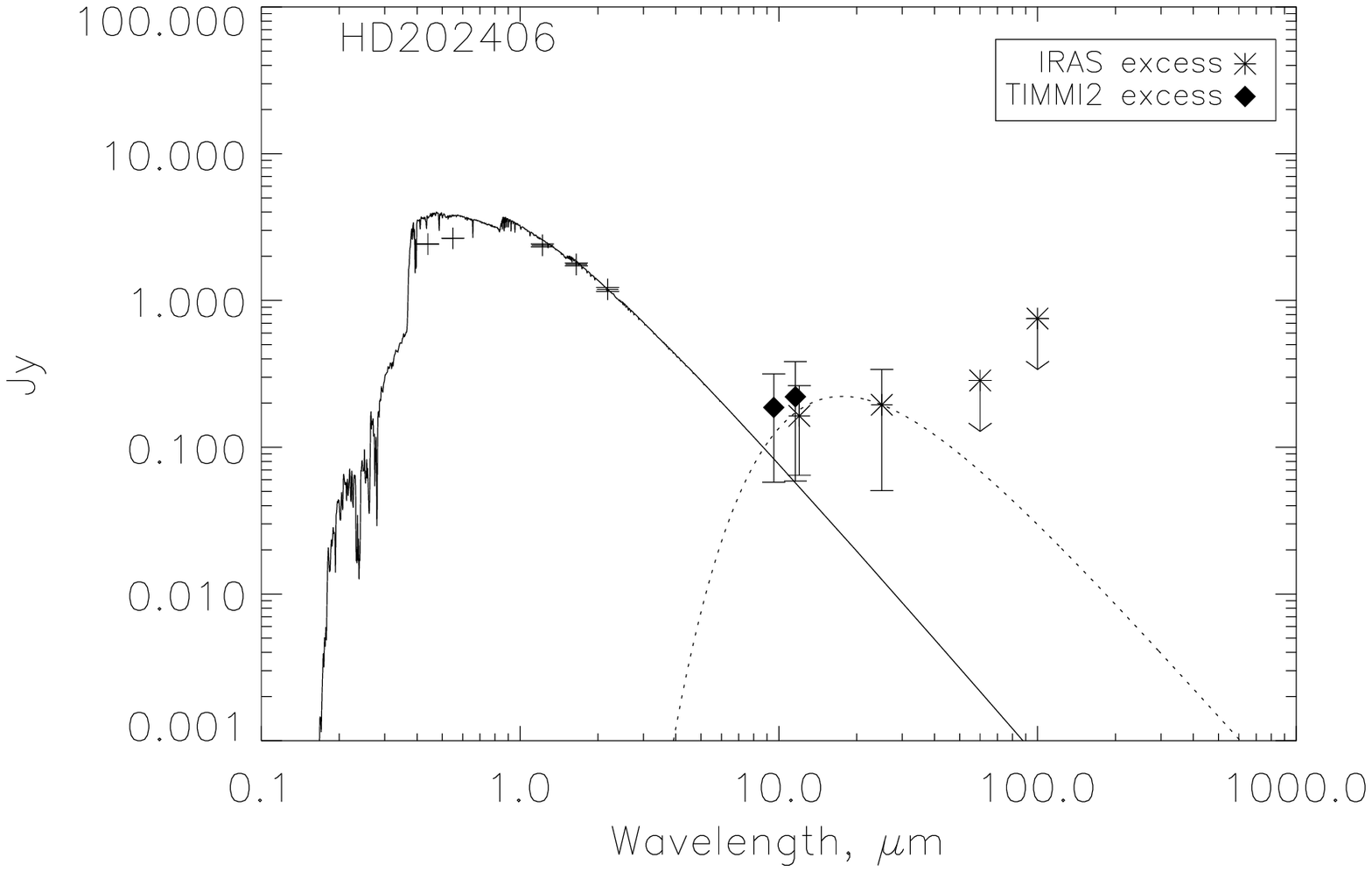}
\end{minipage}
\caption{\label{nothot} The SEDs of the two young confirmed excess emission
  sources. The solid line gives the
  photospheric emission modelled by Kurucz atmospheres, and the
  flux levels plotted at $>10\mu$m are measurements of excess after
  the subtraction  of the photosphere.  Limits and error bars are 3
  sigma.  Dotted  lines are single temperature blackbody fits to the
  excess.  The fits  are described in Table \ref{results}.}
\end{figure*}

Two of the sample are also confirmed to have hot excess
emission. However, on further investigation these are revealed not to
be main-sequence stars of a similar age to the rest of the sample. 

\emph{HD145263:} The star was originally proposed as a debris disk
hosting candidate in Mannings and Barlow (1998).  It has an IRAS
excess at 12 $\mu$m of 422 $\pm$ 50 mJy and at 25
$\mu$m of 583 $\pm$ 35 mJy (see Table
\ref{sample}). It was also studied by
Honda et al. (2004) using Subaru/COMICS from 8-13 $\mu$m. No pointing
error is quoted by Honda et al. (2004), but the blind pointing
accuracy of the Subaru Telescope is less than 1\arcsec \,, and so it can
be assumed that the crystalline silicate grains with a broad feature with
shoulders at 9.3 and 11.44 $\mu$m seen in their spectrum are from a
disk around the star. 
HD145263 is a member of the Upper Scorpius association, whose age is
estimated to be 8-10Myr. It is close to the zero-age main sequence in
the H-R diagram (Sylvester and Mannings 2000). The fractional
luminosity as measured using the fits to the IRAS detections is
$L_{IR}/L_\ast = 0.014$, smaller than is typical for T Tauri and
HAeBe stars but larger than debris disk hosts (Honda et al. 2000 and
references therein).  Thus Honda et al. (2000) suggest this star could be
considered a young Vega-like star.  

The excess at 8.6 $\mu$m is confirmed with the TIMMI2 data, finding a
flux of 426 $\pm$ 57 mJy (expected photospheric emission at this
wavelength is 37 mJy).  This result is consistent with the IRAS fluxes
and also the spectra of Honda et al. (2004).  The data place a limit on
undetected background or companion sources of less than 64 mJy. Since
the stellar photosphere would not have been detected, we can only confirm
that the source is centered on the star to within 1\arcsec, the
accuracy of the pointing. No extension is detected in the image of this
source. Applying a blackbody fit to the excess
emission gives a temperature of 290K (see Figure \ref{nothot} left), 
and at this wavelength an
$R_\lambda $ of 0.88.  Though the disk flux is bright, the radial
offset of the dust is predicted to be 1.8AU, which at the distance of
this star as measured by its parallax is only 0\farcs015 on-sky.  Such
a small disk is beyond the resolution limits of even the 8m class
telescopes, and could only be resolved using interferometry (see
e.g. Ratzka et al. 2007 for an example of a T Tauri star resolved
using interferometry).  The extension limits from these
observations are only very weak (see Table \ref{results}).

\emph{HD202406:}  Oudmaijer et al. (1992)  
identified this object in a survey of SAO stars for IRAS excess. 
Its luminosity class in the Hipparcos catalogue
is identified as IV/V.  The parallax of this object is quite
uncertain (2.33 $\pm$ 1.44 mas), and gives a distance to this object
of $430^{+410}_{-142}$ pc, but assuming the star has the
luminosity of a main sequence F2 star (2.9$L_\odot$) would imply a
distance of only 63 pc, which is incompatible with the Hipparcos parallax.
It is likely to be
a subgiant or pre-main sequence object.  There is no information in
the literature about rotational velocity or spectral lines for this
object to enable us to make a distinction between these two
possibilities.  However it does lie in the direction of a group of
molecular clouds M46, M47 and M48, which lie at a distance of $\gtrsim
290 $ pc (Franco 1989).  The proximity to
this cloud region suggests the star is more likely to be a pre-main
sequence star.  We assume a distance of 300pc to be consistent with
the molecular clouds implying that $L_\ast = 65 L_\odot$ which we
adopt in the following discussion.
Using the stellar models of Siess et al. (2000) and taking an effective
temperature of 7000K (appropriate for an F2 star) a
likely age for this star is 1.6 Myr.  This is in agreement with
the evolutionary tracks of Palla \& Stahler (1993) which suggest an
age of 3 Myr for this object.
\begin{figure*}
\begin{minipage}{8cm}
\includegraphics[width=8cm]{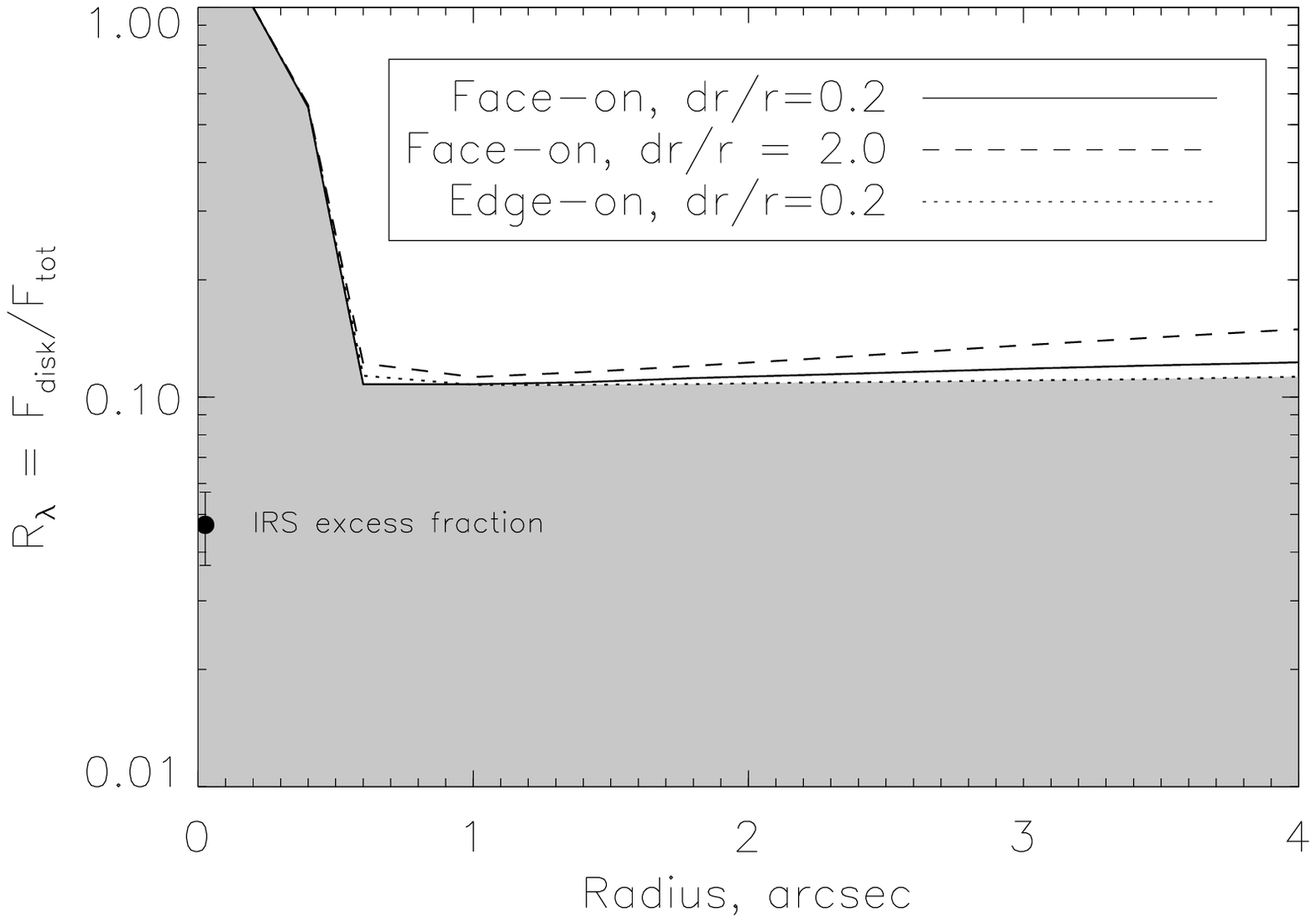}
\end{minipage}
\hspace{0.2cm}
\begin{minipage}{8cm}
\includegraphics[width=8cm]{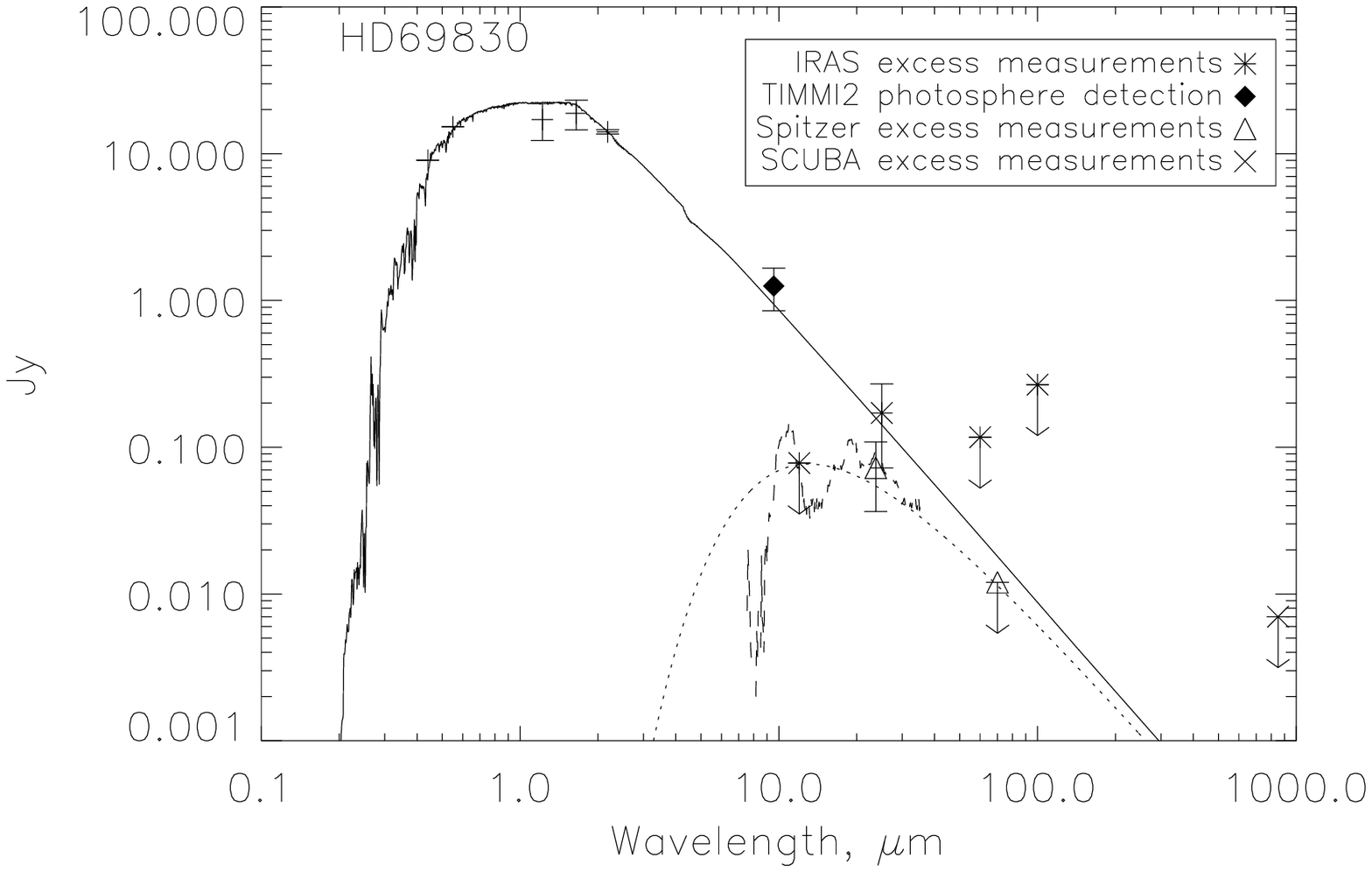}
\end{minipage}
\caption{\label{ext69830} Observations results for HD69830.
  \emph{Left:} The extension testing limits for the
  observation of HD 69830.  Note that at the measured level of
  fractional excess no limits can be placed on possible extension
  (fractional excess level and errors taken from IRS spectra). The
  predicted disk size is shown by an asterisk with error bars
  marking the 3 sigma photometric errors. The shaded area shows the
  possible disk location. Parameters used in determining these limits
  are $S_\star = 39$, $\theta$ = 0\farcs88 and $d\theta/\theta$ = 0.1.
  \emph{Right:} The SED of this object, with excess
  measurements shown after the subtraction of the photospheric
  contribution.  The blackbody fit to the excess shown by the dotted
  line and gives the predicted disk size shown in the left-hand
  Figure. The dashed line shows the photosphere subtracted Spitzer IRS
  spectra obtained by Beichman et al. (2005). Note the strong silicate
  features are obvious from this plot.}
\end{figure*}

The TIMMI2 observations of HD202406 detect the excess emission centred
on the photosphere at above 4 $\sigma$ at 9.56 and 11.59 $\mu$m.  The
detected levels of flux at these wavelengths are 270 $\pm$ 43 mJy and
278 $\pm$ 54 mJy (photosphere expected to be 83 and 57) respectively.
A limit of less than 30 mJy can be 
placed on any undetected background object at 9.56 $\mu$m, and
less than 43 mJy at 11.59 $\mu$m.  Fitting the excess emission with a
blackbody gives a temperature of 290K (see Figure \ref{nothot},
right) which corresponds to a
dust location of 7.4AU (0\farcs025). Note that should we have chosen a
different stellar distance, the dust offset in arcseconds would be the
same (due to an increased luminosity and thus radial offset of dust
for the same temperature blackbody fit at increased distance). 
Given this small predicted size, it is unsurprising that no extension
was detected in the images. Indeed the limit set from extension
testing is less than 0\farcs33 radius for a thin ring around this
source at $R_{11.59\mu\rm{m}} = 0.81$, corresponding to a radius of 99 AU. 
The shape of the emission here has been modelled by a
blackbody.  However, at the level of 3 $\sigma$ significance a simple
power-law would fit this excess flux equally well. Thus we
require limits on excess at shorter wavelengths to determine a
grouping according to the scheme of Meeus et al. (2001) and
differentiation between a flat 
and flared disk geometry.  This in turn may indicate evolutionary
status, as a dip around 10 $\mu$m is thought to develop and widen with
age (see e.g. van den Ancker et al. 1997).  It should be noted however
that we derive a $L_{\rm{IR}}/L_\ast$ of 0.00371, which as for
HD145263, is lower than typical T Tauri stars for which values of
$L_{\rm{IR}}/L_\ast \sim 0.1$ are more typical (see e.g. Padgett
et al. 2006). This may indicate that these objects are in a
transitional stage.

\subsection{Constraints on hot dust sources}

\begin{figure*}
\begin{minipage}{8cm}
\includegraphics[width=8cm]{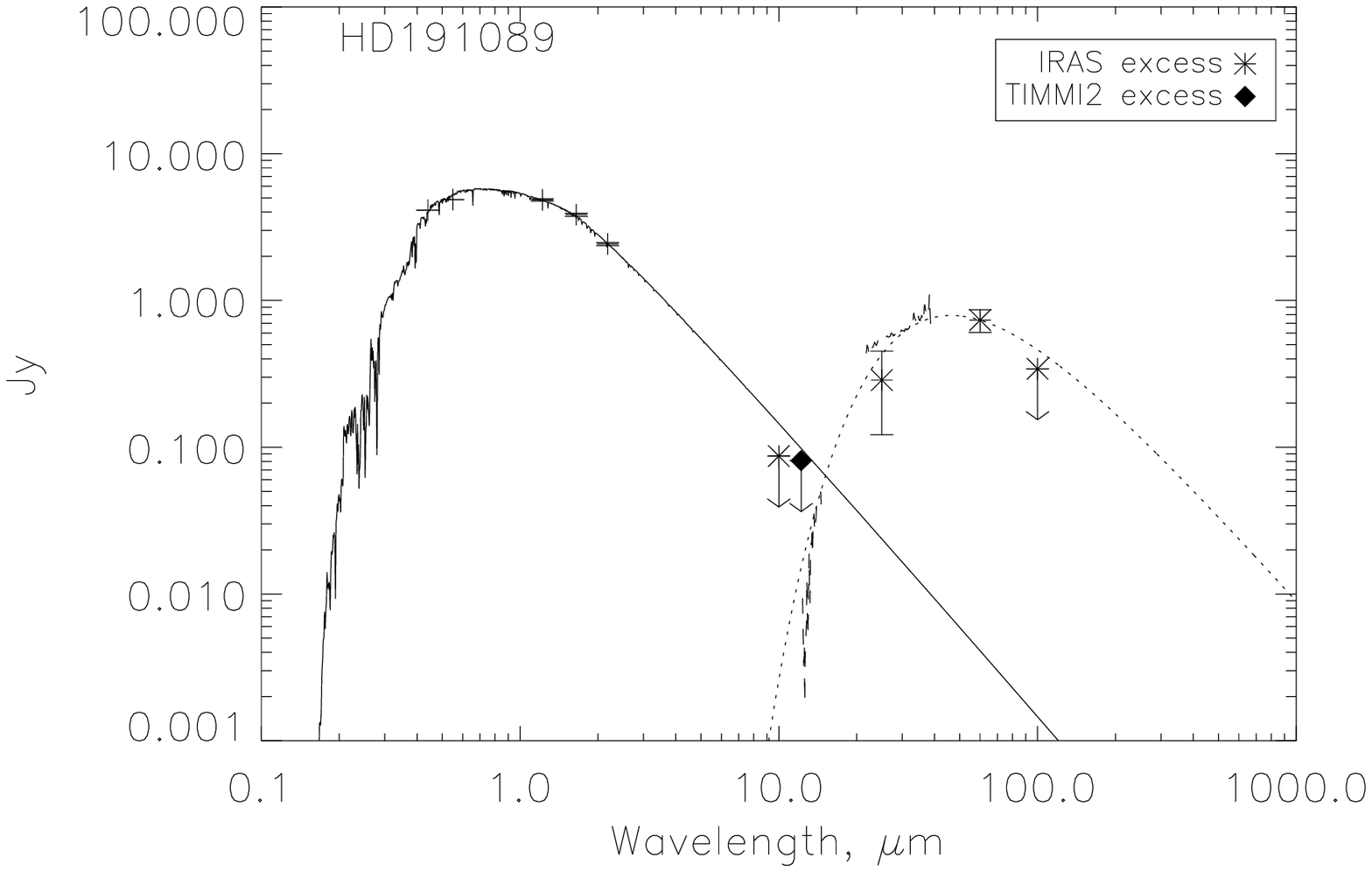}
\end{minipage}
\hspace{0.2cm}
\begin{minipage}{8cm}
\includegraphics[width=8cm]{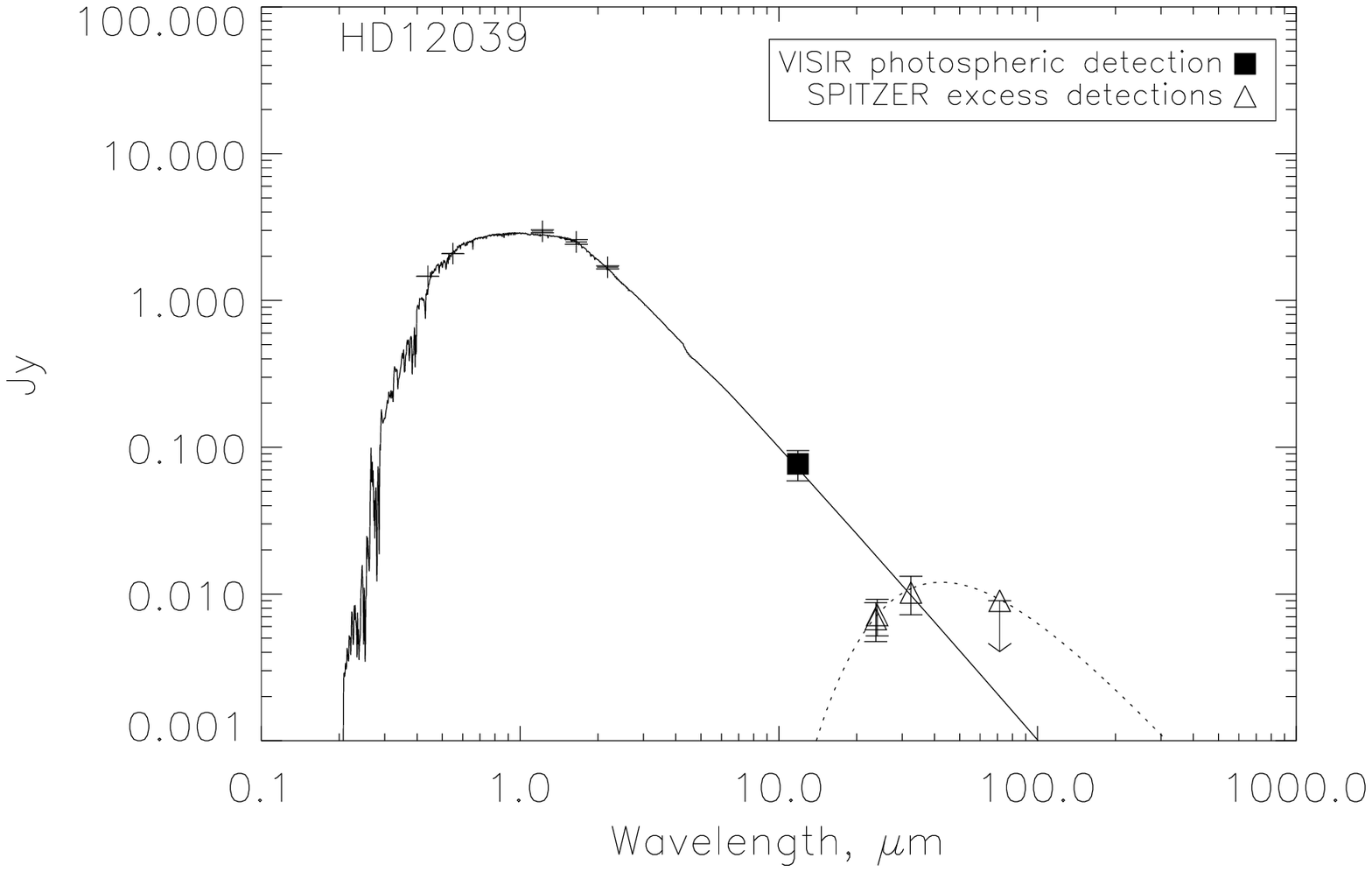}
\end{minipage}
\caption{\label{confirmed} The SED fits of objects with hot dust
  confirmed in the literature.  For both objects the solid line is
  photospheric emission as  modelled by a Kurucz profile.  Symbols
  representing the excess  measurements are the measured flux minus
  the photospheric emission as modelled by the Kurucz profiles. Error bars and
  upper limits are at the 3 sigma level. The dashed line on the plot
  of HD191089 is the publicly available low-resolution IRS spectra
  after photospheric subtraction originally presented in Chen et
  al. (2006). The dotted lines are blackbody fits
  to the dust emission with parameters described in Table 3.}
\end{figure*}

\emph{HD69830:} Mannings and Barlow (1998) used the IRAS database to
identify an excess around HD69830 in the 25 $\mu$m band, at the level
of 5 $\sigma$ (142 mJy photosphere, excess 171 $\pm$ 33 mJy, see Table
\ref{sample}).  There is no detection of excess at longer wavelengths,
and an insignificant excess at 12 $\mu$m. SCUBA observations limit the
excess at 850 $\mu$m to $< 7$ mJy (Matthews et al. 2007). 
Beichman et al. (2005) observed this object with the IRS and
MIPS instruments on Spitzer and found further evidence for excess at
24 $\mu$m with MIPS, and between 8 and 35 $\mu$m with IRS.  No excess
was found at 70 $\mu$m.  At 24 $\mu$m the excess was measured to be 70
$\pm$ 12 mJy (aperture 15\arcsec radius). The IRS spectra between 8-35
$\mu$m reveals the presence of crystalline silicates (see dashed line
Figure \ref{ext69830}, right). Interest in this source has
intensified since the discovery of 3 Neptune mass planets at $< 1$ AU
(Lovis et al. 2006). 

Unfortunate conditions mean the measures of the N band emission of
this object are non-photometric. The object is detected at a S/N of 39, 
and find a calibrated flux of 1255 $\pm$ 135 mJy using just the
standard observations immediately before and after the science
observation for calibration.  As conditions were very changeable
over the course of the night, this may mean the the errors are
under-estimated.  At this level of flux we are within 3 $\sigma$ of the
predicted photosphere at this wavelength (941 mJy at 9.56 $\mu$m).  
A 3 $\sigma$ limit of 84 mJy can be placed on any background/companion
object in the field of view, making it highly unlikely that the
Spitzer photometric excesses obtained in a larger aperture are
due to any such object. Thus we can
be confident that the excess emission is centered on the star.

The source did not exhibit extension.  The extension testing
procedures were applied to this observation and the resulting
detectability limits are shown in Figure \ref{ext69830} (left). 
The limits show that a minimum extended contribution of $R_\lambda =
0.107 $ is necessary to place spatial constraints on the disk
flux. The SED 
fit of a blackbody at 390K translates to a disk radius of 0.33AU
(0\farcs026), with a fractional contribution to the excess of
$R_\lambda =0.05 \bf{\pm 0.01}$ at the wavelength of this
observation seen in the 
IRS spectrum of this source (see Figure
\ref{ext69830}, right). This predicted disk model is shown on Figure
\ref{ext69830} (left). Beichman et al (2005) suggest a disk radius of
0.5AU (0\farcs04).
Also Lisse et al. (2007) model the IRS spectrum in detail 
and find a dust radius of $\sim$ 1AU
(0\farcs08). However given the expected fractional flux contribution
at N is only 5\%, it is unsuprising that the disk is unresolved.  The
small spatial scale suggested by these models would require
mid-infrared interferometry to resolve the emission (see section 4.1.4). 

\emph{HD191089:} HD191089 was identified by Mannings and Barlow (1998)
as a debris disk candidate based on its IRAS photometry.  This source
has excesses  of 287 mJy at 25 $\mu$m and 735 mJy at 60 $\mu$m at the
5 and 17 $\sigma$ levels respectively (as noted in Table
\ref{sample}). At shorter wavelengths there was no excess detected by IRAS.

This object was observed at 12.21 $\mu$m with TIMMI2. 
The photosphere was detected at a signal to noise of
5.75.  The photometry is consistent with the predicted photospheric
emission (92 $\pm$ 27 mJy calibrated flux; Kurucz model profile
predicts 98mJy from the photosphere). No other source was detected in the
field, placing a limit on undetected objects of less than 43
mJy. There are no bright 2MASS sources within the IRAS error lobe of
14\arcsec of this star 
which could be responsible for IRAS confusion. These limits suggest it
is highly likely that the excess detected at longer wavelengths is
indeed centered on the star HD191089. Publicly available
Spitzer IRS low resolution spectra (originally presented in Chen et
al. 2006) is shown after photospheric subtraction on the SED of this object
by a dashed line (see Figure \ref{confirmed}).  This spectra shows
that at less than 12 $\mu$m there is  no excess, which allows us to
place limits on the minimum radius and maximum temperature of the dust
around the star of no hotter than 110K (11.5AU, 0\farcs21) as fit by a
blackbody curve (see Figure \ref{confirmed}). The IRS
data shows good agreement with the blackbody fit at longer
wavelengths (20-40 $\mu$m), but a steeper cut-off at the
short-wavelength end (8-15 $\mu$m), which may be an effect of grain
properties such as chemical composition and size.  The predicted size
and flux level of this disk makes it an ideal candidate for imaging at
25 $\mu$m with an 8m telescope to determine the true size and nature
of this disk. 

The age of this source is subject to some uncertainty.  Isochrone
fitting has given an age of 3Gyr (Nordstrom et al. 2004) or 1.6Gyr
(Chen et al. 2006).  However using X-ray and lithium abundance data among
other techniques, Zuckerman \& Song (2004) put the age of this source at
$\leq$ 100Myr.  Mo\'{o}r et al. (2006) also suggested this source is a
possible member of the $\beta$ Pictoris moving group, giving HD191089
a likely age of 12Myr. As membership of this moving group is not
yet confirmed, we have chosen to adopt an age of 100 Myr for this
source. (The age of the system will have a bearing on the calculation
of $f_{\rm{max}}$ described in section 6.2; note that a younger age
would increase the value of $f_{\rm{max}}$ and so make the
interpretation of this source's emission as possibly steady-state
even stronger. ) 

\emph{HD12039:}
This star was identified by Hines et al. (2006) as having an excess at 24
$\mu$m of 7 mJy (3 $\sigma$ detection) and no excess at 70 $\mu$m. 
The target aperture used in the Spitzer observations was 14\farcs7 at
24 $\mu$m. Further IRS spectra were taken with Spitzer, with a 0\farcs4
1 $\sigma$ uncertainty radius in the spectrograph slit. This spectra
shows the infrared emission departing from the photosphere at 12-14
$\mu$m (see Figure 4 of Hines et al. 2006). 

HD12039 was studied with VISIR in the N band.  At the N band this
source is detected with S/N of 26, and calibrated flux of 77 $\pm$ 3
mJy; this is within 2 $\sigma$ of the predicted photospheric
emission. We place an upper limit on the excess at 11.85 $\mu$m of 14
mJy.  No other source was detected within the field of view and we can
place a limit of $\le$ 2mJy on undetected sources. Our data agrees
well with the Spitzer data in Hines et al. (2006); the Spitzer
photometry limits excess to less than 32 mJy at 13 $\mu$m.  The
pointing accuracy achieved in the Spitzer observations, the lack of
detection of additional sources within the field, and the agreement
between the VISIR  photometry and that of Spitzer suggests that the
IRS spectra and MIPS photometry are indeed measuring an excess
centered on the star.  

SED fitting to the MIPS detections suggests a dust temperature of
120K, corresponding to an offset of 5AU (0\farcs12) from the star (see
Table \ref{results} and Figure \ref{confirmed}, right for SED).
This is in good agreement with a model for the emission proposed by
Hines et al., which adopts blackbody grains at 4-6AU from the star.
However, as pointed out in Hines et al. (2006), an alternative model
of a power-law distribution of grains with radii between
0.4-1000$\mu$m located between 28 and 40 AU from the star provides an
equally good fit to the data.  

\subsection{No dust detection}

We now consider the members of the sample which were erroneously
identified as having excess emission.  Five of these objects have
companion or background sources which are responsible for the IRAS
detection of excess; one shows no evidence for current excess
emission. 

\begin{figure}
\begin{minipage}{8cm}
\includegraphics[width=8cm]{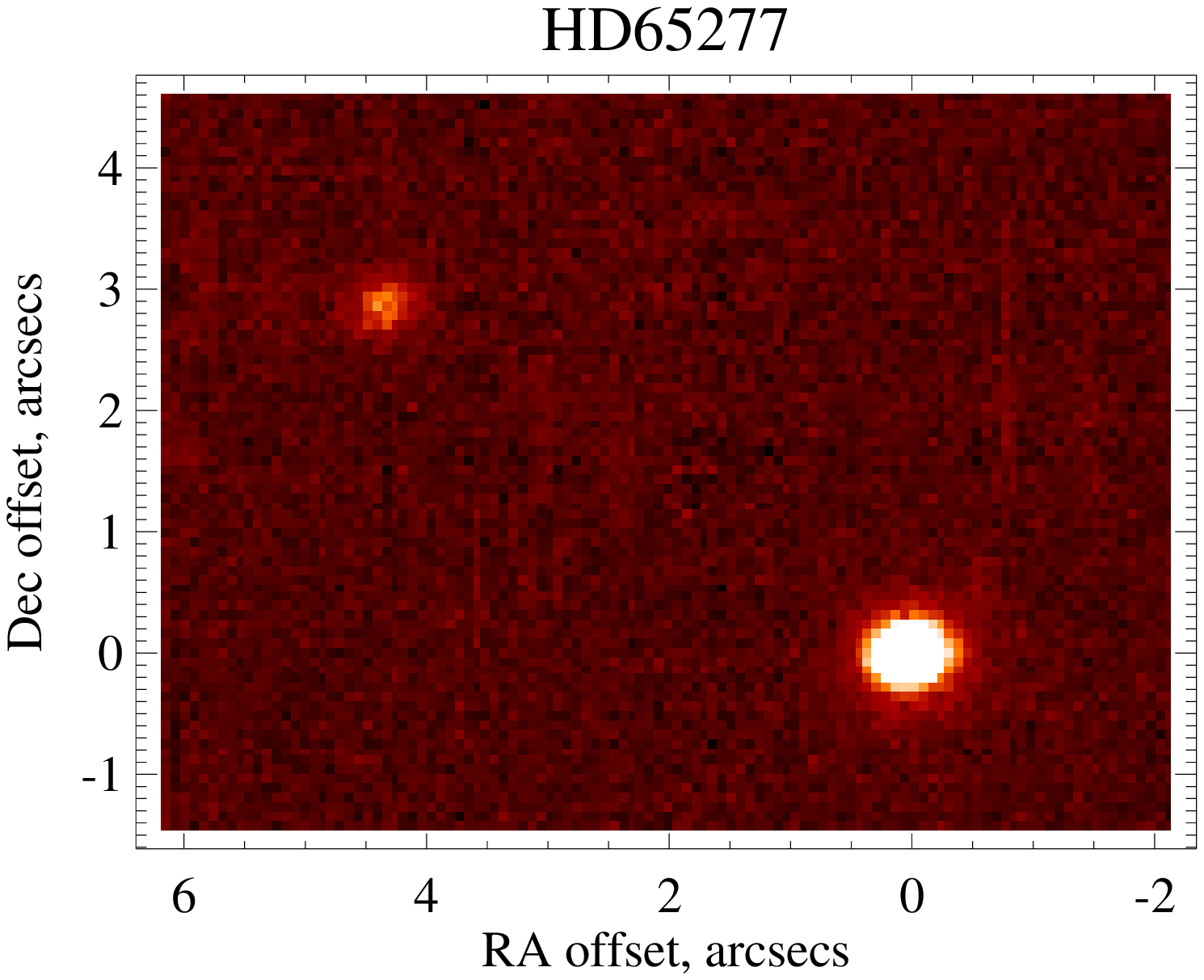}
\end{minipage}
\hspace{-1.5cm}
\begin{minipage}{8cm}
\includegraphics[width=8cm]{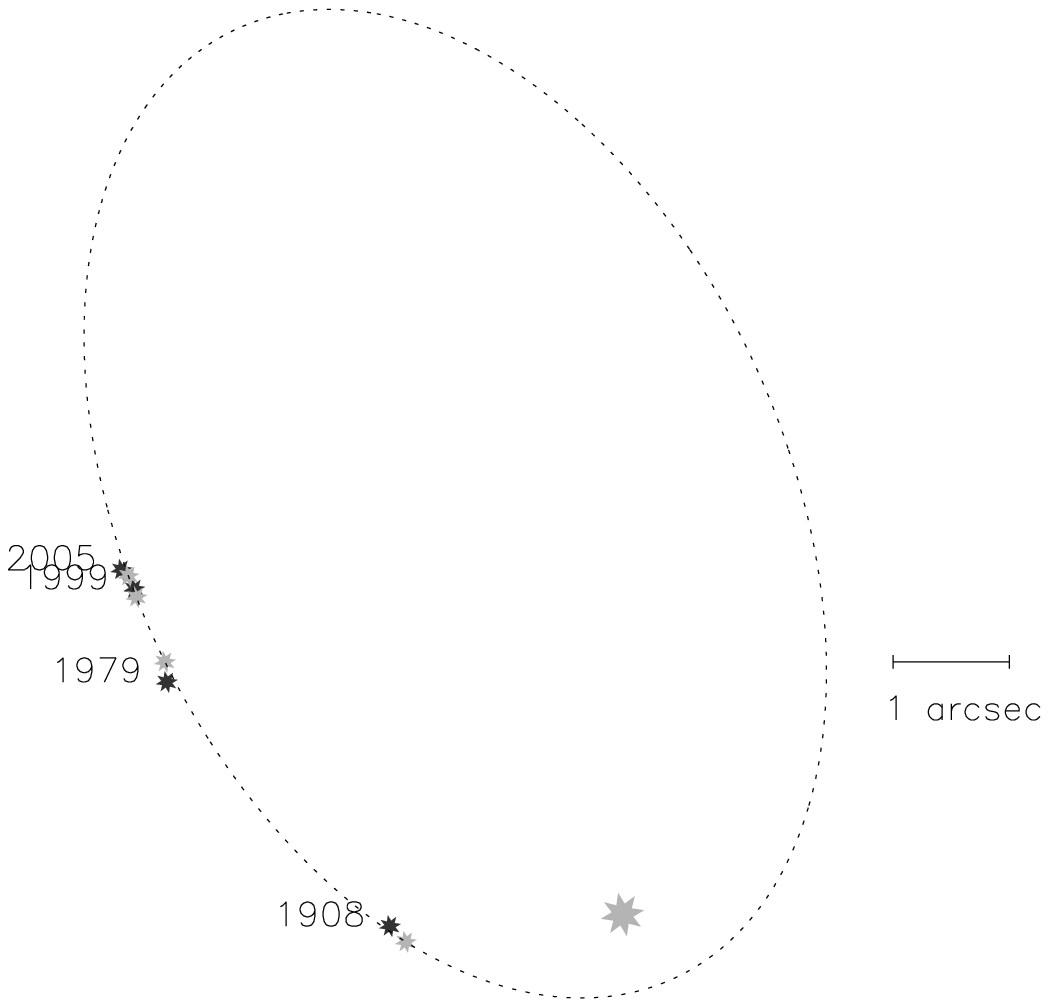}
\end{minipage}
\hspace{1cm}
\caption{\label{vlt65277} The companion of HD65277 (images are
    North up, East left).
  Top: The N band VLT image of HD65277 and its
  binary companion offset by 5\farcs2 at position angle $56^\circ$
  East of North. Bottom: A fit to the orbit of this companion, with the
  VLT data shown as 2005. Black star symbols represent measured
  offsets and grey symbols the position of the companion predicted by
  the orbital fit.  HD65277 is shown by the large light grey star
  symbol. See text for full details of the orbit. } 
\end{figure}

\emph{HD65277:} This star has a 25$\mu$m
excess at the 2.8 $\sigma$ level, and no significant
excess at 12 $\mu$m (see Table \ref{sample}). In the 2MASS catalogue
there is an additional object 2MASS 07575807-0048491 (which for
brevity in the following discussion shall be referred to as HD65277b) at
a separation of $\sim$5\arcsec \,. 

The primary object is detected at 182 $\pm$ 3 mJy and 78
$\pm$ 14 mJy in N and Q on VISIR. HD65277b is detected in the M and N
band images of TIMMI2 at 5 $\sigma$ and 3 $\sigma$ respectively.  It
is strongly detected in N by VISIR, with a calibrated flux of 32 $\pm$
4 mJy and is detected at Q at the 3 $\sigma$ level (14 $\pm$ 6 mJy
including calibration errors, see Table \ref{observations}).  
The N band VISIR image is shown in Figure \ref{vlt65277}. The
companion is at $\Delta$ra = 4\farcs32 $\pm$ 0\farcs09, $\Delta$dec =
2\farcs91 $\pm$ 0\farcs085.  The observations place constraints on
additional undetected objects within the VISIR field of view of 2 and
10 mJy at 12 and 18 $\mu$m respectively.   

\begin{figure*}
\begin{minipage}{8cm}
\includegraphics[width=8cm]{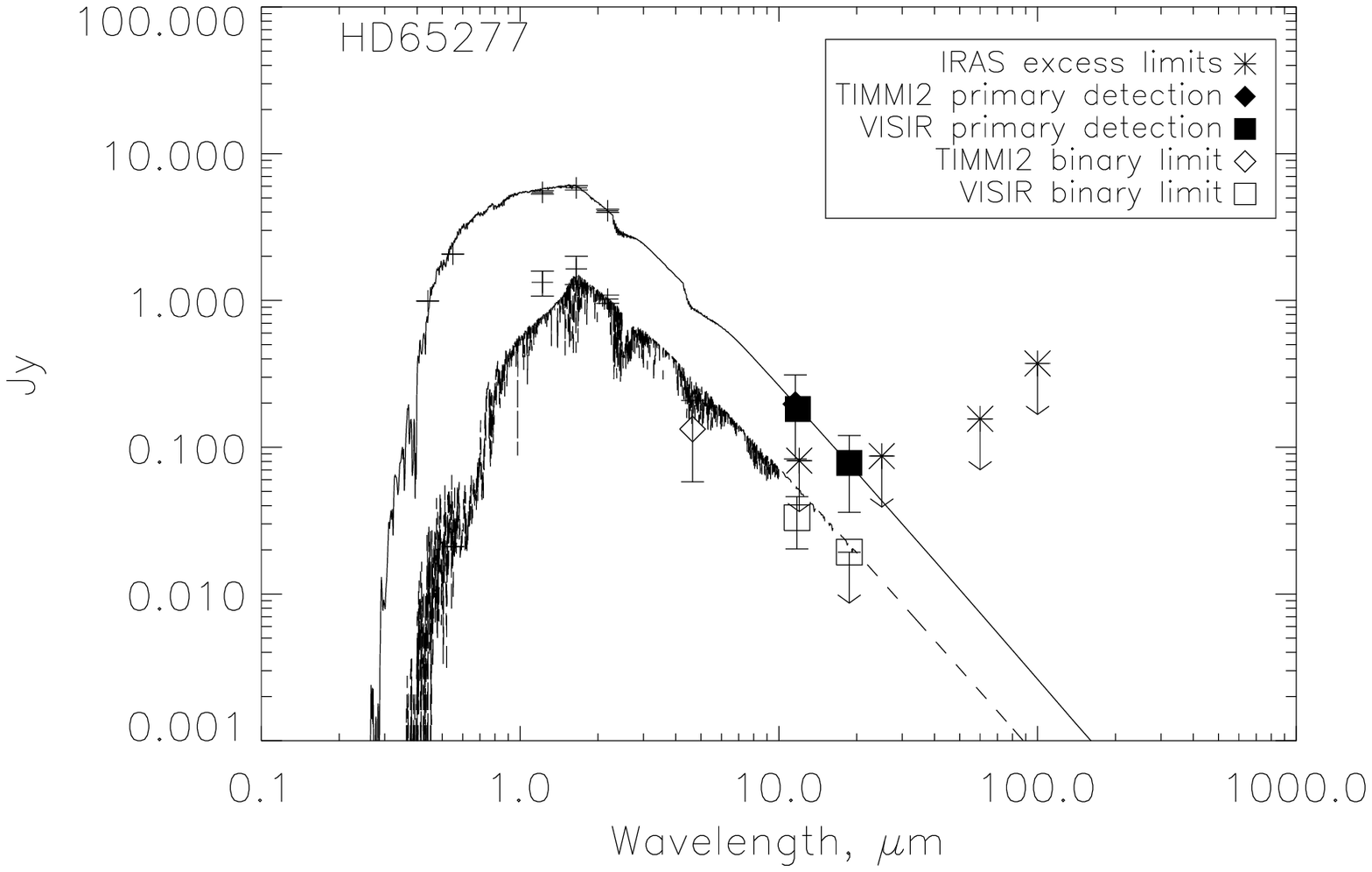}
\end{minipage}
\hspace{1cm}
\begin{minipage}{8cm}
\includegraphics[width=8cm]{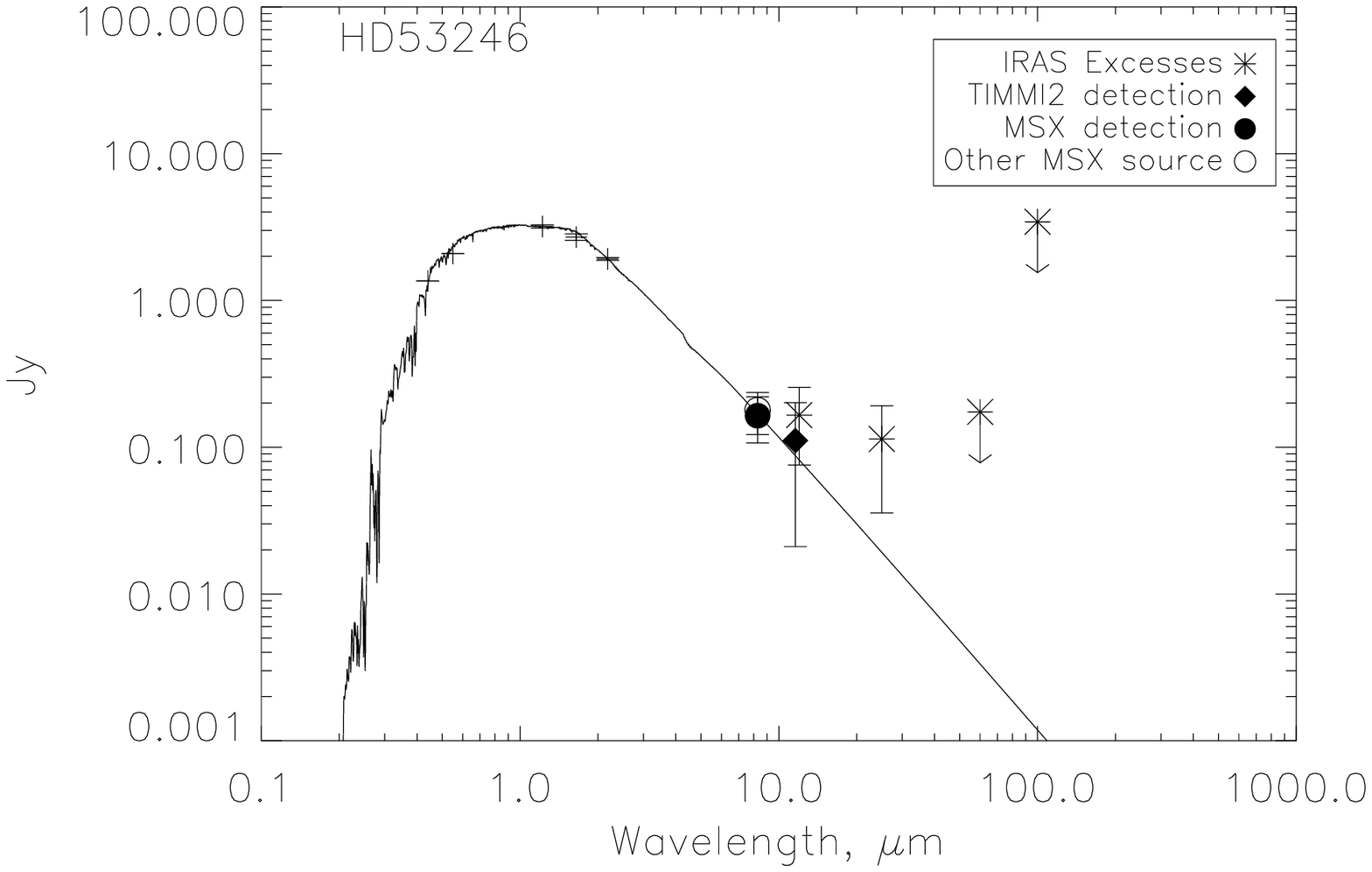}
\end{minipage}
\vspace{1cm}
\begin{minipage}{8cm}
\includegraphics[width=8cm]{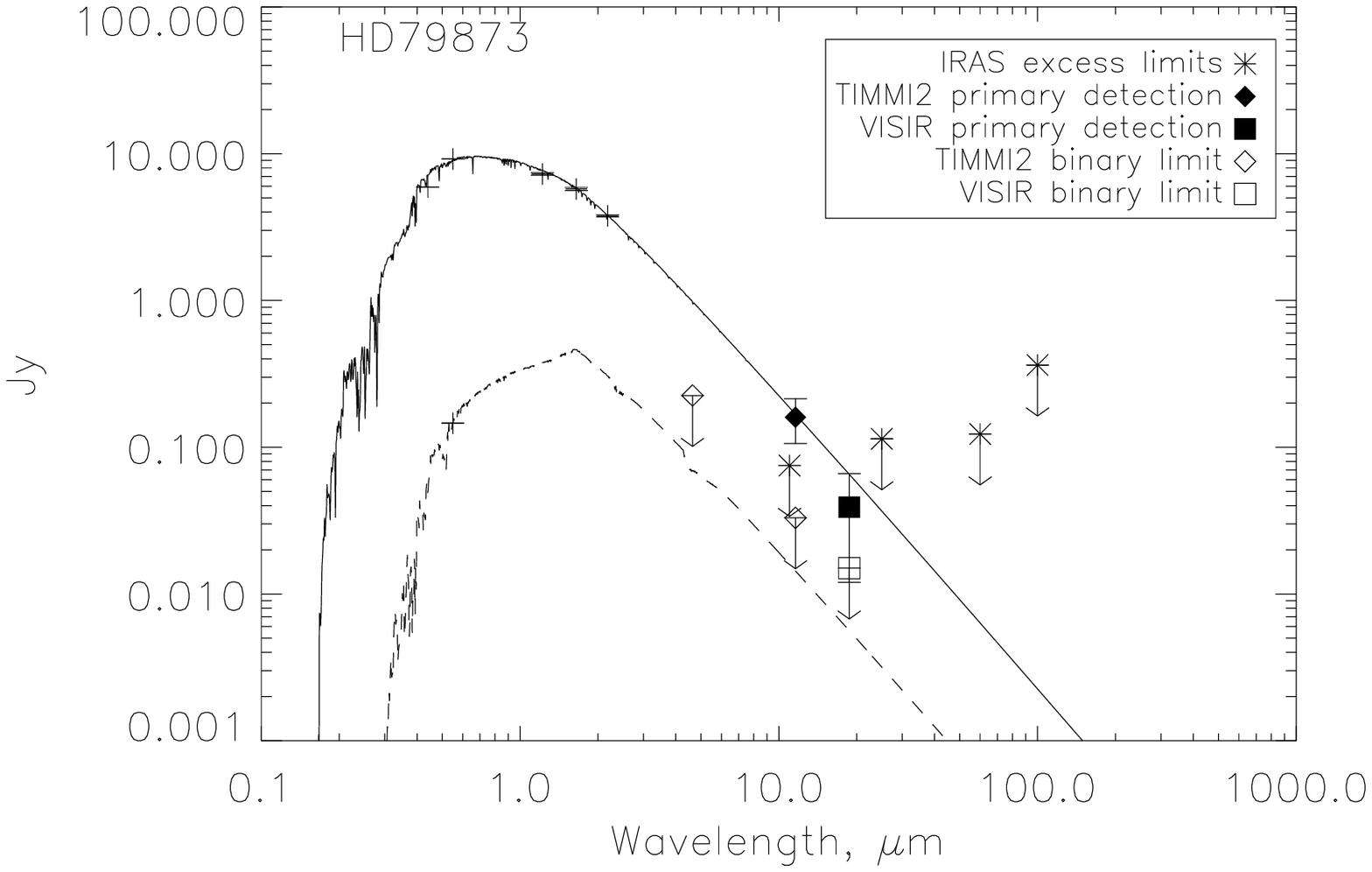}
\end{minipage}
\hspace{1cm}
\begin{minipage}{8cm}
\includegraphics[width=8cm]{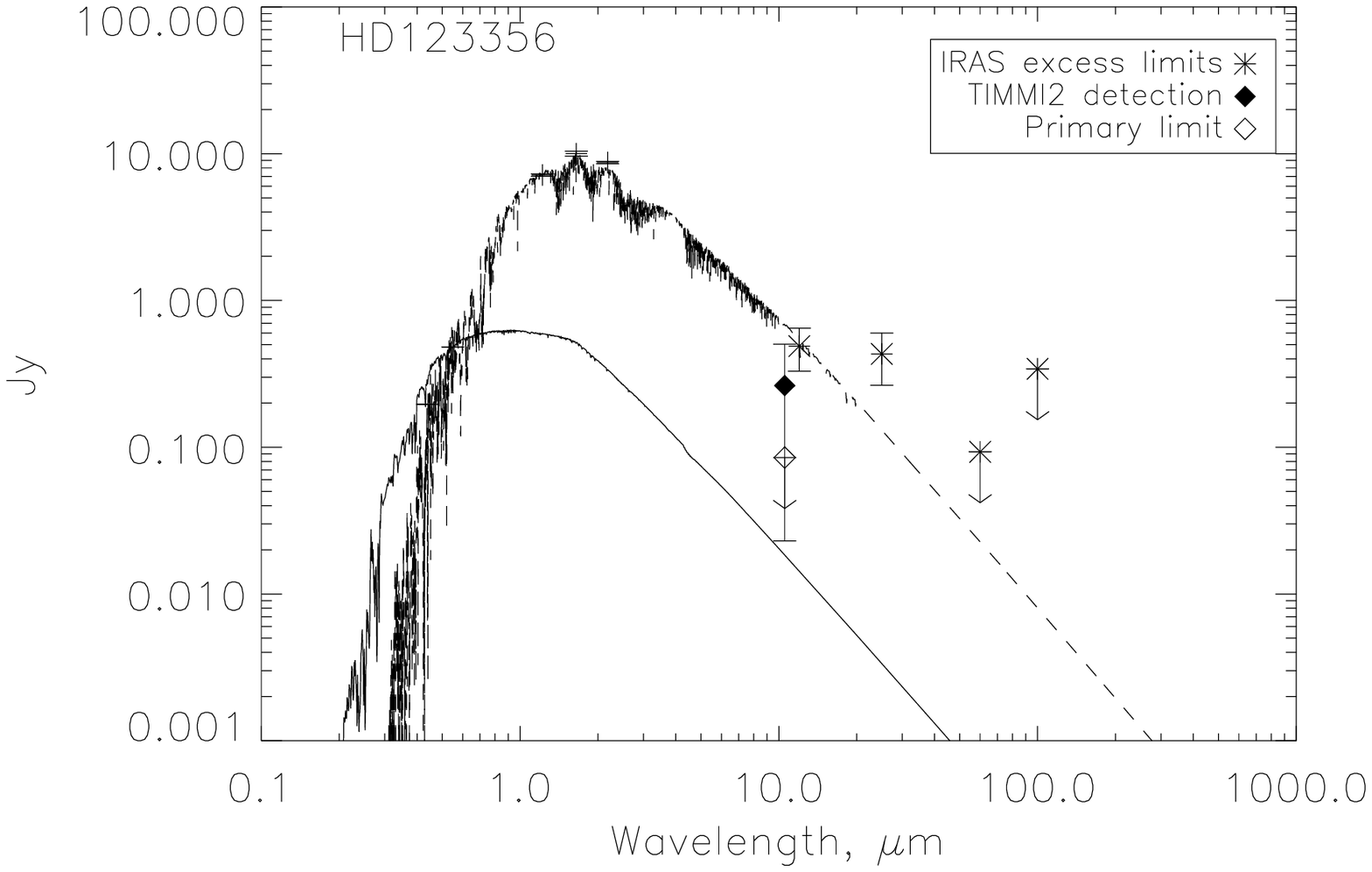}
\end{minipage}
\vspace{1cm}
\begin{minipage}{8cm}
\includegraphics[width=8cm]{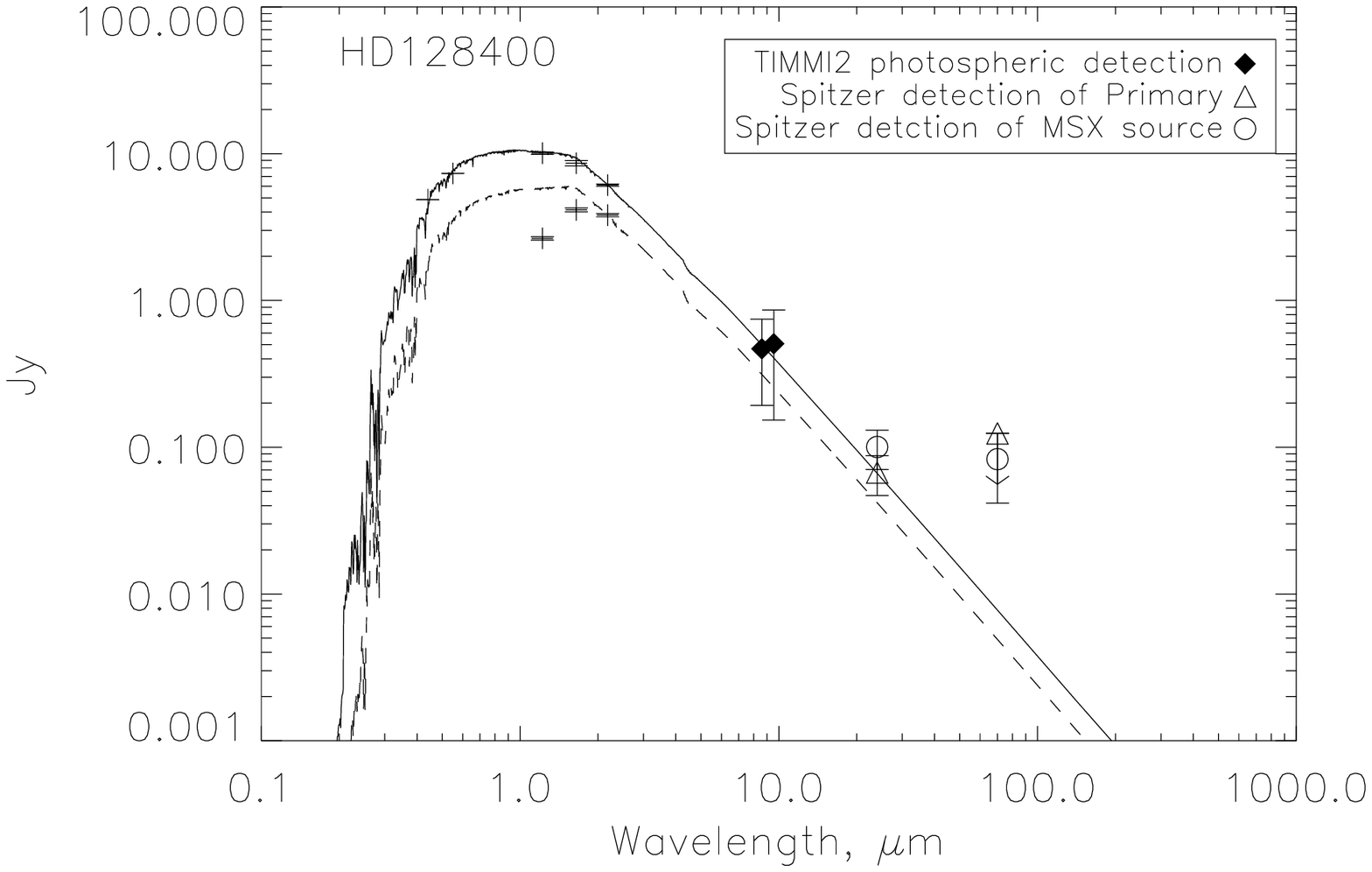}
\end{minipage}
\hspace{1cm}
\begin{minipage}{8cm}
\includegraphics[width=8cm]{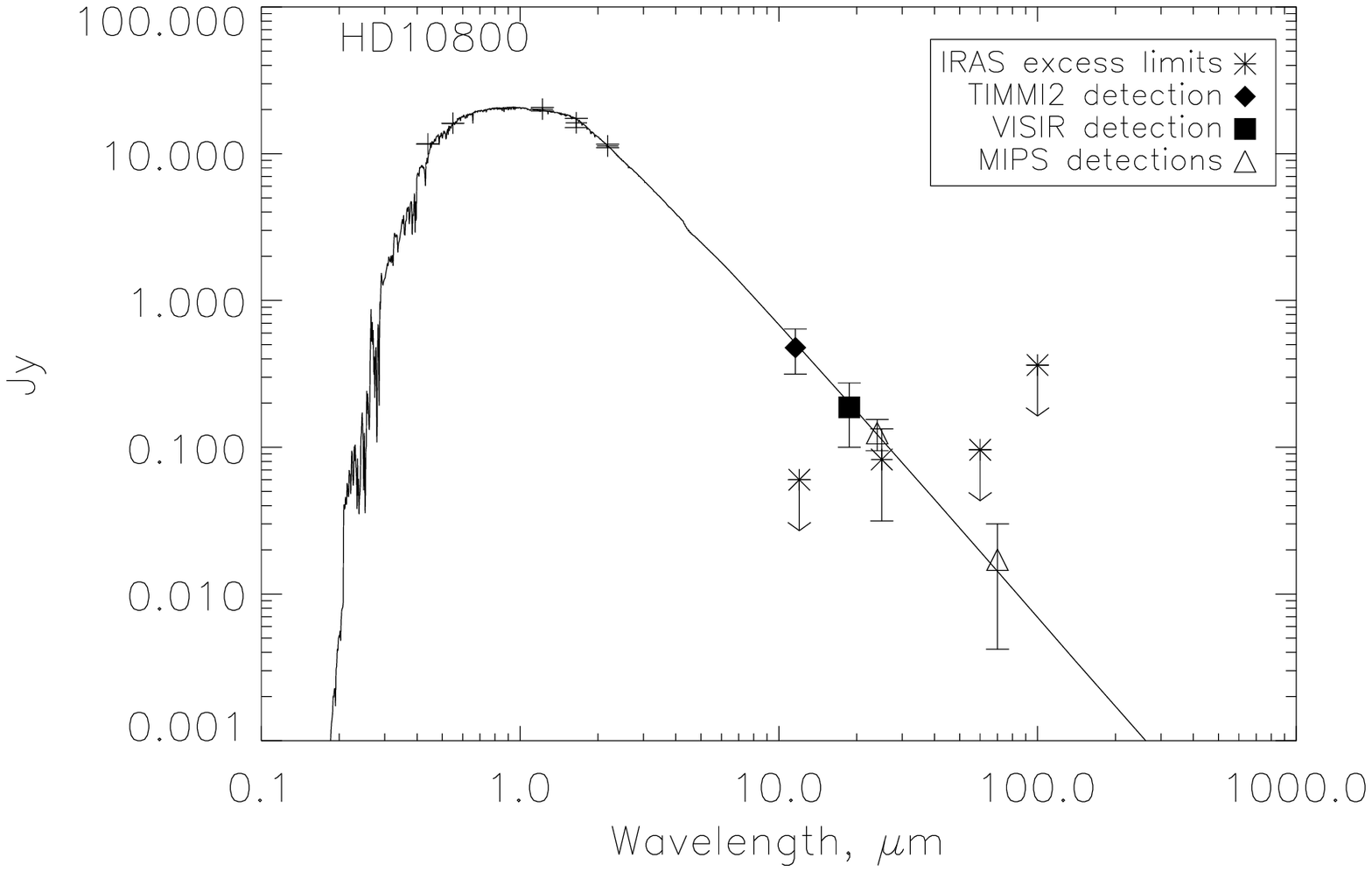}
\end{minipage}
\caption{\label{nodust} The SED fits and limits for objects without
  confirmed hot dust.  Photospheric emission as modelled by Kurucz
  atmospheres are shown as a solid line.  Dashed lines are the
  photospheric models of the binary (modelled occasionally using
  NextGen spectra - see text).  Errors are 3 sigma, and upper
  limits are also 3 sigma.} 
\end{figure*}

The measured levels of flux for the primary are consistent with the
expected photospheric emission (see Figure \ref{nodust}, top left).  
We use the K band magnitude of the secondary as listed in the 2MASS
catalogue and assume a common distance of 17.5 pc with the primary to
fit the spectral type of the companion as M4.5.  Note that this
spectral type was found to be the best fit to the currently
available data but remains subject to great uncertainty. The model profile
is designed to be representative of the possible SED of the source
only.  The profile is modelled
with a NextGen model atmosphere appropriate to this spectral type
(Hauschildt, Allard \& Baron 1999). The M band detection of the
secondary is calibrated to the expected flux of the primary
photosphere and is measured as 133 $\pm$ 25 mJy, a little low compared
to the expected 244 mJy which may be the result of a large filter
width ($\Delta \lambda = .69 \mu$m) and the TiO absorption features
seen in M-type stars at around this wavelength. The VLT N band flux of
HD65277b is also a little low, but scaling to the expected primary
flux the difference is not significant above the 4 $\sigma$ level.  

Additional data available for this object allows us to make a
preliminary estimate of the orbit for HD65277b.  This orbit is shown
in Figure \ref{vlt65277}. The VISIR data is the point marked as 2005 (exact
epoch 2005.935).  The data from 1999 is the 2MASS catalogue data
(observed 12-01-1999).  The earlier data are listed in the Washington
Double Star Catalogue (Worley \& Douglass 1997).   The orbital fit has
the following parameters: 
$a = 95AU; e = 0.85; I = 35^\circ; \bar \omega = 290^\circ; \Omega =
100^\circ; $ with the last pericenter pass in 1885.  The masses of the 
stars are taken to be 0.69 $M_\odot$ for the primary and 0.23
$M_\odot$ for HD65277b, as appropriate to their spectral types.  
The predicted flux of the binary at 25 $\mu$m is 12 mJy, and
subtracting this from the IRAS measurements leaves an excess of only
59 $\pm$ 29 mJy, an insignificant detection.  Thus we conclude that
the IRAS detection of excess is caused by inclusion of the binary and
is not indicative of circumstellar disk emission.   

\emph{HD53246:}
This star has an excess at 12 $\mu$m of 293 mJy at the 9.8 $\sigma$
level, and at 24 $\mu$m of 143 mJy at the 5.5 $\sigma$ level (Table
\ref{sample}) based on the IRAS catalogue.  This star is detected in the MSX
catalogue at 8.28$\mu$m, with flux 164 $\pm$ 19 mJy. This detection is
consistent with the expected photosphere at this wavelength (168 mJy).

In the observations presented here a source is detected within
1\arcsec of the expected 
source location at a signal/noise of 4.5, but calibration errors
introduce high uncertainty in the photometry. The calibrated flux is
111 $\pm$ 30 mJy (expected photospheric emission from HD 53246 is 87
mJy).  However there is no evidence for excess as the fluxes are in
line with that expected from the photosphere, and limit any undetected
excess to less than 114mJy (see Figure \ref{nodust}, top right, for
SED).  The possibility of a companion within the TIMMI2 field of view
of above 62 mJy is ruled out at the 3 $\sigma$ level.    

We attribute the significant excess emission to an additional MSX
source (G234.4643-07.5741) at 89 \arcsec (position angle -11$^\circ$)
detected at 8.28$\mu$m at a level of 179 $\pm$ 19 mJy.  The IRAS Point
Source Catalogue position for this object is between HD53246 and the
MSX source, offset from HD53246 by 31\arcsec \, at a position angle of
94$^\circ$. The error ellipse given in the catalogue is 44\arcsec \,
by 10\arcsec \, (with position angle 101). This is larger than average
for the IRAS catalogue (estimated to be 16\arcsec \, in the cross-scan
direction and 3\arcsec \, in the in-scan direction, Beichman et
al. 1988), suggesting that the IRAS fluxes could be contaminated by
emission falling outside the TIMMI2 field of view.  We believe that
confusion caused by the nearby MSX source is the likely origin for the
excess.  This MSX source has a very similar level of emission to the
star, with a flux of 170 $\pm$ 19 mJy at 8.6 $\mu$m compared to
HD53246 with a flux of 164 mJy, but no other published detections and
so a spectral type cannot be ascribed. As the star is in the galactic
plane ($b=-7.6^\circ$), it is likely to be a background source.
Assuming the same flux as the star at the IRAS wavelengths reduces the
excess emission to 165 mJy and 114 mJy at 12 and 25 $\mu$m
respectively, with significance of 5.5 and 4.4 $\sigma$ respectively,
however the additional uncertainty of having no information on the MSX
source and thus only estimated emission at the IRAS wavelengths means
it is quite possible that the MSX source has higher flux at the IRAS
wavelengths and thus we cannot view the IRAS photometry as evidence of
excess emission.

\emph{HD79873:}
HD79873 has a marginally significant excess at 25 $\mu$m 
but no significant excess at shorter or longer
wavelengths. The 25 $\mu$m excess was 71 mJy at just below 
the 2 $\sigma$ level (Table \ref{sample}). 
This star also has a companion with V band
magnitude of 11 in the Visual Double Star catalogue at a separation of
2\farcs1. (The primary has a corresponding Vmag of 6.5.)  It is not
resolved in 2MASS. 

The primary is detected in the TIMMI2 and VISIR observations at N and Q,
and find levels of emission consistent with the expected photospheric
emission (160 $\pm$ 18 mJy and 39 $\pm$ 9 mJy at N and Q, expecting
167 and 65 mJy from photosphere, see Table \ref{results}). The star
was also observed in the M band filter of TIMMI2, in which the
secondary was detected at the 2.6 $\sigma$ level. The object is offset
by 2\farcs55 $\pm$ 0\farcs25 at position angle -28 $\pm 6 ^\circ$. 
The flux ratio of the primary to the secondary at M is
191 $\pm$ 20.  The N band detection at the location of the secondary
is not significant, at only 1.5 $\sigma$, and the flux limits shown on
the SED of the binary object (in Figure \ref{nodust}, binary plotted
with dashed line and limits with open circles) are those scaled to the
photosphere of the primary using the ratio of fluxes. In the Q band we
find no detection of this object, and place a limit on its emission
accordingly.  The V band magnitude of the binary object and the
assumption that the object is at the same distance as the primary
(68.9 pc) are used to fit the spectral type as K5.  

The photometry of the primary is consistent with photospheric emission
only.  The IRAS excess is at the limits of significance, and once the
secondary emission is taken into account the excess falls to 68 $\pm$
39 mJy, a non-significant level. Thus we attribute the excess detected
in the IRAS observations to the inclusion of the secondary object in
the beam.

\emph{HD123356:}
Detections in the IRAS database of the star HD123356 suggest this
object has excess emission at 12 and 25 $\mu$m of 1270 and 615 mJy
respectively (detections of excess are 24 and 11 $\sigma$ respectively, Table
\ref{sample}). This star has an additional object within 2\farcs5
identified in the WDS and 2MASS catalogues (2MASS 14073401-2104376,
for brevity this shall be called HD123356b in the following discussion).
HD123356b is far brighter in the J, H and K bands (taken from the
2MASS database), although it is fainter in the visual than HD123356 (12.2mag
compared to 10mag).   Sylvester \& Mannings (2000) observed HD123356
at UKIRT using a low resolution spectrometer.  The aperture used for
the UKIRT spectroscopy is 5\farcs5, meaning that the companion object
is at the edge of the measured region.  They found around half the
level of flux that was expected from the IRAS detections. The authors
suggested that all the excess emission may be centered on H123356b.

This source was observed at 10.54 $\mu$m only, and an object detected at
207 $\pm$ 78 mJy (S/N on source excluding calibration uncertainties is
3.2). As this object could not be acquired at M due to saturation
of the filter, the pointing accuracy is reduced to 5-10\arcsec \ here
and so it cannot be confirmed which object was detected. A limit
of 164 mJy can be placed on any undetected sources within the field of
view.  

Given the expected flux from HD123356b from extrapolation of the 2MASS
observations is 681 mJy it
is extremely likely that we observed the secondary
source HD123356b.  The limit placed on undetected objects in this
observation is consistent with the non-detection of the primary. We
show the SED of these two objects in Figure \ref{nodust}, with an M5
NextGen model atmosphere shown as a representative fit to the
secondary, although with so little information available on HD123356b
an identification of its nature cannot be made. The confusion
created by the presence of this object is however the likely source of
the IRAS excesses since although subtracting the M star fit to
HD123356b does not account for all the IRAS flux, it is likely this
source could be a reddened background object and so have higher
infrared flux than is suggested by the M star profile. Otherwise the
excess emission of HD123356 would have to be $L_{IR}/L_\ast = 0.17$, far
brighter than any known debris disk source. 

\emph{HD128400:}
HD128400 has an IRAS excess at above 7 $\sigma$ at 12 $\mu$m of 178
mJy (Table \ref{sample}). Gaidos (1999) suggests an age of 300Myr
based on the star's likely membership of the Ursa Major moving group.  

Poor conditions meant that photometry could not be performed from the
TIMMI2 observations of this object. The star was at 469 $\pm$
41 (92) mJy and 507 $\pm$ 61 (118) mJy at 8.6 and 9.56 $\mu$m
respectively (parentheses indicate inclusion of calibration error).  We
detected no additional sources within the 64\arcsec \, x 48\arcsec \, field
of view of the TIMMI2 instrument.  This limits undetected background
objects to less than 109 mJy at 8.6 $\mu$m.  However, there is an
additional object in the 2MASS catalogue at 83\arcsec \, (2MASS
14421386 - 7508356, in the following discussion this object shall be
called HD128400b for brevity). The source listed in the IRAS Point
Source Catalogue is at a distance of 23\arcsec \, from HD128400.
Pointing errors for this observation are listed as 28\arcsec \, in
major axis, 9\arcsec \, in minor axis, with the major axis at position
angle 117$^\circ$. The 2MASS source is at a position angle of nearly
99$^\circ$, almost exactly along the axis of greatest error. 

Publicly available Spitzer data analysed using the MOPEX package
(Makovoz and Marleau, 2005; Makovoz and Khan, 2005) indicates that
HD128400b emits at a similar level to HD128400 at 24 $\mu$m, with the
primary having a flux of 55 $\pm$ 6 mJy and HD128400b a flux of 87
$\pm$ 9 mJy. At 70 $\mu$m the secondary is detected at a level of 35 $\pm$ 5
mJy, but HD128400 itself is not detected giving an upper limit of 14
mJy.  The emission spectra of HD128400b at $\leq 24\mu$m is best
fitted by a spectral type of M7, implying that level of emission from
HD128400b at 70 microns is much higher than expected (predicted $<$ 1
mJy) and so presumably has its own excess. If the source was a main
sequence star it would be at 3 pc, making it a truly remarkable
object.  However given that it is close to the galactic plane
($b=-13^\circ$) we conclude that it is likely to be a reddened
background object

Given the photometric results presented here, the longer wavelength Spitzer
photometry and the size of the pointing error in the IRAS data, we
believe that confusion with the 2MASS source is the  cause for the
excess identification of HD128400, as is confirmed by IRS spectra
showing photospheric emission only at 12 $\mu$m (Beichman et al. 2008,
in prep.).

\emph{HD10800:}
HD10800 was reported as having an excess at 25$\mu$m in the
IRAS database of 82 mJy (4.5 $\sigma$ detection, see Table
\ref{sample}).  This source was
observed with MIPS by Bryden et al. (2006) at 24 and 70
$\mu$m and no excess found, with a 3$\sigma$ upper limit to excess of
33 and 16 mJy respectively.

Emission centered on the stellar location to within 1\arcsec
of 477 $\pm$ 54 mJy at 11.59 $\mu$m and 186 $\pm$ 29 mJy at 18.72
$\mu$m is detected.  The predicted stellar photosphere at these
wavelengths is 513 and 200 mJy respectively, thus there is no evidence
for excess emission in these observations which place upper limits on
excess of less than 126 mJy at 11.56 $\mu$m, and less than 73 mJy at
Q. The detections and those of Bryden et al. are shown on the SED plot
(Figure \ref{nodust}). Furthermore the results can place limits on
possible background sources of less than 14 mJy in the Q band (39 mJy
in N); the IRAS excess is therefore not due to an unseen companion
within a 19\farcs2 square of the source (field of view of
VISIR). There are no bright 2MASS sources within the pointing errors
of the IRAS observation likely to be the source of the additional IRAS
flux (as for HD53246 or HD128400).  Thus there is no evidence that
this source currently has an associated excess.  It is possible that
this source has evolved in the terrestrial regions since the epoch of
the IRAS observations, and so the emission has disappeared beyond the
detection limits of these observations.  
Alternatively it may be that this object is a statistical anomaly,
as the detection of excess from the IRAS catalogue is at only a
moderately significant level.

%

\section{Discussion}
\label{s:disc}

\subsection{Results summary}
\subsubsection{Hosts of mid-infrared excess}

In this study we have confirmed the presence of warm dust around three
stars, $\eta$ Corvi, HD145263 and HD202406.  The last two of these
sources are young, around a few million years old, and may be still
forming planetary systems, although it is notable that these sources
have relatively low $L_{IR}/L_\ast$ compared to typical T Tauri
stars (Padgett et al. 2006) and so these may be transitory objects
(transitioning between proto-planetary and debris disk stages). 
$\eta$ Corvi, on the other hand, is around
1.3 Gyr old, at an age where we would expect any planetary system to
have finished forming (see e.g. de Pater and Lissauer, 2001).  
For three other sources
we have placed stringent limits on the possible level of any
background/companion object within the fields of view of the
instruments.  From these limits and the photometry of the IRAS
catalogue and published Spitzer data, we have concluded that the
excesses in the mid-infrared, originally determined from the IRAS
catalogue, are highly likely to be centered on the stars for HD12039,
HD69830, and HD191089.  

\subsubsection{Background exclusion and the importance of confirmation}

Five of the sources in the sample turned out to be the result of
source confusion in the IRAS beam. For HD65277, HD79873 
and HD123356 the source could be identified in the TIMMI2 and VISIR images
(albeit without a detection of the primary in the case of HD123356).
For HD53246 and HD128400, the source responsible for the excess
measured in analysis of the IRAS catalogue was $\sim$ 80\arcsec \, away,
and so beyond the field of view of TIMMI2.  These
examples show the dangers of trusting the IRAS catalogue without full
and detailed analysis of all pertinent catalogue data and follow-up
observations. Indeed out of an initial sample of 11 sources believed to be
hosts of mid-infrared excess, only 3 were confirmed in this study, and
a further 2 by other authors. 

HD10800 was shown to have no excess and no other source which is
likely to be responsible for the levels of the IRAS detections.  The
significance of the excess as judged from the IRAS measurements is not
high, at 4.5 $\sigma$.  Thus it is possible that this object never had
an excess and is an illustration of the potential errors to be found
when searching close to the significance limit for excess (Song et
al. 2002).  

The need for confirmation of debris disk candidates has
also been found by Rhee et al. (2007), who combined data from the IRAS
database, the Hipparcos catalogue and the 2MASS catalogue to search
for excess sources, finding a total of 153 sources.  Included in this
paper are 97 sources rejected for reasons including contamination by
additional sources or cirrus, pointing inaccuracy of the IRAS
measurements, and follow-up with Spitzer showing photospheric emission
only. Additional source contamination is a particular issue for sources in
the galactic plane.  In this study HD53246 and HD128400 are in and
near the galactic plane respectively, and have been found to have been
erroneously identified as hosts of debris.  HD155826, identified by
Lisse et al. (2002) as being a bogus disk due to source confusion also
lies in the galactic plane at $b = -0.1$.  Removal of bogus disks is
important when attempting to perform a statistical analysis on disk
populations.  Greaves and Wyatt (2003) include HD128400 as a disk
host. Removing this disk changes their statistics from 4/22 to 3/22 young G
stars hosting a disk (a total of 11/177 G star systems possess a disk
as opposed to their quoted 12/177). Though this is only the removal of
a single disk the sample size involved is not
particularly large, and so the removal of only a few sources can be
significant and the additional uncertainty from bogus disks should be
born in mind when considering statistical studies (such as analysing disk
evolution over time or dependence on stellar spectral type or
environment) needing large samples.  Fortunately the Spitzer Space
Telescope has greater resolution (as illustrated by HD128400) and 
is now providing more reliable large disk samples (see
e.g. Meyer et al. 2006).

\subsubsection{Extension limits}

Our new methods of testing extension limits have quantified how for small
disks the variation and subsequent uncertainty in the PSF will
provide the greatest restrictions in the ability to detect the disk
extension in a particular observation whereas for large disks
detection is limited by the S/N that can be achieved on the disk
(which has decreasing surface brightness with increased angular
size). The optimal size of a disk to be detected (i.e. the disk size
requiring the least bright disk to be detected as an extended source) 
is one with a radius approximately equal to the FWHM of the PSF (for
disks at 18 $\mu$m the FWHM on an 8m telescope $\simeq$ 0\farcs6 which
translates to a disk offset of 12 AU for a systems at 20pc). 

Analysis of the observations presented in this paper including:
comparison of FWHM fits to image profiles; analysis of surface
brightness profiles; and simple subtractions of PSFs (determined from
standard star observations) from science images and examination of the
residuals,  has revealed no evidence for extension around any of the
observed objects.  A new technique of extension limits testing can
give quantifiable constraints on which disk models can be ruled out
and at what level of certainty with such data. 
The extension testing limits have been used to constrain the possible
disk populations of $\eta$ Corvi (see section 5.1). The limits suggest
that model A, in which the mid-infrared emission comes from a single
temperature component is more likely at a 2.6 $\sigma$ level,
however a deeper Q band image should either resolve or rule out the
mid-temperature ($\sim 120$K) component of model B (the three 
temperature fit), as described in detail in section 5.1.  The hot
components of both dust models (at 0\farcs09 and 0\farcs07 for models
A and B respectively) are comparable to the pixel
scales of the detectors of VISIR and MICHELLE (0\farcs075 and
0\farcs099 respectively).  Disks on these scales cannot be resolved
using these single aperture 8m instruments (see 5.1 for further
discussion), and will require the resolving power of an interferometer
to be resolved.

This extension testing method can be applied to future observations of
these and other potential disk sources to determine what limits can be 
placed on unresolved disks.  Furthermore, the predictions of this
modelling, as shown in section 3.2.3, can be used to determine which
sources, with predicted disk flux and radii, will be the most fruitful
sources for imaging with single large-aperture telescopes.  Work
exploring this exciting aspect of the technique is underway and the
results shall be presented in a forthcoming paper (Smith and Wyatt,
in prep.). For now we note that this technique provides more
quantitative limits on the location of dust, and note that the
possibility of detecting extended emission is strongly affected by
whether the dust is confined to a single radius (temperature) or in a
more broad distribution with multiple temperatures. 

\subsection{The nature of mid-infrared excess sources}

Four recent papers have looked at the statistics of mid-infrared
excess around Sun-like stars: Gaidos (1999); Laureijs et al. (2002);
Hines et al. (2006); and Bryden et al. (2006). All of these surveys
found hot emission to occur around 2$\pm$2\% of FGK-type stars,
with Trilling et al. (2008) finding 24 $\mu$m excess around 4
  $^{+2}_{-1.1}$\% of Sun-like stars observed with Spitzer, although
  it is worth noting that these surveys are limited by their
  photometric accuracy and therefore there may be a larger population
  of hot disks that are more tenuous and thus have a fractional luminosity
below the current levels of detectability in these surveys.  The
sample of objects in the study presented in this paper were chosen
deliberately to be the 
objects thought to have excess following analysis of the IRAS
catalogue results, and so does not represent an unbiased sample.  Our
detection rate cannot be compared with these statistical results.  For
any star observed in the survey papers mentioned above and included in
this paper the conclusions regarding the presence of excess emission
are the in agreement with the exception of
HD128400, which was included in the work by Gaidos (1999) as a
positive detection of excess.  As shown in section 5.4, the results show
no evidence for excess, and a nearby 2MASS source is likely to
be the source of confusion in the IRAS results.  This result does not
change the validity of the 2$\pm$2\% statistic however, as for Gaidos
(1999) it reduces the detected excesses to 0/36 (giving a hot emission
occurrence of 0$\pm$3\% from this paper alone).  

Many disks have been observed around T Tauri and Herbig Ae/Be stars
(see e.g. Meeus et al. 2000).  Massive proto-planetary disks have been
observed around stars up to 10Myr (see e.g. Meyer et al. 2007), at
which point the disks rapidly disappear to leave at best a low
fractional luminosity dust belt. The disks of HD145263 and HD202406
lie at an intermediate evolutionary stage, having ages of 9 and 2 Myr
respectively, and exhibit a relatively high
fractional excess (see Table \ref{results}) for debris disks, but
these values are low compared to disks around typical T Tauri stars
(Padgett et al. 2006).  Recent work with Spitzer on clusters of
similar ages to these two sources have indicated that mid-infrared
excess emission may be the result of planet building processes in the
terrestrial region (see e.g. Currie et al. 2007). Fitting the excess
emission of HD145263 and HD202406 with a blackbody suggests that the
dust lies in the terrestrial region, even with a 3 times underestimate
of the dust location for HD145263 (see table \ref{results}). It is
therefore possible that the dust is the result of planet building and
not the evolution of a small Kuiper belt.  Further studies of these
sources may help to elucidate their nature. 

Analytical modelling by Wyatt et al. (2007) has
demonstrated that there exists a maximum fractional excess which can
be expected from a belt of planetesimals in a steady-state collisional
cascade. This is because more massive disks which could potentially
produce more emitting dust grains process themselves more quickly.
The equation given for this prediction is $f_{\rm{max}} =
L_{\rm{IR}}/L_\ast = 0.16 \times 10^{-3} r^{7/3} t_{age}^{-1} $. The
application of this model to the stars with confirmed infrared excess
is shown in the last column of Table \ref{results}.  Within the
uncertainties of this model, a disk with $f_{\rm{IR}} > 1000
f_{\rm{max}}$ is unlikely to be evolving in a steady state collisional
cascade.  Within these limits, HD145263 and HD202406 could be
steady-state disks given their young ages.  Their fractional excesses
are high compared to typical debris disks however and it is possible
these disks are in a transitional phase from proto-planetary to debris
disk (see e.g. Calvet et al. 2005).  As shown in Wyatt et al. (2007),
HD69830 and $\eta$ Corvi (assuming the simple single mid-infrared
component, see later in this section) have excess
emission at a much higher level than would be expected for
collisionally evolving disks given their age and radius, and thus it
is expected that there is a transient source for some of the emitting
material.  

There have been several suggested sources of transient emission put
forward in the literature.  One possible source of this emission would
be the recent collisional destruction of two (or more) massive bodies
\cite{song05}.  In our own asteroid belt
a collision large enough to more than double the emission from the belt 
occurs approximately every 20 million years \cite{durda}.  The recent
analytical modelling of Wyatt et al. (2007) has shown that for the
systems with disks that are assumed to be transient ($f_{obs}/f_{max}
>> 1000$) the single massive collision hypothesis is 
highly unlikely to be able to account for such a massive excess.  
It may be that these systems have recently undergone some dynamical
stirring (orbital migration of a massive planet, recent stellar fly-by
etc.) that has triggered a Late Heavy Bombardment-like period
\cite{gomes}. The Late Heavy Bombardment was a period approximately
3.8-4 Gyr ago when the inner planets of the solar system experienced a
greatly enhanced rate of asteroidal collision, possibly due to the
orbital migration of Jupiter. 
The extreme excess emission found around BD+20 307 (a star possessing
mid-infrared excess not included in this study's sample) is
thought to have come from the excitation of a belt resulting in
massive or frequent collisions \cite{song05}.  As noted by these
authors, this system has an extremely high fractional excess and would
therefore be in an extreme state of collisional destruction. 
The recent sublimation of a massive comet would also produce a
transient peak in infrared excess.  Beichman et al. (2005) have
performed spectroscopy of the HD69830 system. 
The resulting spectra showed marked similarities to the emission
spectra of the Hale-Bopp comet, with several peaks of crystalline
olivine identified. Further work by Lisse et al. (2007) has shown that the
spectra is more similar to that of a disrupted P or D-type asteroid.
Spectral analysis may be the most useful tool to analyse the
possibility of cometary sublimation or asteroid disruption for such
systems.  

However, the transient interpretation 
is highly dependent on the radial location of the dust as can be seen
in the above equation for $f_{max}$, ($f_{max} \propto r^{7/3}$).  
In fitting the photometric results of excess emission we have made
assumptions of grains emitting as blackbodies at a single
temperature.  Such an assumption may lead to
an underestimation of disk size by up to a factor of three, as
emitting grains are typically small and hotter than blackbody (see
e.g. Schneider et al. 2006). Further a more extended dust distribution
could lead to an over-estimation of the disk size by assuming a single
size and temperature for the emitting grains. The uncertainty
remaining in the SED fits of these objects can only be avoided by
direct observational confirmation of the size of the emitting region.
The example of $\eta$ Corvi is an appropriate illustration of this
issue.  In model A the mid-infrared emission cannot be explained by a
steady-state evolution (see discussion above and Table \ref{results}).
In model B the
hot dust component at 360K is also likely to be transient, however the
mid-temperature component at 120K (12 AU) can be explained by a
collisionally evolving disk at 12 AU (Table \ref{results}).  Indeed
this population of dust lies in an appropriate location to be a
possible parent planetesimal belt to the hot dust emission according
to Figure 4 of Wyatt et al. (2007). This would require a radial
transport mechanism that would move the dust from the 12 AU belt to a
1.3 AU location, which is not well modelled or understood, but could
be analogous to the inward scattering of planetesimal material into
the terrestrial regions during the Late Heavy Bombardment period
initiated by the resonance crossing of Jupiter and Saturn (see Wyatt et
al 2007 discussion for elaboration on this scenario).  Thus the
model with a total of three components (model B) could represent two
steady state populations (the 12 AU mid-infrared component and the
large sub-mm disk) and a transient component, the source of which is
currently unknown.  This model is ruled out by current data at the
2.6$\sigma$ level, although it is important to note that whichever
model we adopt for $\eta$ Corvi there is a transient component, and so
we are unlikely to find a fit to this system which does not require
some transitory contribution to the excess emission. However, as 
highlighted in this study, a confirmed radial location is key
to understanding the nature of this system, and the hot dust
populations as a whole.

%

\section{Conclusions}
\label{s:conc}

We have presented an observing programme focussing on main sequence
F, G and K stars purported to have infrared excess.  The findings can
be summarised as follows:
\begin{itemize}
\item We have confirmed the excess emission to be both real and
  centred on the star for 3 objects, all of which have excess emission
  within the terrestrial regions as fitted by SED modelling.  Two of
  these objects are believed to be pre-main sequence stars. 
\item For 5 further objects, the dust was found to be from a
  companion/background source, and not associated with the star.  This
  demonstrates the importance of high resolution imaging as a tool to
  confirm IRAS sources. 
\item One object was found to have no associated excess nor any object
  nearby likely to be responsible for the levels of flux in the IRAS
  measurements.
\item Our new method of testing extension limits has enabled us to
  place limits on the radial extent of some disk populations and shown
  that for some others, single aperture imaging with current 8m-class
  telescopes will not be able to resolve the extent of the disk. 
\item The extension limits testing suggests a fit to the $\eta$ Corvi
  emission spectrum using a single mid-infrared component at around
  320K and a cool component at 40K is more likely to represent the
  true dust distribution than a fit using two mid-infrared components
  at 360K and 120K, together with the cool 40K dust at the 2.6
  $\sigma$ level (or lower or higher significance depending on the
  geometry of the dust belts).  
\end{itemize}

Sources of hot dust emission fall into distinct groupings.  Either the
sources are young and possibly transitional, in which case the dust
can be primordial,  or the result of steady-state evolution
(e.g. HD145263 and HD202406), or they 
are old and sources of transient emission ($\eta$ Corvi and HD69380),
or they are old and have relatively low radius steady-state
planetesimal belt (HD12039 and HD191089, and possibly the
mid-temperature component of $\eta$ Corvi).  

The rare hot dust in main sequence systems may be transient as
suggested by comparison to collisional modelling.
However, uncertainties inherent from the SED modelling
process mean that only by resolving the location of the emitting
region can we deprive these systems of their enigmatic status.  Our
new method of extension testing allows us to constrain dust locations
much more tightly than a simple comparison with the PSF. 
Application of these techniques to further observations and
other sources is one way to determine the radial extent of the
dust emission and thus begin to determine the nature of these hot
dust sources.

\begin{acknowledgements}
      RS is grateful for the support of a PPARC studentship.  MCW is
      grateful for the support of a Royal Society University
      Fellowship. The authors are grateful to Ben Zuckerman for
      directing our attention to the Spitzer observations of
      HD128400, and to Geoff Bryden for fruitful discussions on these
      observations. Furthermore the authors wish to extend their
      thanks to Christine Chen for providing the IRS spectra
      of $\eta$ Corvi, and to Chas Beichman for the spectra of
      HD69830.  Based on observations made with ESO
      Telescopes at the La Silla and Paranal Observatories under
      programme IDs 71.C-0312, 72.C-0041 and 74.C-0700. Also based on
      observations obtained at the Gemini Observatory, which is
      operated by the Association of Universities for Research in
      Astronomy, Inc., under a cooperative agreement with the NSF on
      behalf of the Gemini partnership: the National Science
      Foundation (United States), the Particle Physics and Astronomy
      Research Council (United Kingdom), the National Research Council
      (Canada), CONICYT (Chile), the Australian Research Council
      (Australia), CNPq (Brazil) and CONICET (Argentina).  
\end{acknowledgements}

\end{document}